\documentclass[11pt]{article}
\pdfoutput=1

\usepackage{jheppub} 

\usepackage{verbatim} 
\usepackage{float}
\restylefloat{table}
\usepackage[all]{xy}
\usepackage{pdflscape}
\usepackage{tikz}
\usepackage{array}
\usepackage{youngtab}
\usepackage{multirow}
\usepackage{color}
\usepackage{ulem}
\usepackage{cleveref}

\makeatletter

\newcommand{\be}{\begin{equation}}
\newcommand{\ee}{\end{equation}}
\newcommand{\ba}{\begin{aligned}}
\newcommand{\ea}{\end{aligned}}

\newcommand{\bea}{\begin{eqnarray}}
\newcommand{\eea}{\end{eqnarray}}

\def\Tr{\mathop{\mathrm{Tr}}\nolimits}
\def\tr{\mathop{\mathrm{tr}}\nolimits}

\def\unit{{1\kern-.65ex {\rm l}}}
\def\1{{1\kern-.65ex {\rm l}}}

\makeatother

\begin{document}

\numberwithin{equation}{section}  



\vspace*{0.8cm} 
\begin{center}
 {\LARGE Chiral 2d Theories  from $N=4$ SYM with Varying Coupling \\}

 \vspace*{1.8cm}
{Craig Lawrie$\,^1$, Sakura Sch\"afer-Nameki$\,^2$, and Timo Weigand$\,^1$}\\

 \vspace*{1.2cm} 
{\it $^1$ Institut f\"ur Theoretische Physik, Ruprecht-Karls-Universit\"at,\\
 Philosophenweg 19, 69120 Heidelberg, Germany }\\
 {\tt {gmail:$\,$craig.lawrie1729}}\\
 {\tt t.weigand\phantom{@}thphys.uni-heidelberg.de}\\
  
\bigskip
{\it $^2$ Mathematical Institute, University of Oxford \\
 Woodstock Road, Oxford, OX2 6GG, UK}\\
  {\tt {gmail:$\,$sakura.schafer.nameki}}\\
\vspace*{0.8cm}
\end{center}
\vspace*{.5cm}
%
\noindent 
We study 2d chiral theories arising from 4d $N=4$ Super-Yang Mills (SYM) with varying coupling $\tau$. 
The 2d theory is obtained by dimensional reduction of $N=4$ SYM on a complex curve with a partial topological twist that accounts for the non-constant $\tau$. 
The resulting 2d theories can preserve $(0,n)$ with $n=2, 4, {6}, 8$ chiral supersymmetry, and have a natural realization in terms of strings from wrapped D3-branes in F-theory.  
We determine the twisted dimensional reduction, as well as the spectrum and anomaly polynomials of the resulting strings in various dimensions. 
We complement this by considering the dual M-theory configurations, which can
either be realized in terms of M5-branes wrapped on complex surfaces, or
M2-branes on curves that result in 1d supersymmetric quantum mechanics.


\pagenumbering{gobble}
\newpage
\pagenumbering{arabic}

\tableofcontents

\section{Introduction}

Two-dimensional chiral supersymmetric and superconformal theories, or strings, have enjoyed revived interest in recent years. A useful starting point for generating new theories of this type are compactifications of higher-dimensional superconformal theories  in a suitable background. 
For instance 4d  $N=2$ or $N=4$ SYM on a complex curve with a partial topological twist have been studied in  \cite{Bershadsky:1995qy,Benini:2013cda,Benini:2015bwz}. In this paper we provide a new class of such 2d theories obtained from $N=4$ SYM by compactification along a curve where the complexified coupling $\tau$ varies. 
The resulting 2d theories have chiral supersymmetry, and in the limit of the curve volume going to zero size, they give rise to new superconformal theories. The goal of this paper is to present a setup where such theories can be naturally studied and classified, and to derive the dimensional reductions including the spectra.

There are several topological twists of $N=4$ SYM with constant coupling, known as Vafa-Witten \cite{Vafa:1994tf}, Geometric-Langlands \cite{Kapustin:2006pk} and half-twist \cite{Yamron:1988qc}. Allowing the coupling to vary requires changing the background for the $N=4$ SYM theory. This can be realized, whilst preserving supersymmetry, by the so-called  topological duality twist, introduced in \cite{Martucci:2014ema} for the abelian case and generalized in a non-abelian manner in \cite{Assel:2016wcr}.  This twist combines both an R-symmetry background field, and one for the so-called bonus-symmetry $U(1)_D$
\cite{Intriligator:1998ig,Intriligator:1999ff, Kapustin:2006pk} of the abelian $N=4$ SYM theory.

As often, such non-trivial backgrounds for supersymmetric gauge theories have a brane realization. 
As we consider $N=4$ SYM with varying coupling, the natural setting are D3-branes in IIB backgrounds where the axio-dilaton, which is identified with $\tau$ on the brane world-volume, varies.  Such backgrounds go by the name F-theory \cite{Vafa:1996xn,Morrison:1996na,Morrison:1996pp}. In fact we will show that F-theory on elliptically fibered Calabi-Yau spaces provides supersymmetric backgrounds for the $N=4$ SYM theory which preserve chiral supersymmetry in 2d. 
There are several possibilities for twisting the $N=4$ SYM theory including the $U(1)_D$ bonus-symmetry, each of which will have an interpretation in terms of an F-theory background.

The compactification space of F-theory is an elliptic Calabi-Yau manifold,
  with the complex structure of the elliptic fiber representing the
  axio-dilaton of IIB string theory.  The gauge sector is realized in terms of
  7-branes. In addition, D3-branes can be mutually supersymmetric and provide
  a new and interesting sector of the theory.  The $SL(2,\mathbb Z)$ symmetry
  of Type IIB theory is identified with the Montonen-Olive duality of the
  $N=4$ SYM theory on the worldvolume of the D3-brane,
  acting on the complexified coupling constant $\tau$.  
  In this paper we will study D3-branes which are wrapped on complex curves
  $C$, and in the low energy limit thus correspond to strings in $d$
  dimensions.  As the curve is part of the base of the elliptic fibration, the
  D3-coupling  $\tau$ indeed varies over $C$, as long as $C$ has
  transversal intersections with the 7-branes, which source the
$\tau$-monodromy.  Such strings not only result in new 2d chirally
supersymmetric theories, which become conformal as the volume of $C$ vanishes,
but also form important additional sectors in F-theory compactifications.

Specifically in compactifications to 2d, the D3-brane sector is crucial for the consistency of the F-theory
vacuum.  Indeed, recently 2d string compactifications have
attracted renewed interest
\cite{Franco:2015tna,Franco:2015tya,Schafer-Nameki:2016cfr,Franco:2016nwv,Apruzzi:2016iac,Franco:2016qxh,Apruzzi:2016nfr}.
In the F-theory realization of 2d $(0,2)$ string vacua on Calabi-Yau
five-folds \cite{Schafer-Nameki:2016cfr, Apruzzi:2016iac}, consistency of
the theory forces the introduction of space-filling D3-branes wrapping a
curve in the compactification manifold. Thus understanding the D3 sector is
paramount to fully characterizing such chiral 2d string vacua.

In higher dimensions, strings usually play the role of additional sectors of
the theory which pinpoint loci in the moduli space with interesting physics.
The prime example for this are strings in 6d $N=(1,0)$ theories, which
have recently seen a resurgence of interest, following the seminal paper
\cite{Heckman:2013pva}. Supersymmetric self-dual strings are an important
subsector of these theories
\cite{Ganor:1996mu,Seiberg:1996vs,Witten:1996qb,Klemm:1996hh}: In their
tensionless limit, corresponding to wrapped curves of zero volume, they are
indicators of superconformal invariance of the underlying 6d theories.
Multiple examples of strings in 6d have been investigated from various
points of view, including \cite{Minahan:1998vr,Haghighat:2013gba,
Kim:2014dza,Haghighat:2014vxa,Gadde:2015tra,Haghighat:2015ega,DelZotto:2016pvm} and
references therein.

Strings in 4d $N=1$ compactifications have been comparatively less explored. In 4d, strings are dual to instantons. Again when they become tensionless, interesting physics can be expected in the 4d theories. However, unlike in 6d, these are not BPS objects since the string vacuum only preserves four real supercharges. Thus their tension is a priori not protected and may receive, albeit possibly small, quantum corrections \cite{Mayr:1996sh}. Nevertheless they are an inevitable part of the rich landscape of 4d string compactifications, which deserve further study.

The strings we consider in this work preserve $(0,p)$
supersymmetry along their worldvolume, with the value of $p$ depending on the
dimensionality of the transverse space as summarized in table \ref{tab:SUSYS}.
We will determine the string action, as well as the BPS equations and
zero-mode spectrum in each dimension and utilize these to study anomaly
cancellation on the strings.  For strings from single D3-branes, as studied in
this paper, the variation of $\tau$ has the interesting effect of breaking
the $U(1)$ gauge symmetry in the 2d effective field theory along the string.
This is in fact counter to naive expectations from the weakly coupled Type IIB
limit, which we describe in more detail  in section \ref{IIBorientifolds}.
The breaking of the abelian gauge group is owed to non-perturbative dynamics localised in the vicinity of
the orientifold plane. Such effects induce a quantum Higgsing of the
perturbative $U(1)$ in F-theory. As we will explain, the order parameter of
the quantum Higgsing corresponds to the distance between the two mutually
non-local 7-branes into which the orientifold plane splits in F-theory.
Another interesting aspect of the duality twist which we will find is that it
considerably modifies the BPS equations, and hence the Hitchin system, along
the D3-brane, already in the abelian case.  The general solutions of the gauge
field components along $C$ can be viewed as duality-twisted flat connections on $C$. It would be interesting to study the properties of
this duality-twisted Hitchin system further, including the non-abelian
formulation.

The duality twist as defined in \cite{Martucci:2014ema} has the shortcoming
that it only strictly applies to the abelian $N=4$ SYM theory, as the $U(1)_D$
does not survive in the non-abelian generalization
\cite{Intriligator:1998ig,Intriligator:1999ff}. Nevertheless, the expectation
is that a stack of D3-branes wrapped on the base of an F-theory
compactification should be a consistent setup and thus an analogue of the topological twist
should exist also in the non-abelian theory. It was subsequently pointed out
in \cite{Assel:2016wcr} and shown that the duality twist in fact, once mapped
to M-theory via M/F duality, becomes a standard (geometric) topological twist
of the M5-brane theory and a non-abelian generalization of this was proposed.
The D3-brane wrapped on a cycle in the base of the Calabi-Yau maps to an
M5-brane wrapping in addition the elliptic fiber of the Calabi-Yau, and the
$U(1)_D$ becomes the symmetry associated to the elliptic fibration. This
topological twist has a clear generalization to the non-abelian case and for
the D3-branes wrapping four-cycle in the base of the Calabi-Yau this was
presented in \cite{Assel:2016wcr}. 

For strings the M-theory dual description also provides a way to first of
all test the duality twist, in addition to
providing a generalization to non-abelian
strings. There are two options how to dualize the D3-brane, either to an M5
or an M2. We will utilize both points of view and show agreement between the
resulting theories.  The M2-brane approach leads to a 1d
supersymmetric Quantum Mechanics (SQM) theory, which is the dimensional
reduction of the string on a circle. These theories have so far received
comparatively little attention, but recently for M2-branes on curves in K3
they have been studied in \cite{Okazaki:2014sga, Okazaki:2015pfa}.  The BPS
spectrum which we find on M2 branes wrapping a curve $C$ in the base of a
general elliptic fibration is matched with the spectrum on the dual D3-brane
setup. As we will explain, this makes use of an interesting speciality of
supersymmetric theories in one dimension known as automorphic duality
\cite{Gates:2002bc,Bellucci:2005xn}.  Again, similar to the M5-brane
approach, this point of view allows for a
generalization to the BLG theory \cite{Gustavsson:2007vu, Bagger:2007jr},
and thus a non-abelian version of the dimensionally reduced string. 

So far we have only discussed the D3-brane in terms of the $N=4$ SYM degrees
of freedom. However, the key players of a generic F-theory compactification,
the 7-branes, intersect the D3-brane world-volume in complex codimension-one
loci. From the point of view of the $N=4$ SYM coupling $\tau$, the 7-branes
are the loci where the coupling diverges. For definiteness let us assume the
F-theory fibration has a section, and thus a presentation in terms of a
Weierstrass model $y^2 = x^3 + fx + g$. The so-defined variety is singular
along $\Delta=0$ in the base of the fibration, where $\Delta = 4 f^3 + 27 g^2$
is the discriminant. The discriminant characterizes precisely the loci where
the 7-branes wrap the base of the fibration. The intersection of the D3-branes
with $\Delta=0$ are, from the point of view of the $N=4$ SYM theory, loci
where $\tau$ undergoes monodromy in $SL(2,\mathbb{Z})$. In fact the loci are
so-called duality defects, which have been studied in \cite{Martucci:2014ema,
Assel:2016wcr} for the D3-brane with duality twist wrapping a surface in the
base of the  elliptic fibration. In that case the duality defects are 2d with chiral  $3$--$7$ strings localized on the defect, and as observed in \cite{Assel:2016wcr} these duality defects themselves intersect along points. 

In the present case where the D3-brane wraps a curve $C$ in the base of the Calabi-Yau, the duality defects are point-like in $C$, but again there are degrees of freedom localized on them, which correspond to $3$--$7$ strings. 
These give rise to
a number of left-moving degrees of freedom, 
which we show is {\it universally} given in terms of 
\be\label{eqn:37contribution}
n_{37}= 8 \, \text{deg}(\mathcal{L}_D) = 8\, c_1(B_{n-1}) \cdot C \,,
\ee
where $\mathcal{L}_D$ is the line bundle associated to the duality $U(1)_D$.
This contribution can be corroborated in terms of the dual M5-brane picture.
Directly counting these strings, however, is in general a formidable task,
which we address only in passing, but the technology for determining the
spectra is readily available, see for instance \cite{Gaberdiel:1997ud,
DeWolfe:1998zf, Mikhailov:1998bx,Grassi:2014ffa,Grassi:2016bhs}. For Type
IIB orientifolds we provide a discussion of this sector in section
\ref{IIBorientifolds}. The general anomaly polynomial, which we analyse in
section \ref{sec:anomalies}, places further interesting constraints on the
spectrum of $3$--$7$ strings. Specifically, demanding that the anomaly
  polynomial arising via inflow from the bulk in which the string propagates is
  cancelled by the spectrum on the string requires that there are additional
  states contributing as in (\ref{eqn:37contribution}).

An important off-spring of the present work is a full characterization of the gravitational anomalies along the string theories in various dimensions.
As noted already, in the 2d setting of \cite{Schafer-Nameki:2016cfr,Apruzzi:2016iac} the addition of extra D3-branes is in fact imperative due to tadpole constraints as the strings are spacetime-filling. 
Tadpole cancellation in turn guarantees that the complete spectrum, consisting
of the 7-brane sector, the 2d $(0,2)$ supergravity sector and the D3-brane
sector, is anomaly free. For the gauge anomalies this has already been
investigated and related to M-theory Chern-Simons terms in
\cite{Schafer-Nameki:2016cfr}, but in absence of a complete understanding of
the D3-brane sector no such verification was possible. In \cite{MINI} we
combine this article's results on the D3-brane sector in 2d $(0,2)$ theories
together with a derivation of the supergravity spectrum to show cancellation
of all gravitational anomalies in F-theory compactifications on Calabi-Yau five-folds.

Finally, it is of considerable interest to explore the theory on the strings from wrapped D3-branes as 2d super-conformal field theories by themselves. 
The central charges can be computed for each given base and would yield interesting new holographic setups in IIB with varying axio-dilaton. Other computations, which naturally have a starting point with the results in this paper are the computation of elliptic genera for these theories as already started in \cite{Haghighat:2015ega} for the 6d case, but looking ahead for the $N=(0,2)$ theories, the starting points would be the general expression in \cite{Benini:2013xpa}.

For ease of locating the results the various spectra can be found in the
following locations in the paper.  An overview of the general setup for
strings from D3, M5, and M2-branes is given in section \ref{sec:Overview}.
The D3-branes with topological duality twist along $C$ are
discussed in section \ref{sec:StringsfromD3}. The spectra in 6d, 4d and
2d are summarized in the tables \ref{tbl:CY3fcmultiplets}, \ref{eqn:CY4D3zm}
and \ref{tab:CY5D3zm}, respectively. The 8d case is special in that the
D3-brane wraps the full base of the elliptic K3, and can be found in appendix
\ref{app:K3}. The
remainder of the paper then discusses the M5 and M2-duals in section
\ref{sec:M5M2}. The anomalies of the strings in all dimensions can be found in
section \ref{sec:anomalies}. Appendices providing shelter for conventions and
details of the string actions as well as cohomological computations can be
found at the end of the paper.


\section{Duality Twists of $N=4$ SYM and Brane Realizations} 
\label{sec:Overview}

In this section, we study 4d $N=4$ SYM on a curve $C$ along which the coupling constant $\tau$ varies,
and determine the possible duality twists which retain supersymmetry in the non-compact two dimensions. 
We then present three brane-setups related to this, in terms of D3-branes, M5-branes and M2-branes, respectively.


\subsection{Duality Twisted $N=4$ SYM}
\label{sec:D3sector}

We determine all possible duality twisted theories and the supersymmetry that they preserve in the non-compact two dimensions. For constant coupling the analog of this analysis has been performed in \cite{Benini:2013cda}. 
For $N=4$ SYM with varying coupling, the twists have to include the structure group $U(1)_C$ along the curve, an R-symmetry factor $U(1)_R \subset SU(4)_R$ and the duality (or `bonus symmetry') $U(1)_D$ of the abelian $N=4$ SYM theory \cite{Intriligator:1998ig,Intriligator:1999ff}. 
Indeed, the variation of $\tau$ can be understood in terms of a non-trivial line bundle whose structure group $U(1)_D$ forms a bonus symmetry of the underlying abelian $N = 4$ SYM theory. 
If under the $SL(2,\mathbb Z)$ transformation the complexified gauge coupling 
\begin{equation}
  \tau = \frac{\theta}{2\pi} + \frac{4 \pi i }{g^2}  = \tau_1 + i \tau_2 
\end{equation}
is mapped to 
\be
\tau \ \rightarrow \ \frac{a \tau + b}{c \tau + d} \,,\qquad ad-bc= 1\,,\qquad a,b,c,d \in \mathbb{Z} \,,
\ee
 then a  field of $U(1)_D$ charge $q$ transforms as \cite{Kapustin:2006pk}
\begin{equation} \label{dualitychargeq}
  \Phi \rightarrow e^{i q  \alpha} \,   \Phi \qquad {\rm with} \qquad e^{i
  \alpha} = \frac{c \tau + d}{|c \tau + d|} \,.
\end{equation}
We shall briefly recap the spectrum of $N=4$ SYM theory in four
dimensions, including the charges of the fields under the $U(1)_D$ symmetry.
The field content consists of a vector multiplet containing a gauge field
$A_\mu$, six scalars $\phi_i$, and fermions $\Psi_\alpha^I$ and
$\widetilde{\Psi}_{\dot{\alpha}I}$. The indices transform in various different
representations of the symmetry group $SO(1,3)_L \times SU(4)_R$: $\mu$ in the
vector of $SO(1,3)_L$, $\alpha$ ($\dot{\alpha}$) in the fundamental of the
left (right) $SU(2)$ of $SO(1,3)_L$, $i$ in the ${\bf 6}$ of $SU(4)_R$, and
finally $I$ in the ${\bf 4}$ or ${\bf \overline{4}}$ of $SU(4)_R$.  The  field
content transforms under the total symmetry group
\begin{equation}
  G_{\rm total} = SO(1,3)_L \times SU(4)_R \times U(1)_D \,,
\end{equation}
as  
\begin{align} \label{N4content}
  A_\mu \,&:\, {\bf (2,2,1)}_{*} & \phi_i \,&:\, {\bf (1,1,6)}_0 & 
  \Psi_\alpha^I \,&:\, {\bf (2,1,4)}_1 & \widetilde{\Psi}_{\dot{\alpha}I} \,&:\,
  {\bf (1,2,\overline{4})}_{-1} \,,
\end{align}
where the subscript indicates the $U(1)_D$ representation. 
The gauge field does not itself form a $U(1)_D$ eigenstate and thus has no
well-defined $U(1)_D$ charge, but the self-dual and
anti-self-dual components of the field strength have definite $U(1)_D$ charges
\begin{equation}
  \begin{aligned}
q_D=+2:\qquad     \sqrt{\tau_2} F^+ &= - \frac{i}{2\sqrt{\tau_2}} (F_D - \bar{\tau}F) \cr
q_D=-2:\qquad     \sqrt{\tau_2} F^- &= \frac{i}{2\sqrt{\tau_2}} (F_D - {\tau}F) \,,
  \end{aligned}
\end{equation}
where $F_D$ is the duality transformed field strength
\be \label{Fdual1}
F_D = \tau_1 F + i \tau_2 \star F \,.
\ee
The sixteen supersymmetries $Q_{\alpha I}$ and
$\widetilde{Q}_{\dot{\alpha}}^I$ transform under $G_{\rm total}$ as  \cite{Kapustin:2006pk}
\begin{align}\label{eqn:QN4}
    Q_{\alpha I} \,&:\, {\bf (2,1,\overline{4})}_{1} &
    \widetilde{Q}_{\dot{\alpha}}^I \,&:\, {\bf (1,2,4)}_{-1} \,,
  \end{align}
and, equivalently, the supersymmetry transformation parameters are given by
 \begin{align} \label{epsilon4dN4}
    \epsilon_{\alpha I} \,&:\, {\bf (2,1,{4})}_{-1} &
    \widetilde{\epsilon}_{\dot{\alpha}}^I \,&:\, {\bf (1,2,\overline{4})}_{1} \,.
  \end{align}

Associated to the $U(1)_D$ symmetry is the complex line bundle $\mathcal{L}_D$
defined over the subspace of spacetime, $C$, where $\tau$ varies. 
A field of $U(1)_D$ charge $q$ transforms
as a section of the $q$-th power of this $U(1)_D$ bundle $\mathcal{L}_D$.
The one-form connection on the bundle is given by
\begin{equation} \label{Lconnection1}
   {\cal A} = \frac{d \tau_1}{2 \tau_2} \,.
\end{equation}
The connection can be decomposed into $(0,1)$ and $(1,0)$ forms on $C$ as
\begin{equation}
  \mathcal{A} = \mathcal{A}^{(0,1)} + \mathcal{A}^{(1,0)} \,,
\end{equation}
where these forms define\footnote{We use an identical notation in
  $\mathcal{L}_D$ for both the complex and holomorphic line bundles merely to
prevent a proliferation of notation. Where not clarified the particular line
bundle of interest at any point should be unambiguous by context.}
respectively a holomorphic line bundle, $\mathcal{L}_D$, and an
anti-holomorphic line bundle, $\bar{\mathcal{L}}_D$. To write expressions in
an appropriately covariant way we shall also need to introduce covariant
derivatives, on $C$, with respect to the connections on these $U(1)_D$ bundles. Given a
field with $U(1)_D$ charge $q_D$ these are defined via
\begin{equation}\label{eqn:covderivs}
  d_{\cal A} = d + i q_D^{\rm twist} {\cal A} = 
  \frac{1}{2}(\partial + i q_D {\cal A}^{(1,0)}) + 
  \frac{1}{2}(\bar\partial + i q_D {\cal A}^{(0,1)}) 
  = \frac{1}{2} (\partial_{\cal A} + \bar \partial_{\cal A}) \,,
\end{equation}
where $\partial$, $\bar{\partial}$ are the standard holomorphic and
anti-holomorphic derivatives on $C$.

Since the 4d spacetime of the $N=4$ SYM theory is the product space  $\mathbb{R}^{1,1}
\times C$,  the $SO(1,3)_L$ Lorentz symmetry is broken to
$SO(1,1) \times U(1)_C$. The relevant representations decompose as
\begin{align}\label{eqn:LorentzDecomp}
  SO(1,3)_L &\ \rightarrow\  SO(1,1) \times U(1)_C \cr
  {\bf (2,2)} &\ \rightarrow\  {\bf 1}_{2,0} \oplus {\bf 1}_{-2,0} \oplus{\bf
  1}_{0,2} \oplus{\bf 1}_{0,-2} \cr
  {\bf (2,1)} &\ \rightarrow\  {\bf 1}_{1,1} \oplus {\bf 1}_{-1,-1} \cr
  {\bf (1,2)} &\ \rightarrow\  {\bf 1}_{1,-1} \oplus {\bf 1}_{-1,1} \,.
\end{align}
The topological twist requires to turn on a $U(1)$ R-symmetry background gauge field. We are therefore interested in decompositions of the R-symmetry $SU(4)_R$ of $N=4$ SYM which contain a $U(1)$ factor. These are given by the following decompositions:
\begin{align}
  &\left. \begin{aligned}
   \label{CY3RSym}
    SU(4)_R &\ \rightarrow\  SO(4)_T \times \underline{U(1)_R} \cr
    {\bf 4} &\ \rightarrow \ {\bf (2,1)}_1 \oplus {\bf (1,2)}_{-1} \cr
    {\bf 6} &\ \rightarrow \ {\bf (1,1)}_2 \oplus {\bf (1,1)}_{-2} \oplus {\bf
    (2,2)}_0
  \end{aligned} \,\,\right\rbrace \text{CY$_3$ Duality-Twist}\\
  &\left. \begin{aligned}
  \label{CY4RSym}
    SU(4)_R &\ \rightarrow\  SU(2)_R \times \underline{U(1)_R} \times SO(2)_T \cr
    {\bf 4} &\ \rightarrow\  {\bf 2}_{0,1} \oplus {\bf 1}_{1,-1} \oplus {\bf
    1}_{-1,-1} \cr
    {\bf 6} &\ \rightarrow\  {\bf 1}_{0,2} \oplus {\bf 1}_{0,-2} \oplus {\bf
    2}_{1,0} \oplus {\bf 2}_{-1,0}
  \end{aligned} \right\rbrace \text{CY$_4$ Duality-Twist} \\
  &\left. \begin{aligned}
   \label{CY5RSym}
    SU(4)_R &\ \rightarrow\  SU(3)_R \times \underline{U(1)_R} \cr
    {\bf 4} &\ \rightarrow\  {\bf 1}_3 \oplus {\bf 3}_{-1} \cr
    {\bf 6} &\ \rightarrow\  {\bf 3}_2 \oplus {\bf 3}_{-2} 
  \end{aligned} \qquad\qquad\,\,\, \right\rbrace \text{CY$_5$ Duality-Twist} \,.
\end{align}
 As we will show later on, these particular decompositions are those induced when the curve $C$ is a
  complex curve in the base of a Calabi-Yau $n$-fold in F-theory. The case where the R-symmetry remains unbroken $SU(4)_R$ corresponds to  compactification on CY$_2$, i.e. K3, and is discussed in appendix \ref{app:K3}. 
  {There is an additional case where the $SU(4)_R$ breaks to $U(1)^3$, with decomposition 
\begin{align}
   \left.\begin{aligned}
 \label{CY5PrimeRSym}
    SU(4)_R &\ \rightarrow\  U(1)_1 \times U(1)_2 \times U(1)_3 \cr
    {\bf 4} &\ \rightarrow\  {\bf 1}_{1,1,1,} \oplus {\bf 1}_{-1,1,1}\oplus {\bf 1}_{1,-1,1}\oplus {\bf 1}_{1,1,-1}\cr
    {\bf 6} &\ \rightarrow\   {\bf 1}_{\pm 2, 0, 0}\oplus {\bf 1}_{0, \pm2, 0}\oplus {\bf 1}_{0, 0, \pm2} 
  \end{aligned} \quad  \right\rbrace \text{CY$_5$' Duality-Twist}
\end{align}
}
The standard topological twist combines the R-symmetry $U(1)_R$ in one of the decompositions   in (\ref{CY3RSym}) - (\ref{CY5RSym}) with the internal Lorentz symmetry $U(1)_C$ in (\ref{eqn:LorentzDecomp}).
As the supercharges  transform non-trivially under the $U(1)_D$ symmetry it is necessary to combine  this standard topological twist of $U(1)_C$ with $U(1)_R$ with an additional one including 
 $U(1)_D$, as follows from (\ref{eqn:QN4}). This 
is the duality twist, discussed in \cite{Martucci:2014ema} for
Euclidean D3-branes wrapping a K\"ahler surface on a base $B_3$ and generalized to non-abelian theories in \cite{Assel:2016wcr}. Throughout we shall twist $U(1)_D$ with the same $U(1)_R$, as we have twisted  $U(1)_C$ with. 
The resulting theories then preserve chiral supersymmetry in 2d as summarized in table 
\ref{tab:SUSYS}. 
The twist in (\ref{CY5PrimeRSym}) with the R-symmetry taken as the diagonal $U(1)_{\rm diag} = U(1)_R$ results also in a $(0,2)$ theory. 
Finally, we should note that the decomposition in (\ref{CY5RSym}) can in addition give rise to an $N= (0,6)$ in 2d, which will be discussed in detail in section \ref{sec:Nis6}.

As pointed out above, the 4d vector field $A_\mu$, which is insensitive to the R-symmetry and so
will have the same decomposition regardless of the dimension of the
compactification space, does not have a single
well-defined $U(1)_D$ charge. We shall write the decompostion of this vector
under the 4d Lorentz and duality group as
\begin{equation}\label{eqn:4dvecdecomp}
  \begin{aligned}
SO(1,3)_L \times U(1)_D&\quad  \rightarrow \quad SO(1,1)_L \times U(1)_C \times U(1)_D  \cr 
  {\bf (2,2)}_* &\quad \rightarrow\quad  {\bf 1}_{2,0,*} \oplus {\bf
  1}_{-2,0,*} \oplus{\bf
    1}_{0,2,*} \oplus{\bf 1}_{0,-2,*} \,,
  \end{aligned}
\end{equation}
and label the four resulting fields respectively as $v_+$, $v_-$, $\bar{a}$,
and $a$. We shall identify these fields in terms of the components of $A_\mu$
by determining which combinations of components have the same charges as
listed. Further we will find that the twisted supersymmetry variations, which
relate bosonic fields of unspecified $U(1)_D$ charge to fermionic
fields of known $U(1)_D$ charge, require that the objects 
\begin{equation}
  \sqrt{\tau_2} a \quad \text{ and } \quad \sqrt{\tau_2}\bar{a} \,,
\end{equation} 
do indeed have a precise $U(1)_D$ charge. The $v_\pm$ fields do
not have this feature, however, they do not give rise to any massless fields
in the compactification to two dimensions, and so the ambiguity in the charge
of $A_\mu$ does not translate into an ambiguity in the field content of the
low energy theory.

\begin{table}
$$
  \begin{array}{|c|c|c|c|c|} 
\hline
      {\rm Spacetime \ dim\ } d & 8 & 6 & 4 & 2 \cr\hline
    \text{CY}_{n} & 2 & 3 & 4 & 5 \cr\hline
    \text{2d \ supersymmetry} & (0,8) & (0,4) & (0,2) & (0,2) \cr\hline
    \text{1d\  supersymmetry} & 8 & 4 & 2 & 2 \cr \hline
  \end{array} \label{nSUSYM2}
$$
 \caption{ The number of supersymmetries preserved by 4d $N=4$ SYM on $C\times
   \mathbb{R}^{1,1}$ with duality twist as defined in (\ref{CY3RSym}) -- (\ref{CY5RSym}). 
The brane realization of this setup corresponds to D3-branes wrapping a holomorphic curve $C$ inside a Calabi-Yau $n$-fold in F-theory, giving rise to strings in $d= 12-2n$ dimensions. 
Equivalently, this is the supersymmetry of the strings from 
M5-branes wrapped on the elliptic surface $\widehat{C} \times \mathbb{R}^{1,1}$. We also list the supersymmetries  
of the 1d SQM arising from an M2-brane wrapping $C\times \mathbb{R}$.  }\label{tab:SUSYS}
\end{table}

\subsection{Brane Realizations}\label{sec:setup}

Naturally, the setup so far has a realization in terms of D3-branes in F-theory, which will be briefly discussed now. We consider F-theory on an elliptically fibered Calabi-Yau $n$-fold $Y_n$, which gives rise to a $(12-2n)$-dimensional theory. The D3-brane worldvolume is  $\mathbb R^{1,1} \times C$, where $C$ is a holomorphic curve in the base $B_{n-1}$, and the 2d theory obtained by reduction along $C$ represents a string in the bulk non-compact directions. 
We only consider situations where the curve $C$ contains transversal intersection points with the 7-brane locus
$\Delta$ in the base F-theory $B_{n-1}$. As one encircles the 7-brane loci,
the coupling $\tau$ of the underlying $N=4$ $U(1)$ SYM theory on the D3-brane undergoes a monodromy.

In the F-theory description,  the complex line
bundle $\mathcal{L}_D$ is nothing other than the anti-canonical bundle of the
base of the fibration  \cite{Bianchi:2011qh,Martucci:2014ema}
\begin{equation} \label{LDanticano}
  \mathcal{L}_D \cong K_{B_{n-1}}^{-1}|_C \,.
\end{equation}
Indeed the anti-canonical bundle $K_{B_{n-1}}^{-1}$ fundamentally defines the Calabi-Yau
elliptic fibration $Y_n$ over $B_{n-1}$ via the associated Weierstrass model 
\begin{equation}
  y^2 = x^3 + f x z^4 + g z^6 \,,
\end{equation}
where $f$ and $g$ transform according to $f \in H^0(B, K_{B_{n-1}}^{-4})$ and $g \in
H^0(B, K_{B_{n-1}}^{-6})$. Similarly the restriction of the elliptic fibration to $C$
describes itself a (generically non-Calabi-Yau) elliptic fibration
$\widehat{C}$, defined precisely by the anti-canonical bundle restricted to $C$, i.e. $\mathcal{L}_D$. The existence of the elliptic
fibration $\widehat{C}$ guarantees that $\mathcal{L}_D$ is ample, and further
$\mathcal{L}_D$ is trivial if and only if $C$ does not intersect the discriminant locus
of $Y_n$. Appendix \ref{app:ellipticsurfaces} explains such attributes of the
elliptic surfaces $\widehat{C}$.

In view of the D3-brane realizations, the decomposition of the $SU(4)_R$-symmetry underlying the twists (\ref{CY3RSym}) - (\ref{CY5RSym}) corresponds to remnant rotation symmetries of the string in the transverse directions. 
If we were to consider flat space, the R-symmetry of the $N=4$
SYM theory living on the D3-brane would decompose as
\begin{equation}
  SU(4)_R \simeq SO(6)_R \rightarrow SO(10 - 2n)_T \times SO(2n-4)_R .
\end{equation}
The first factor in the decomposition is the rotation group of the
non-compact spacetime directions transverse to the D3-brane, and the second
factor comes from the structure of the normal bundle of $C$ in $B_{n-1}$. 
The K\"ahler nature of $B_{n-1}$ further reduces the structure from 
\begin{equation}
  SO(2n-4)_R\quad  \rightarrow \quad U(n-2)_R \simeq SU(n-2)_R \times U(1)_R \,.
\end{equation}
In this way we identify the twists in  (\ref{CY3RSym}) -- (\ref{CY5RSym})  with the embeddings into Calabi-Yau elliptic fibrations. 
{The additional case of the CY$_5$' twist in  (\ref{CY5RSym}) can be thought of as an F-theory compactification 
on an elliptic Calabi-Yau five-fold where the base is the sum of three line-bundles over $C$.}

\begin{figure}
\centering
\includegraphics[width=9cm]{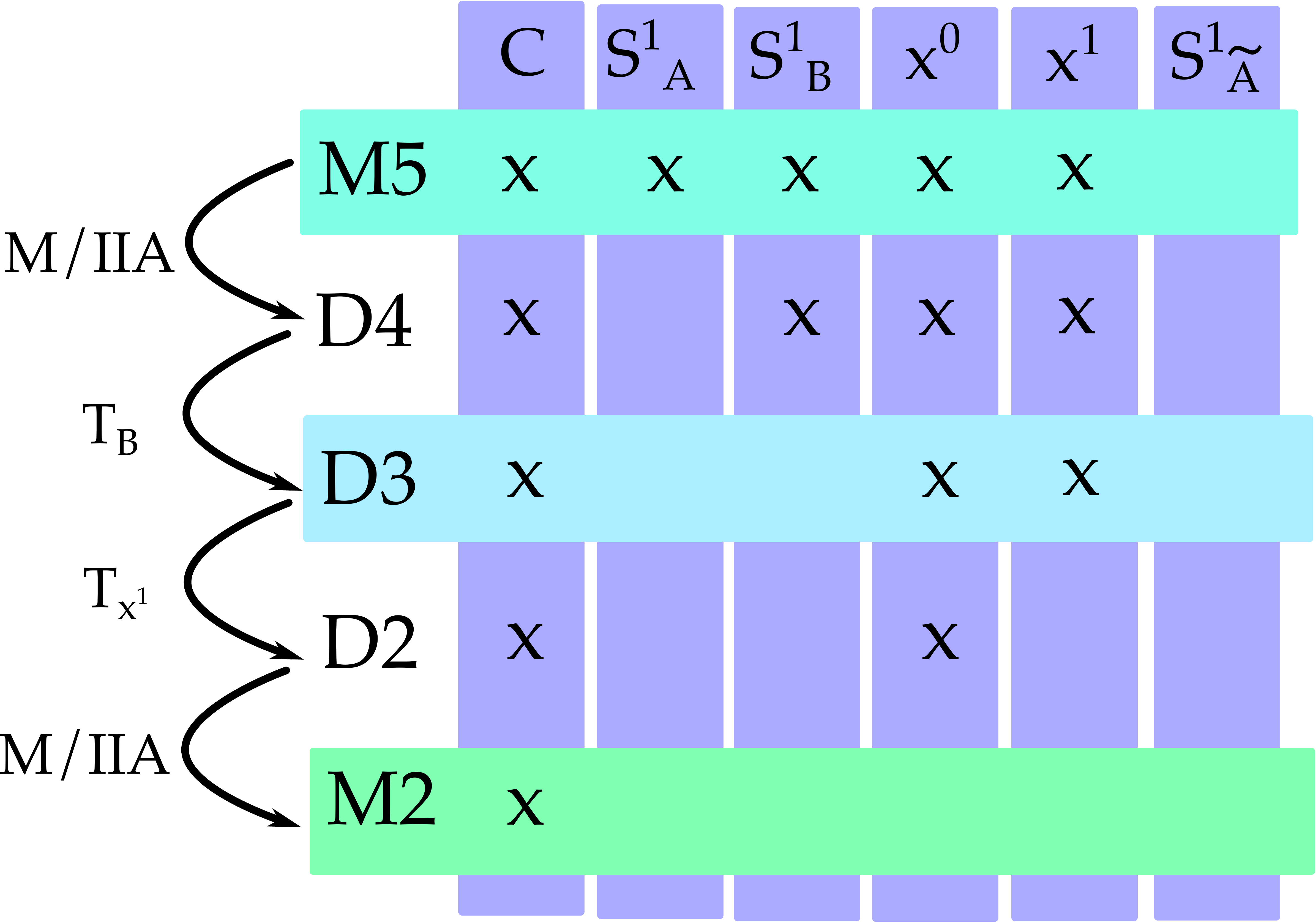}
\caption{Overview of setup and duality maps between D3-branes and their M-theory dual description in terms of  M2- and M5-branes wrapping the curve $C$ or the elliptic surface with base $C$, respectively. }\label{fig:dualities}
\end{figure}

One can dualise this
D3-brane to an M5-brane wrapping the elliptic surface $\widehat{C}$ formed by restricting the fibration to $C$ in $Y_n$:  A T-duality transverse to the D3-brane first leads to a
D4-brane, which is then uplifted to an M5-brane in M-theory. Alternatively one can 
T-dualise along one of the extended directions of the D3-brane, giving rise
to a D2-brane which is uplifted to an M2-brane in M-theory. The interrelations
between the three different viewpoints are shown in figure
\ref{fig:dualities}.

The $\tau$-monodromies in the D3-brane F-theoretic setup are geometrised in the M5-brane picture as the M5-brane also
wraps the elliptic fiber above $C$, which documents the variation of the
coupling as already observed in \cite{Assel:2016wcr}. As such the only topological twist necessary is that mandated by the
curvature of the elliptic surface $\widehat{C}$, and there is no additional
duality twist.  
The only instance when the M5-brane point of view is not applicable is the case of strings in 2d, where the 
low energy effective theory of M-theory on the  Calabi-Yau five-fold 
compactification is a 1d Super Quantum Mechanics (SQM). 
In this instance,  the only  window into the worldvolume theory in M-theory arises from the point
of view of the M2-brane.

Generally, the  M2-brane point of view gives rise to a  1d  SQM
which is the circle compactification of the 2d theory living on the
worldvolume of the string from the D3-brane. Such SQM theories have had some appearance in the literature, see
\cite{Okazaki:2014sga, Okazaki:2015pfa}. 
By analysing the structure of the
supersymmetric multiplets in the 1d theory and how they arise from the 2d
theory before the $S^1$ reduction one can match the content of the two
theories. 
The number of supersymmetries preserved in the 1d SQM arising from an M2-brane wrapping $\mathbb{R} \times C$ for $C$ a holomorphic
curve inside a Calabi-Yau $n$-fold is known for $n \leq 5$
\cite{Gauntlett:2001qs} and is listed in table \ref{tab:SUSYS}.

We shall consider only situations with  strictly this much supersymmetry
preserved, and ignore special cases where the supersymmetry is
enhanced further due to non-generic choices of $C$ or $Y_n$. By the
highlighted dualities the same number of supersymmetries should be preserved
when a D3-brane wraps $\mathbb{R}^{1,1} \times C$ with $C \subset B_{n-1}$ in F-theory. Indeed one can
observe that it is necessary to consider a non-trivial $\tau$-profile to
preserve exactly the same number of supersymmetries as in table \ref{tab:SUSYS};
without 7-brane insertions the elliptic fibration is trivial, $Y_n = B_{n-1} \times T^2$, and thus the base $B_{n-1}$
is also Calabi-Yau. In such a situation the number of supersymmetries
preserved after the partial twist is known \cite{Bershadsky:1995vm} to be
precisely double the number shown in table \ref{tab:SUSYS} realized in a non-chiral fashion\footnote{When $n=2$ the
  Calabi-Yau $Y_2$ is a K3-surface, which is a special case in this analysis
  and is described in appendix \ref{app:K3}.}.

\subsection{M5-branes on Elliptic Surfaces}
\label{sec:M5}

Duality with M-theory maps a D3-brane wrapped on $C \subset B_{n-1}$ to 
the theory of a single M5-brane on $\mathbb{R}^{1,1} \times
\widehat{C}$, where $\widehat{C}$ is a non-trivial elliptically fibered K\"ahler 
surface obtained by restricting the F-theory  elliptic Calabi-Yau $n$-fold $Y_n$
to the fiber over a curve $C$ inside the base. 

The theory living on a single M5-brane is known as the 6d abelian tensor
multiplet theory which preserves $N=(0,2)$ supersymmetry. It has an $Sp(4)$ R-symmetry in addition to the $SO(1,5)$ Lorentz
symmetry. The abelian tensor multiplet contains a self-dual two-form field
strength, $B_{\mu\nu}$, scalars, $\Phi^{ij}$, and symplectic Majorana-Weyl fermions $\rho^i$. Each of
these fields transforms in the following representations of the $SO(1,5)\times Sp(4)$ symmetry group:
\begin{align}
  B_{\mu\nu} \,&:\quad  ({\bf 15, 1}) & \Phi^{ij} \,&:\quad  ({\bf 1,5}) & \rho^i \,&:\quad 
  ({\bf  \overline{4}, 4}) \,.
\end{align}
The field $B_{\mu\nu}$ is self-dual so there are not, in flat space, fifteen
independent degrees of freedom contained in the field. Often it shall be
clearer to consider the field strength $H = dB$, which by self-duality,
\begin{equation}
  H = \star_6 H \,, 
\end{equation}
transforms as 
\begin{equation}
  H \,:\quad  ({\bf \overline{10}, 1}) \,,
\end{equation}
under the symmetry group\footnote{In our conventions a self-dual
  (resp. anti-self-dual) three-form in six dimensions transforms in the ${\bf
    \overline{10}}$ (resp. ${\bf 10}$) representation of $SO(1,5)$.}. The 
supercharges of the theory transform in the same representation as the fermions, 
\begin{equation}
  Q^i \,:\quad  ({\bf \overline{4}, 4}) \,.
\end{equation}
Once we consider the theory on the geometry $\mathbb{R}^{1,1} \times
\widehat{C}$ and further given that $\widehat{C}$ is K\"ahler,  the 6d Lorentz group
$SO(1,5)_L$ is broken to the subgroup
\begin{equation}\label{eqn:M5SO6decomp}
  \begin{aligned}
    SO(1,5) &\quad \rightarrow \quad SU(2)_l \times SO(1,1)_L \times U(1)_l \cr
    {\bf 4} &\quad \rightarrow \quad  {\bf 2}_{1,0} \oplus {\bf 1}_{-1,1} \oplus {\bf
    1}_{-1,-1} \cr
    {\bf \overline{10}} &\quad \rightarrow \quad  {\bf 3}_{-2,0} \oplus {\bf 1}_{2, \pm 2} \oplus {\bf
    1}_{2, 0} \oplus {\bf 2}_{0, \pm 1} \cr
    {\bf 15} &\quad \rightarrow \quad  {\bf 1}_{0,0} \oplus {\bf 3}_{0,0} \oplus {\bf 1}_{0,\pm 2}
    \oplus {\bf 1}_{0,0} \oplus {\bf 2}_{\pm 2,\pm 1} \,,
  \end{aligned}
\end{equation}
where all combinations of signs must be summed over. Here $SU(2)_l \times U(1)_l$ is
the holonomy of the K\"ahler  surface $\widehat{C}$ and $SO(1,1)_L$ is the
Lorentz rotations of $\mathbb{R}^{1,1}$.

As the two-form potential $B_{\mu\nu}$ is unconcerned with the particular
decomposition of the R-symmetry it is also insensitive to the dimension $n$ of
the compactification space $Y_n$. Instead of imposing the self-duality condition on
  $B_{\mu\nu}$ post-decomposition we can consider the decomposition of the $({\bf
  \overline{10}}, {\bf 1})$ as in (\ref{eqn:M5SO6decomp}) to understand the
field strengths in evidence when the M5-brane is placed on the product of 
$\mathbb{R}^{1,1} \times \widehat C$. Indeed by studying the
relationship between the decomposition of the $({\bf 15, 1})$ and the $({\bf
\overline{10}, 1})$ one can see that the resulting fields are 
\begin{equation}
  {\bf 3}_{0,0} \,,\quad {\bf 1}_{0,\pm 2} \,,\quad {\bf 1}_{0,0} \,,\quad
  {\bf 2}_{{\pm 2}, \pm 1} \,,
\end{equation}
at least on the level of the zero-modes of such fields as the derivative does
not change the representation content of the internal symmetry groups.

The decomposition of the R-symmetry $Sp(4)$ will depend on the
dimension of the Calabi-Yau $n$-fold containing $\widehat{C}$. Similarly to the decomposition for the D3-brane,
\be
{ Sp}(4) \rightarrow SO(9 - 2n)_T \times U(n-2)_R \,,
 \ee
where $SO(9 - 2n)_T$ is the group of 
rotations in the non-compact directions transverse to both the M5-brane and
the Calabi-Yau, a $U(n-2)_R$ is the holonomy group associated with the
directions internal to the Calabi-Yau that are transverse to the M5-brane. It
is clear from this discussion that it will not be possible to study the
2d theory arising from the compactification on a Calabi-Yau five-fold, since
there are not two non-compact directions transverse to the five-fold for the
M5-brane to fill -- in this case we shall instead compare with the M2-brane
theory expounded on in section \ref{sec:M2over}.

The M5-brane theory has the feature that the $SL(2, \mathbb{Z})$ self-duality
of the D3-brane theory is manifest -- this means that we do not have to
include the bonus symmetry $U(1)_D$ explicitly. The appearance of the D3-brane
line bundle $\mathcal{L}_D$ can be seen as follows. To the elliptic surface
$\widehat{C}$ is associated a line bundle as part of its Weierstrass data (see
appendix \ref{app:ellipticsurfaces} for more details) and this line bundle is
exactly the restriction of the anti-canonical bundle of $B_{n-1}$ to $C$, which is
precisely $\mathcal{L}_D$. In this way
one can see that the duality becomes encoded in the elliptic surface
$\widehat{C}$ through its defining Weierstrass line bundle -- as expected from
the geometrization of the $SL(2,\mathbb{Z})$ symmetry when uplifting from $N =
4$ super-Yang-Mills to the 6d $(0,2)$ theory
\cite{Witten:1995zh,Witten:2007ct}.


\subsection{M2-branes on Curves and Super-QM}\label{sec:M2over}

Our final description of the effective string theories we consider is in terms 
of an M2-brane wrapping a curve $C$ in the base of an elliptically fibered Calabi-Yau $Y_n$.
The resulting supersymmetric Quantum Mechanics (SQM) is related to the 2d effective field theory 
on a D3-brane wrapping the same curve $C$ by a circle reduction.  

Since the compactified theory turns out to be quite subtle, let us first recall the well-known theory on a single M2 brane extended along ${\mathbb R}^{1,2} \subset \mathbb R^{1,10}$.
The effective theory on an M2-brane in flat space is a 3d  $N=8$ superconformal theory with an $SO(8)_R$ symmetry group corresponding to the rotational group of the transverse space. 
The supersymmetry parameters $\epsilon$ and fields $\Phi_i$ and $\rho$ of the theory transform as follows:
\be
\begin{array}{c|c}
 & SO(3)_L \times SO(8)_R \cr \hline
\epsilon &  ({\bf 2}, {\bf 8}_s)\cr 
\Phi_i  &  ({\bf 1}, {\bf 8}_v) \cr 
\rho &  ({\bf 2}, {\bf 8}_c) \cr 
\end{array}
\ee
As is well-known, this field content can be derived from the field content of a Type IIA D2-brane along ${\mathbb R}^{1,2} \subset \mathbb R^{1,9}$:
The transverse fluctuations of the D2-brane within $\mathbb R^{1,9}$  give
rise to seven real scalar fields; in addition the D2-brane carries a $U(1)$ gauge field potential, which in $\mathbb R^{1,2}$ carries one dynamical degree of freedom and which can be dualised into a single real dynamical scalar field. From the M2-perspective, this gauge-scalar is interpreted as the fluctuation scalar of the M2-brane in the direction of the M-theory circle and thus completes the set of scalars to transform in the $({\bf 1}, {\bf 8}_v)$ of $SO(8)_R$.

In our setup the M2-brane worldvolume takes the form $C\times \mathbb{R}$, giving rise to an $N=4$ or $N=2$ SQM. Compactification on the holomorphic curve $C$ breaks the $SO(1,2)_L$ symmetry to the structure group $U(1)_L$ of the tangent bundle to $C$, and also breaks the conformal symmetry for finite volume of the curve $C$.
The ${\bf 2}$ of $SO(1,2)_L$ clearly decomposes as ${\bf 2} \rightarrow {\bf 1}_{1} \oplus {\bf 1}_{-1}$, while the decomposition of the R-symmetry group depends on the dimensionality of $Y_n$ and will be discussed for the various situations in subsequent sections.

It is our desire to compare the 2d $(0,2)$ or $(0,4)$ theories from the
worldvolumes of the D3- and M5-branes to the 1d $N = 2$ or $N = 4$ SQM from
the M2-brane. It is known, and is explained in appendix \ref{app:SQM}, that
under the circle reduction the multiplets transform according to the rule
\begin{equation}
  \begin{aligned}
   & (0,2) \text{ chiral multiplet} &\rightarrow & \quad N = 2B \,\, (2,2,0) \text{ multiplet } \cr
    &(0,2) \text{ Fermi multiplet } &\rightarrow &\quad N = 2B \,\, (0,2,2) \text{ multiplet } \,,
  \end{aligned}
\end{equation}
where the three entries in the $N=2B$ multiplet specify, respectively, the number of real
scalars, fermions, and auxiliary fields.
A similar decomposition holds for the $(0,4)$ multiplets following the construction of $(0,4)$
multiplets in terms of $(0,2)$ multiplets in appendix \ref{app:SUSYS}. 
However, in order to match the circle reduction of the 2d (0,2) and (0,4) theories to the M2 SQM, it is important 
to make use of a special property of  SQM called 
automorphic duality.
Automorphic duality, as described in more detail in appendix \ref{app:SQM},
allows one to replace physical scalar fields in a 1d supersymmetric sigma model that are not a dependency of the
moduli space metric with auxiliary fields. In effect this transforms e.g. a $(2,2,0)$
multiplet into a $(0,2,2)$ multiplet.\footnote{In fact, this duality is also required to match the 2d $(0,2)$ supergravity spectrum from F-theory compactified on a Calabi-Yau five-fold \cite{MINI}  to the dual M-theory $N=2$ SQM of \cite{Haupt:2008nu}.}

For the 2d theory from the D3-brane
compactification we shall always observe that there exist Wilson line scalar fields  $a$, $\bar{a}$ from
the reduction of the 4d gauge field $A_\mu$ along $C$. These scalars thus inherit a shift
symmetry, and the moduli space metric cannot depend on them. The same is true for the resulting $(2,2,0)$ multiplets obtained by circle reduction to SQM.
Hence for the $(2,2,0)$ multiplets associated with the shift-symmetric Wilson line scalars we shall always be able to invoke automorphic duality to
transform the $(2,2,0)$ multiplet into a $(0,2,2)$ multiplet. Such a situation also holds for the
$N = 4B$ SQM from the compactification of the 2d $(0,4)$ theory on $S^1$,
where now we can exchange the $(4,4,0)$ multiplet containing $a$ and $\bar a$ (along with two more real scalar fields)
with a $(2,4,2)$ multiplet by dualising the Wilson line degrees of freedom only.

As we shall see the zero-modes of $a$ and $\bar{a}$ are always counted by the 
same cohomology group
\begin{equation}
  h^0(C, K_C \otimes \mathcal{L}_D) = g - 1 + \text{deg}(\mathcal{L}_D) \,,
\end{equation}
as they come from the 4d Lorentz sector and do not
couple to the R-symmetry.
We shall thus always be able to transform this
many $(2,2,0)$ or $(4,4,0)$ multiplets in the compactification on $S^1$ to
$(0,2,2)$ or $(2,4,2)$ multiplets respectively. Indeed
agreement between the spectra from the D3-brane theory compactified on $S^1$
and the M2-brane is reached only due to this possible reinterpretation.

Furthermore we expect there to be states analogous to the $3$--$7$ strings in
the F-theory picture, which have to be considered in addition on the M2-brane
side.  While it would be interesting to understand the microscopic origin  of
these states explicitly in the M2-brane picture, it is clear by duality that
we must add the usual number of $8 c_1(B_{n-1}) \cdot C$ complex fermions to account for these modes.


\section{Strings from Duality-Twisted $N=4$ SYM}

\label{sec:StringsfromD3}

We now  determine the dimensional reduction of the D3-branes wrapped on a complex curve in the base of an F-theory elliptic Calabi-Yau compactification. Our analysis makes crucial use of the duality twist as introduced in \cite{Martucci:2014ema} and applied recently in \cite{Haghighat:2015ega, Assel:2016wcr}. This not only  allows us to work out the 2d action for the abelian theory, but also the  spectrum, multiplet structure, and the BPS equations. The latter give rise to
generalized Hitchin equations including a $\tau$-dependence. 
The following sections discuss the strings in 6, 4, and 2 dimensions, respectively. 
Each dimension is somewhat different and requires its own study, as the transverse symmetry groups depend on the dimension. 

\subsection{Strings in 6d $N=(1,0)$ Theories}\label{sec:CY3D3}

First we consider a 6d $N=(1,0)$ F-theory compactification on an elliptic Calabi-Yau three-fold $Y_3$
with a D3-brane along $\mathbb R^{1,1} \times C$. Here $C$ is a curve inside the base $B_2$ of $Y_3$. 
The resulting string, which
propagates in the non-compact six dimensions transverse to $Y_3$, has $(0,4)$ supersymmetry, and can be described in terms of a duality-twisted $N=4$ SYM theory, which we now derive. 

\subsubsection{Duality Twist for 2d $N=(0,4)$}

To determine the duality twist, we first recall from section \ref{sec:D3sector} that
the background geometry breaks the $SU(4)_R$-symmetry of the $N=4$ SYM theory on the D3-brane according to
\be
\ba
 SU(4)_R &\quad \rightarrow \quad  SO(4)_T \times U(1)_R \cr
    {\bf 4} &\quad \rightarrow \quad  {\bf (2,1)}_1 \oplus {\bf (1,2)}_{-1} \cr
    {\bf 6} &\quad \rightarrow \quad {\bf (1,1)}_2 \oplus {\bf (1,1)}_{-2} \oplus {\bf
    (2,2)}_0\, .
\ea\ee
Here $SO(4)_T$ and $U(1)_R$ represent the rotation groups in the four external directions normal to the D3-brane and, respectively, the two directions normal  to $C$ inside $B_2$.
The Lorentz symmetry decomposes as in (\ref{eqn:LorentzDecomp}).
Under the reduced symmetry group, the supercharges of the D3-theory transform as
\begin{equation}
  \begin{aligned}
G_{\rm total} &\quad \rightarrow \quad  SO(4)_T    \times SO(1,1)_L \times U(1)_R \times U(1)_C \times U(1)_{D} \cr 
({\bf 2}, {\bf 1}, \overline{\bf 4})_1 &\quad \rightarrow \quad  
({\bf 2}, {\bf 1})_{1;-1,1,1} \oplus ({\bf 2}, {\bf 1})_{-1;-1,-1,1} \oplus
({\bf 1}, {\bf 2})_{1;1,1,1} \oplus ({\bf 1}, {\bf 2})_{-1;1,-1,1} 
 \\ 
({\bf 1}, {\bf 2}, {\bf 4})_{-1}&\quad \rightarrow \quad  
({\bf 2}, {\bf 1})_{1;1,-1,-1} \oplus ({\bf 2}, {\bf 1})_{-1;1,1,-1} \oplus
({\bf 1}, {\bf 2})_{1;-1,-1,-1} \oplus ({\bf 1}, {\bf 2})_{-1;-1,1,-1}  \,.
  \end{aligned}
\end{equation}
According to the discussion in section \ref{sec:setup}, a D3-brane wrapping a curve in the base of a Calabi-Yau three-fold is expected to 
preserve $(0,4)$ supersymmetry in 2d. This is accomplished by the topological twist
\begin{align}
  T_C^{\text{twist}} &= \frac{1}{2}(T_C + T_R) \cr
  T_D^{\text{twist}} &= \frac{1}{2}(T_D + T_R) \,,
\end{align}
where $T_C$, $T_D$, and $T_R$ are, respectively, the generators of $U(1)_C$, $U(1)_D$,
and $U(1)_R$.
The decomposition of the supercharges with respect to these twisted symmetry
groups is
\be
\ba
G_{\rm total}&\quad \rightarrow \quad SO(4)_T \times   SO(1,1)_L \times U(1)_C^{\rm twist} \times U(1)_{D}^{\rm twist}  \cr  
({\bf 2}, {\bf 1}, \overline{\bf 4})_1 &\quad \rightarrow \quad   
({\bf 2}, {\bf 1})_{1; 0,0} \oplus ({\bf 2}, {\bf 1})_{-1; -1, 0} \oplus ({\bf
1}, {\bf 2})_{1;1,1} \oplus ({\bf 1}, {\bf 2})_{-1; 0,1} 
 \cr 
({\bf 1}, {\bf 2}, {\bf 4})_{-1}&\quad \rightarrow \quad  
({\bf 2}, {\bf 1})_{1; 0,0} \oplus ({\bf 2}, {\bf 1})_{-1; 1,0} \oplus ({\bf
1}, {\bf 2})_{1;-1, -1 } \oplus ({\bf 1}, {\bf 2})_{-1; 0,-1} 
 \,.
  \end{aligned}
\end{equation}
After the twist one thus ends up with four positive chirality scalar supercharges
in 2d,
\begin{align}
  Q_{B+} \,&:\, {\bf (2,1)}_{1;0,0} & \widetilde{Q}_{B+} \,&:\, {\bf
  (2,1)}_{1;0,0} \,,
\end{align}
where $B$ is an index in the left $SU(2)$ inside $SO(4)_T$. 
A similar decomposition of the supersymmetry parameters (\ref{epsilon4dN4}) identifies these as 
\begin{align}
  \epsilon_{B-} \,&:\, {\bf (2,1)}_{-1;0,0} & \widetilde{\epsilon}_{B-} \,&:\, {\bf
  (2,1)}_{-1;0,0} \,.
\end{align}
From the field content (\ref{N4content}) of the original $N=4$ SYM theory on the D3-brane we determine the spectrum of fields of the topologically half-twisted theory on $\mathbb R^{1,1} \times C$:
\begin{equation}\label{DecompFieldsCY3}
  \begin{aligned}
    &\qquad SO(4)_T \times SO(1,1)_L \times U(1)_C^{\text{twist}} \times
    U(1)_D^{\text{twist}} \times U(1)_R \cr
    A \,&:\, \quad 
    ({\bf 1,1})_{2,0,*,0} \oplus ({\bf 1,1})_{-2,0,*,0} \oplus ({\bf 1,1})_{0,1,*,0} \oplus ({\bf 1,1})_{0,-1,*,0}  \cr
     \, &  \, \qquad  =   v_+ \oplus v_- \oplus \bar a_{\bar z} \oplus a_{z} \cr
    \Phi \,&:\, \quad ({\bf 1,1})_{0,1,1,2} \oplus ({\bf
    1,1})_{0,-1,-1,-2} \oplus ({\bf 2,2})_{0,0,0,0} \cr
    & \, \qquad  = \bar \sigma_{\bar z} \oplus \sigma_{ z} \oplus \varphi \cr
    \Psi \,&:\, \quad ({\bf 2,1})_{1,1,1,1} \oplus ({\bf 1,2})_{1,0,0,-1}
    \oplus ({\bf 2,1})_{-1,0,1,1} \oplus ({\bf 1,2})_{-1,-1,0,-1} \cr
    & \,  \qquad  = \psi_{+,\bar z}  \oplus \mu_+   \oplus  \lambda_- \oplus \rho_{-,z} \cr
    \widetilde{\Psi} \,&:\, \quad ({\bf 2,1})_{1,-1,-1,-1} \oplus ({\bf
    1,2})_{1,0,0,1} \oplus ({\bf 2,1})_{-1,0,-1,-1} \oplus ({\bf
    1,2})_{-1,1,0,1} \cr
    & \,  \qquad  = \tilde\psi_{+,z}  \oplus \tilde\mu_+   \oplus  \tilde\lambda_- \oplus \tilde\rho_{-,\bar z}    \,.
  \end{aligned}
\end{equation}

\begin{table}
  \centering
  \begin{tabular}{|c|c|c|c|c|c|}
    \hline
    $(q_C^\text{twist}, q_D^\text{twist})$ &  \multicolumn{2}{c|}{Fermions} &
    \multicolumn{2}{c|}{Bosons}      &   Form Type \cr\hline
    $(1,1)$ & ${\bf (2,1)}_1$  &$\psi_{+, \bar{z}}$& ${\bf (1,1)}_0$, ${\bf
    (1,1)}_0$ & $ \bar{a}_{\bar{z}}, \bar{\sigma}_{\bar{z}}$  &
    $\Omega^{0,1}(C) \otimes \Gamma(\mathcal{L}^{-1}_D)$  \cr
    $(-1,-1)$ & ${\bf (2,1)}_1$ &$\tilde{\psi}_{+, z}$& ${\bf (1,1)}_0$, ${\bf
    (1,1)}_0$ & $a_z,\,\sigma_z$ & $\Omega^{1,0}(C) \otimes \Gamma(\mathcal{L}_D)$   \cr\hline
   \multirow{2}{*}{ $(0,0)$} & ${\bf (1,2)}_1$ & $\mu_+ \, $& \multirow{2}{*}{${\bf (2,2)}_{0}$} & \multirow{2}{*}{$\varphi$} & \multirow{2}{*}{$\Omega^{0,0}(C)$} \cr
   &  ${\bf (1,2)}_1$&  $\tilde{\mu}_+$ & & &\cr
   \hline
    $(1,0)$ & ${\bf (1,2)}_{-1}$ &$\tilde{\rho}_{-, \bar{z}}$& &&$\Omega^{0,1}(C)$\cr
    $(-1,0)$ & ${\bf (1,2)}_{-1}$ & ${\rho}_{-, z}$& &&$\Omega^{1,0}(C)$\cr\hline\hline
    $(0,1)$ & ${\bf (2,1)}_{-1}$ & $\lambda_-$& ${\bf (1,1)}_{2}$&
    $v_+$&$\Omega^{0,0}(C) \otimes \Gamma(\mathcal{L}^{-1}_D)$\cr
    $(0,-1)$ & ${\bf (2,1)}_{-1}$ & $\tilde{\lambda}_- $&${\bf (1,1)}_{-2}$ &
    $v_-$ &$\Omega^{0,0}(C) \otimes \Gamma(\mathcal{L}^{-1}_D)$\cr\hline
  \end{tabular}
  \caption{The field content of the $(0,4)$ theory on a D3-brane wrapping a curve $C$ inside a K\"ahler surface $B_{2}$ which is the base of an elliptic CY3 in F-theory. The spectrum is obtained by the partially twisted reduction of  4d $N=4$ SYM. The representations in the second and third double-column refer to $SO(4)_T \times SO(1,1)_L$, and fields are labeled by their $SO(1,1)_L$ spin as well as their form-type along $C$.  }\label{tbl:CY3fc}
\end{table}
At this stage all fields depend both on the coordiates $x^0, x^1$ along
$\mathbb R^{1,1}$ and on the  local (anti-)holomorphic coordinates $z, \bar z$
on $C$. The superscripts $\pm$ of the fermionic fields denote the $SO(1,1)_L$
chirality. 
This field
content  is summarized in table \ref{tbl:CY3fc}.
As a result of the topological twist, the fields assembled in table \ref{tbl:CY3fc} transform as differential forms on $C$.
Their bidegree is fixed by the $U(1)^{\rm twist}_C$ charge, which determines the transformation behaviour of the field with respect to the (twisted) structure group of $C$.  The relation between the topological twist charge and the cotangent bundle of $C$ is
\be
\ba
&q^{\rm twist}_C = +1  &\qquad \longleftrightarrow \qquad    &\Omega^{0,1}(C) \cr
&q^{\rm twist}_C = -1  &\qquad \longleftrightarrow   \qquad  &\Omega^{1,0}(C) \, . 
\ea
\ee
As discussed in section \ref{sec:D3sector}, 
fields carrying in addition duality twist charge transform as form-valued
sections of suitable powers of the duality bundle, $\mathcal{L}_D$, viewed as a complex line bundle on $C$ with connection (\ref{Lconnection1}).
To characterize the relevant sections we need the kinetic operators acting on
the fields. To determine these it is necessary to work out the topologically
twisted action in detail. The sections that the fields transform as listed in
the rightmost column of table \ref{tbl:CY3fc} are the result of this analysis.
This way we will also be able to  determine the massless spectrum in the
effective 2d theory along the string.

\subsubsection{Action and BPS Equations} \label{subsec:Stringaction6d}

The string action and the  supersymmetry variations of the fields follow by
the decomposition of the
 4d $N = 4$ SYM theory, or, equivalently,
via dimensional reduction of the 10d $N = 1$ SYM theory with action
\begin{equation}
 S_{10d} = \int_{\mathbb R^{1,9}}  - \frac{1}{4g^2} {\rm Tr} \hat F_{MN} \hat
 F^{MN}  - \frac{i}{2 g^2} \overline{\hat\Psi} \Gamma^M D_M \hat\Psi \,,
\end{equation}
and supersymmetry variations
\begin{equation} \label{SUSY10d}
  \delta \hat A_M =   -i \overline{\hat\epsilon}  \Gamma_M \hat\Psi, \qquad \quad
  \delta \hat\Psi =  \frac{1}{2} \Gamma^{M N} \hat F_{M N} \hat\epsilon \,.
\end{equation}
Here $\hat \Psi$ and $\hat\epsilon$ denote the 10d Majorana-Weyl spinor and
SUSY parameter. In addition after the dimensional reduction to four dimensions
one adds explicitly a topological term
\begin{equation}
  \frac{i\theta}{2\pi} \text{Tr}F \wedge F \,,
\end{equation}
where $F$ is the 4d reduction of the 10d field strength $\hat F$. 
The 4d abelian ${N}=4$ SYM action for the fermions and scalars is 
\be
S_{\rm fermions} + S_{\rm scalars }= {i\over 2 \pi} \int_{\mathbb{R}^{1,1}
\times C} \sqrt{-g} d^4x \Tr \tilde\Psi \bar\sigma^\mu \partial_\mu \Psi
+ {1\over 4\pi}   \int_{\mathbb{R}^{1,1} \times C} \Tr \star_4 d\Phi\wedge d\Phi\,.
\ee
The gauge fields in the abelian ${N}=4$ SYM theory have the 
action 
\be
S_{F } = {1\over 4\pi} \int_{\mathbb{R}^{1,1} \times C} \tau_2 \Tr F \wedge \star_4 F +   {1\over 4\pi i}  \int_{\mathbb{R}^{1,1} \times C} \tau_1\Tr F\wedge F  \,.
\ee
Via the decomposition (\ref{DecompFieldsCY3}) and including the twist we
can determine the theory in terms of fields along $C$ and
$\mathbb{R}^{1,1}$, and subsequently integrate out the internal degrees of freedom. 

Let us introduce coordinates $x^\pm = x^0 \pm x^1$ along $\mathbb R^{1,1}$, with derivatives $\partial_\pm := \partial_0 \pm \partial_1$, 
 as well as (anti-)holomorphic derivatives along $C$, $\partial \equiv \partial_z$, $\bar\partial \equiv \partial_{\bar z}$.
The internal derivatives of fields with definite $U(1)_D$ transformation
properties will be written in terms of the duality covariant derivatives
$\partial_{\cal A}$ and $\bar\partial_{\cal A}$, which were defined in
(\ref{eqn:covderivs}).
The action for the full theory after the decomposition and twist is
\begin{equation}
\ba\label{TotalAction}
 S_{\rm string} &= \int_{\mathbb{R}^{1,1} \times C} d^4 x \sqrt{|g|}\mathcal{L} =  \int d^4 x \sqrt{|g|} \left(\tau_2 \mathcal{L}_\text{gauge} +
  \frac{2}{3}\tau_1\mathcal{L}_\text{top} + 2i\mathcal{L}_\text{fermion} + 
  \mathcal{L}_\text{scalar}  \right)\cr 
 & = \int_{\mathbb{R}^{1,1} \times C} d^4 x \sqrt{|g|} \left(\tau_2  F_{\mu \nu} F^{\mu \nu} +  \frac{2}{3}\tau_1 \left(\epsilon^{\mu \nu \alpha \beta} F_{\mu \nu}
      F_{\alpha \beta} \right)
 + 2i  \left(\bar\Psi  \Gamma^\mu \partial_\mu \Psi \right)  
 + \sum_i \partial_\mu \phi_i \partial^\mu \phi^i  \right)
  \,.
  \ea
\end{equation}
The gauge field part takes the same universal form for the 2d actions obtained from D3-branes on any Calabi-Yau $n$-fold,
\be
\ba
\label{GaugeAction}
    {\cal L}_\text{gauge}  =&  
      - \frac{1}{2} (\partial \bar a - \bar \partial a)^2  
      - \frac{1}{2}  (\partial_- v_+ - \partial_+ v_-)^2 
      - \partial_- a \partial_+ \bar a   
      - \partial_- \bar a \partial_+ a 
      - \partial v_+ \bar \partial v_- \cr 
      &- \partial v_- \bar\partial v_+
      + \bar \partial v_+ \partial_- a 
      + \partial v_+ \partial_- \bar a 
      + \partial v_- \partial_+ \bar a 
      + \bar \partial v_- \partial_+ a  \\ 
 {2\over 3i}   {\cal L}_{\rm top} =& 
      (\partial v_+\bar\partial v_- 
      - \partial v_- \bar\partial v_+      
      + \partial_- \bar a  \partial_+ a 
      - \partial_- a \partial_+ \bar a  ) \cr 
      &+ ( \bar\partial v_+ \partial_- a 
      - \partial v_+ \partial_- \bar a 
      + \partial \bar a \partial_- v_+ 
      - \bar \partial a \partial_- v_+ ) \cr
      &+  ( -\bar\partial v_- \partial_+ a 
      + \partial v_- \partial_+ \bar a 
      - \partial \bar a \partial_+ v_- 
      + \bar \partial a \partial_+ v_- ) \,.
 \ea
 \ee
The part that is twist specific is the action for the fermions and scalars. It
depends on the embedding of the curve $C$ and consequently on the dimension of
the compactification. For the Calabi-Yau three-fold compactification this part takes the form
\be
\ba
    {\cal L}_\text{fermion} =&
    + \lambda_- \partial_+ \tilde\lambda_- 
    + \psi_+ \partial_\mathcal{A} \tilde\lambda_- 
    - \tilde\lambda_- \partial_+ \lambda_- 
    - \tilde\lambda_- \partial_\mathcal{A}\psi_+ 
    + \mu_+ \partial_- \tilde\mu_+ \cr 
    &- \tilde\mu_+ \partial_-\mu_+ 
    + \rho_- \bar\partial \tilde\mu_+ 
    - \tilde\mu_+ \bar\partial \rho_- 
    - \lambda_- \bar\partial_\mathcal{A} \tilde\psi_+ 
    - \psi_+\partial_- \tilde\psi_+ 
    + \tilde\psi_+ \bar\partial_\mathcal{A} \lambda_- \cr
    &+ \tilde\psi_+ \partial_- \psi_+ 
    - \mu_+ \partial \tilde\rho_- 
    - \rho_-\partial_+ \tilde\rho_- 
    + \tilde\rho_-\partial\mu_+ 
    + \tilde\rho_- \partial_+ \rho_- \cr 
     {\cal L}_\text{scalar} =&
      - \partial_- \sigma \partial_+ \bar \sigma  
      - \partial_- \bar\sigma \partial_+ \sigma 
      + \bar\partial_\mathcal{A} \sigma \partial_\mathcal{A} \bar \sigma 
      + \bar \partial_\mathcal{A} \bar \sigma \partial_\mathcal{A} \sigma
      + \partial_+ \varphi \partial_- \varphi 
      - \partial \varphi \bar\partial \varphi  \,.
  \end{aligned}
\end{equation}
The supersymmetry variations likewise follow by dimensional reduction, starting in 10d.
The result of this decomposition
 of the bosonic SUSY variations is
\be
\label{eqn:CY3bosvar}
\ba
\sqrt{\tau_2 } \delta a &= 2 i \epsilon_- \tilde \psi_{+}   \cr 
 \delta \sigma &= -  2 i \tilde \epsilon_- \tilde \psi_{+}    \cr 
 \sqrt{\tau_2}  \, \delta v_- &= 2 i (\lambda_- \tilde \epsilon_- + \tilde
\lambda_-  \epsilon_-)   \cr 
 \delta \varphi_{A \dot B}  &= - 2 i  (\epsilon_{-A} \mu_{+\dot B} + \tilde
\epsilon_{-A} \tilde \mu_{+\dot B}) \,.
\ea
\qquad 
\ba
\sqrt{\tau_2 }  \delta \bar a  &=   2 i \tilde \epsilon_-  \psi_{+} \cr 
\delta \bar\sigma &=  2  i \epsilon_-  \psi_{+} \cr 
\delta v_+ &= 0\cr 
\cr 
\ea
\ee
Here we have made manifest the transformation of the fields as representations of the transverse $SO(4)_T = SU(2)_{T,1} \times SU(2)_{T,2}$ rotation group, with $A$ and $\dot A$ referring to the two $SU(2)$ factors. More details on the decomposition and our conventions can be found in appendix \ref{app:D3action}. 
For the fermionic variations one finds
\be
\label{eqn:CY3fermvar}
\ba
\delta \psi_{+} &=   \epsilon_- \sqrt{\tau_2}  (- \partial_+ \bar a  +  \, \bar\partial v_+ ) +  \tilde \epsilon_- \partial_+ \bar \sigma \quad    \cr 
 \delta \mu_+^{\dot B} &=  -  \tilde  \epsilon_{- A} \partial_+  \varphi^{A\dot B} \cr 
\delta \rho_-^{\dot B} &=  \tilde \epsilon_{ - A}  \partial_{\cal A}\varphi^{A \dot B}  \cr 
 \delta \lambda_- &= -  \epsilon_-  (\sqrt{\tau_2}  F_{01}  +  {\cal F}_{\cal A})   - \tilde \epsilon_- \ast_C  \partial_{\cal A} \bar \sigma 
 \ea
\qquad 
\ba
 \delta \tilde \psi_{+} &=  \tilde \epsilon_- \sqrt{\tau_2}  (\partial_+ a -  \partial v_+) +  \epsilon_- \partial_+ \sigma\cr 
    \delta \tilde \mu_+^{\dot B} &=   +    \epsilon_{ - A} \partial_+  \varphi^{A \dot B}
   \cr 
     \delta {\tilde \rho}_-^{\dot B} &=  \epsilon_{- A} \bar\partial_{\cal A}  \varphi^{A \dot B}\cr 
      \delta \tilde \lambda_- &= -  \tilde \epsilon_- (\sqrt{\tau_2}    F_{01}  -  {\cal F}_{\cal A} )+  \epsilon_- \ast_C \bar\partial_{\cal A} \sigma   \,.
\ea
\ee
Up to boundary terms the above action (\ref{TotalAction}) is invariant
off-shell  under the supersymmetry transformations given by
(\ref{eqn:CY3bosvar}) and (\ref{eqn:CY3fermvar}). This result depends
crucially on the properties (\ref{partialtau1tau2}) of the
holomorphically varying
axio-dilaton, as is further explained in appendix \ref{app:D3action}.

Note the factor of $\sqrt{\tau_2}$ in front of the external components of the
gauge field; this factor arises by rescaling the 10d supersymmetry variations (\ref{SUSY10d}) such as to comply with our normalization of the 4d $N=4$ SYM action. 
Furthermore, we have defined
\bea
{\cal F}_{\cal A} := \frac{1}{2}\sqrt{\tau_2}(\bar \partial a - \partial \bar a) \,.
\eea
It turns out that this combination can be written entirely in terms of $U(1)_D$ covariant derivatives of the fields $\sqrt{\tau_2} a$ and $\sqrt{\tau_2} \bar a$ with definite $U(1)_D$ charge,
\bea
{\cal F}_{\cal A} = \frac{1}{2}(\bar\partial_{\cal A}(\sqrt{\tau_2}a) -
\partial_{\cal A}(\sqrt{\tau_2}\bar{a})) \,.
\eea
This uses holomorphy of the varying axio-dilaton $\tau$, see equ.  (\ref{partialtau1tau2}).   We will come back to this important point in section \ref{sec:Hitch}.

 Ignoring potential boundary terms, the equations of motion obtained from the variation of the fermionic action are
\begin{align} \label{fermioniceom1}
  \partial_+ \tilde \lambda_- - 2 \bar\partial_{\cal A} \tilde \psi_+ &= 0 
  &\qquad  \partial_+ \lambda_- + \partial_{\cal A} \psi_+ &= 0  \cr
  \partial_- \tilde\psi_+ - 2 \partial_{\cal A} \tilde \lambda_- &= 0  
  &\qquad \partial_- \psi_+ + 2 \bar\partial_{\cal A} \lambda_- &= 0  \cr
 \partial_- \tilde \mu_+ - 2 \partial_{\cal A} \tilde\rho_- &= 0 
 &\qquad \partial_- \mu_+ + 2 \bar\partial_{\cal A} \rho_- &= 0  \cr
 \partial_+ \tilde\rho_- - 2 \bar\partial_{\cal A} \tilde \mu_+ &= 0 
 &\qquad \partial_+ \rho_- + 2 \partial_{\cal A} \mu_+ &= 0 \,.
\end{align}
The  BPS equations that follow from the supersymmetry variations are on the other hand
\begin{align}
  F_{01} = \frac{1}{2}(\partial_-v_+ - \partial_+v_-) &= 0  & 
  \mathcal{F}_\mathcal{A} =
  \frac{1}{2}\sqrt{\tau_2}( \bar\partial a - \partial\bar a) &= 0   \cr
  \bar\partial v_+ - \partial_+\bar a &= 0  &
  \partial v_+ - \partial_+ a &= 0  \cr
  \partial_+\sigma &= 0 & \partial_+\bar\sigma &= 0  \cr
  \partial_\mathcal{A}\bar\sigma &= 0  & 
  \bar\partial_\mathcal{A}\sigma &= 0  \cr
  \partial\varphi^{A\dot{B}} &= 0 & \bar\partial\varphi^{A\dot{B}} &= 0 \cr
  \partial_+\varphi^{A\dot{B}} &= 0 \,. & &
\end{align}
The BPS equations are similar to Hitchin equations for Yang-Mills theory on a Riemann surface, except that the fields here transform as sections of $\mathcal{L}_D$. We shall discuss some aspects of these equations in section  \ref{sec:Hitch}. 
The next task here will be to deduce the cohomology groups counting the spectrum of zero-modes in the 2d effective theory along the string.

\subsubsection{Spectrum}\label{sec:D3CY3spec}

If one performs a dimensional reduction to determine the 2d effective action each field decomposes as 
\be
\Phi(x_\pm,z,\bar z) = \sum_k  \Phi^{(k)}(x_\pm) \otimes \hat\Phi^{(k)}(z, \bar z)
\,,
\ee
with $\Phi^{(k)}(z, \bar z)$ an eigenmode of the internal kinetic operator. 
From (\ref{fermioniceom1}) one finds that the internal part of the 2d zero-modes are characterized by the vanishing of the second term in each equation.
 The external fields in the 2d effective action then satisfy the massless Dirac equation given by the first terms in each equation, in agreement with their chirality.
Consider a field $\hat\Phi_{p,q} (z,\bar z)$ of charge $q_D^{\rm twist}$ and form degree $(p,q)$. If the zero-modes are governed by the internal kinetic operator $\bar\partial_{\cal A}$ in (\ref{fermioniceom1}), i.e. by
\be
\bar\partial_{\cal A} \hat\Phi_{p,q} (z,\bar z) = (\bar\partial + i q_D^{\rm
twist} {\cal A}^{(0,1)})  \hat\Phi_{p,q}(z,\bar z)  = 0 \,,
\ee
then the non-trivial zero-modes correspond to
\bea
\hat\Phi^{(0)}_{p,q} (z,\bar z) \in H^{p,q}_{\bar\partial} (C, {\cal
L}_D^{-q_D^{\rm twist}}) \,.
\eea
If on the other hand the equations of motion involve $\partial_{\cal A}$, i.e.
\be
\partial_{\cal A} \hat\Phi_{p,q} (z,\bar z) = (\partial + i q_D^{\rm twist}
{\cal A}^{(1,0)})  \hat\Phi_{p,q}(z,\bar z)  = 0 \,,
\ee
the zero-modes correspond to 
\bea
\hat\Phi_{p,q}^{(0)} (z,\bar z) \in H^{p,q}_{\partial} (C, \bar{\cal L}_D^{+
  q_D^{\rm twist}  }) = \left(H^{q,p}_{\bar\partial} (C, {\cal L}_D^{+
    q_D^{\rm twist}})\right)^* \,.
\eea
Note the different signs appearing in the powers of the bundles. In the last equation we used  that complex conjugation acts on the (anti-)holomorphic bundles as  $(\bar{\cal L}_D)^\ast = {\cal L}_D$. 

For instance, in view of the $U(1)^{\rm twist}_D$ charges, the 2d gaugino zero-modes $\tilde \lambda_-(x_\pm)$ and $\lambda_-(x_\pm)$ correspond to solutions to
\bea
(\partial - i \tilde {\cal A}^{(1,0)})  \hat{\tilde\lambda}(z,\bar z) = 0,
\qquad \quad (\bar\partial + i \tilde {\cal A}^{(0,1)})  \hat\lambda(z,\bar z)
= 0 \,,
\eea 
given by
\bea
\hat{\tilde\lambda}^{(0)}(z,\bar z) \in H_\partial^{(0,0)}(C, \bar{\cal L}_D^{-1}) =\left( H_{\bar\partial}^{0,0}(C,{\cal L}_D^{-1})\right)^*, \qquad  \hat\lambda^{(0)}(z,\bar z) \in H_{\bar\partial}^{(0,0)}(C, {\cal L}_D^{-1}).
\eea
This systematically leads to the counting of massless fields in the effective
2d $(0,4)$ theory summarized in table \ref{tbl:CY3fcmultiplets}. It is also
the rationale behind the determination of the sections of which bundles the $(2+2)$ dimensional fields transform in table \ref{tbl:CY3fc}. 
The results of our derivation agree with the spectrum stated in \cite{Haghighat:2015ega}.

In order to evaluate the dimensions of these cohomology groups one takes into account that the duality bundle ${\cal L}_D$ on $C$ can be viewed as a bundle on the base $B_2$ which describes the      $SL(2,\mathbb Z)$ monodromies due to the variation of axio-dilaton $\tau$ in F-theory. 
At this stage we recall from the discussion around (\ref{LDanticano}) that \cite{Bianchi:2011qh,Martucci:2014ema}
\be \label{LD=K}
{\cal L}_D = K^{-1}_{B_2} |_{C} \,,
\ee
${\cal L}_D$ is of non-negative degree, 
and ${\cal L}_D$ is trivial 
 if and only if $C$ does not intersect the discriminant locus of $Y_3$. In this case, $\tau$ does not experience any monodromies on $C$ and is therefore constant. 
This has the following consequences:
Unless ${\cal L}_D = {\cal O}$, ampleness of ${\cal L}_D$ implies that $h^0(C,{\cal L}_D^{-1}) = 0$. Therefore the vector multiplet is projected out at the massless level and the 2d effective theory reduces to a sigma-model. Only in the special case that ${\cal L}_D = {\cal O}$ is a vector multiplet retained and the $U(1)$ gauge symmetry unbroken.  This is a rather notable difference to the perturbative description in the Type IIB orientifold limit, where the D3-brane theory 
has a phase with an unbroken $U(1)$ symmetry despite the intersection with the 7-branes. We explain this difference, along with an explanation in terms of non-perturbative effects in the vicinity of the orientifold plane, in section \ref{IIBorientifolds}.

Finally, Serre duality implies that $h^1(C,{\cal L}_D^{-1}) = h^0(C, K_C
\otimes {\cal L}_D)$, and  for ${\cal L}_D \neq {\cal O}$  Riemann-Roch
implies that
\bea
 h^1(C,{\cal L}_D^{-1}) = \chi(C, {K_C} \otimes {\cal L }_D) = g-1 + c_1(B_2) \cdot C,
 \eea
 where we used (\ref{LD=K}) and where $g$ is the genus of $C$. This explains the
 multiplicities in table \ref{tbl:CY3fcmultiplets}, evaluated for the case of
 non-trivial ${\cal L}_D$.

\begin{table}
  \centering
  \begin{tabular}{|c|c|c|c|c|c|c|}
    \hline
    $(q_C^\text{twist}, q_D^\text{twist})$ &  \multicolumn{2}{c|}{Fermions} &
    \multicolumn{2}{c|}{Bosons}     & $(0,4)$ &  Multiplicity \cr\hline\hline
    $(1,1)$ & ${\bf (2,1)}_1$  &$\psi_{+}$& ${\bf (1,1)}_0$, ${\bf (1,1)}_0$
    & $ \bar{a}, \bar{\sigma}$ &\multirow{2}{*}{Hyper} & $h^0(C,K_C \otimes {\cal L}_D)$\cr
    $(-1,-1)$ & ${\bf (2,1)}_1$ &$\tilde{\psi}_{+}$& ${\bf (1,1)}_0$, ${\bf
    (1,1)}_0$ & $a,\,\sigma$ & &$=g - 1 + c_1(B_2) \cdot C$   \cr\hline
    \multirow{2}{*}{$(0,0)$} & ${\bf (1,2)}_1$ & $\mu_+$& \multirow{2}{*}{${\bf (2,2)}_{0}$} & \multirow{2}{*}{$\varphi$} &  Twisted  & \multirow{2}{*}{$h^0(C)$ = 1} \cr
    &  ${\bf (1,2)}_1$ &   $\tilde{\mu}_+$&  &  & Hyper  &  \cr\hline
    $(1,0)$ & ${\bf (1,2)}_{-1}$ &$\tilde{\rho}_{-}$& & &\multirow{2}{*}{Fermi}  &\multirow{2}{*}{$h^1(C) = g$}\cr
    $(-1,0)$ & ${\bf (1,2)}_{-1}$ & ${\rho}_{-}$& &&&   \cr\hline\hline
    $(0,1)$ & ${\bf (2,1)}_{-1}$ & $\lambda_-$& ${\bf (1,1)}_{2}$& $v_+$ &\multirow{2}{*}{Vector} &\multirow{2}{*}{$h^1(C,K_C \otimes {\cal L}_D) = 0$}\cr
    $(0,-1)$ & ${\bf (2,1)}_{-1}$ & $\tilde{\lambda}_- $&${\bf (1,1)}_{-2}$ & $v_-$ &&\cr\hline
  \end{tabular}
  \caption{Massless 2d  (0,4) multiplets of the $(0,4)$ theory on a D3-brane wrapping curve $C$ inside a K\"ahler base $B_{2}$ in F-theory propagating in the bulk of the D3-brane. The specific values for the Betti numbers in the last column refer to a curve $C$ intersecting the disciminant locus of the elliptic fibration in isolated points. The representations refer to $SO(4)_T \times SO(1,1)_L$.}\label{tbl:CY3fcmultiplets}
\end{table}

We now wish to identify the R-symmetries of the various fields.
A general 2d $(0,4)$ theory has an R-symmetry group $SO(4)_R = SU(2)_R \times SU(2)_I$.
If the theory flows to a superconformal theory in the infrared, only a single $SU(2)$ subgroup is preserved by the SCFT \cite{Witten:1993yc}.
Each field must transform in such a way that it is consistent with
the R-charges of the various $(0,4)$ multiplets as explained in section
\ref{app:2dmults}. The supercharges of the $(0,4)$ theory must
transform in the ${\bf 2}$ of  $SU(2)_R$. This suggests 
identifying the $SU(2)_R$ symmetry with the first $SU(2)_{T,1}$ factor in the group
$SO(4)_T = SU(2)_{T,1} \times SU(2)_{T,2}$ of rotations in the space transverse to the string. The remaining $SU(2)_{T,2}$ will act as
an $SU(2)$ current algebra on the worldsheet of the string
\cite{Haghighat:2015ega}. The identification
of  $SU(2)_{T,1}$ as the $SU(2)_R$ symmetry is dependent on the form of the
twist. The second factor $SU(2)_I$ in the R-symmetry group $SO(4)_R = SU(2)_R \times SU(2)_I$ 
is not generally visible from the compact geometry \cite{Haghighat:2015ega}; it is the $SU(2)$ R-symmetry of the 6d
$(1,0)$ theory in which the string lives (see e.g. \cite{Ohmori:2014kda}). The charges of the multiplets in
table \ref{tbl:CY3fcmultiplets} under this $SU(2)_I$ symmetry are fixed by
the structure of the $(0,4)$ supersymmetry as explained in appendix
\ref{app:2dmults}. 

In table \ref{tbl:CY3fcmultiplets}  the states with twist
charges $(\pm 1, \pm1)$ comprise fermions transforming as a  ${\bf 2}$ of $SU(2)_R$ and
corresponding scalars transforming as a ${\bf 1}$. This identifies these states as a 
hypermultiplet, where the supersymmetry forces the fermions and
scalars to transform as the ${\bf 1}$ and ${\bf 2}$ respectively of the
$SU(2)_I$.
Similarly the universal multiplet with twist charges $(0,0)$ has fermions
that transform trivially and scalars which transform in the ${\bf 2}$ of
$SU(2)_R$ --
this is a twisted hypermultiplet. The Fermi multiplets with twist charge
$(\pm 1, 0)$ transform trivially under the R-symmetry, as expected, and
finally the states with twist charges $(0, \pm 1)$ come in the
 R-symmetry representations characteristic of a vector
 multiplet.\footnote{We have seen that for a curve $C$ in generic position to the 7-branes of the F-theory background there are no zero-modes associated with such states. }

It is instructive to collect the left- and right-moving zero-modes. For
definiteness we reiterate that we assume that ${\cal L}_D$ is
non-trivial, equivalent to the assumption that the coupling $\tau$ varies
non-trivially over the curve $C$. 
In this case the only left-handed fermion zero-modes result from the Fermi multiplets and  are counted in the
following way -- with the charges given in the ordering $SO(1,1)_{L}\times U(1)_C^{\rm twist} \times U(1)_D^{\rm twist}$,
\begin{equation}
\hbox{Fermions, L:}\qquad 
  \begin{array}{|c|l|c|l|}
  \hline
\multicolumn{2}{|c|}{\hbox{Representation}}& \hbox{Cohomology Groups} &  \hbox{Multiplicity}\cr \hline
  \rho_{-, z} &  ({\bf 1},{\bf 2})_{-1,-1,0} & H^{1,0}(C, \mathcal{O}) & g \cr
 \tilde\rho_{-, \bar{z}} & ({\bf 1},{\bf 2})_{-1,1,0} & H^{0,1}(C, \mathcal{O}) & g \cr\hline
  \end{array}
\end{equation}
For the right-handed fermion zero-modes we have
\begin{equation}
\hbox{Fermions, R:}\qquad 
  \begin{array}{|c|l|c|l|}
  \hline
  \multicolumn{2}{|c|}{\hbox{Representation}}& \hbox{Cohomology Groups} & \hbox{Multiplicity} \cr \hline
    \psi_{+, \bar{z}} & ({\bf 2},{\bf 1})_{1,1,1} & H^{0,1}(C, \mathcal{L}_D^{-1}) & g - 1 + c_1(B) \cdot
      C \cr
      \tilde\psi_{+, z}&({\bf 2},{\bf 1})_{1,-1,-1} & H^{1,0}(C, \mathcal{L}_D) & g - 1 + c_1(B) \cdot
        C \cr
   \mu_+ &({\bf 1},{\bf 2})_{1,0,0} & H^{0,0}(C, \mathcal{O}) & 1 \cr
   \tilde{\mu}_+ &({\bf 1},{\bf 2})_{1,0,0} & H^{0,0}(C, \mathcal{O}) & 1 \cr\hline
  \end{array}
\end{equation}
Finally one can consider the massless bosonic fields:
\begin{equation}
\hbox{Bosons:}\qquad 
  \begin{array}{|c|l|c|l|}
  \hline
  \multicolumn{2}{|c|}{\hbox{Representation}} & \hbox{Cohomology Groups} &  \hbox{Multiplicity}\cr \hline
 a_z&   ({\bf 1},{\bf 1})_{0,1,1} & H^{0,1}(C, \mathcal{L}_D^{-1}) & g - 1 + c_1(B) \cdot
          C \cr
  \sigma _z  & ({\bf 1},{\bf 1})_{0,1,1} & H^{0,1}(C, \mathcal{L}_D^{-1}) & g - 1 + c_1(B) \cdot
          C \cr
\bar{a}_{\bar{z}}  & ({\bf 1},{\bf 1})_{0,-1,-1} & H^{1,0}(C, \mathcal{L}_D) & g - 1 + c_1(B) \cdot
            C \cr
 \tilde{\sigma}_{\bar{z}} &  ({\bf 1},{\bf 1})_{0,-1,-1} & H^{1,0}(C, \mathcal{L}_D) & g - 1 + c_1(B) \cdot
            C \cr
\varphi   & ({\bf 2},{\bf 2})_{0,0,0} & H^{0,0}(C, \mathcal{O}) & 1 \cr \hline
  \end{array}
\end{equation}

Apart from this sector, extra massless fermionic states arise from strings stretched between the D3-brane and the 7-branes in the F-theory background.
These zero-modes are localised at the intersection of the curve $C$ with the discriminant of $Y_3$, which,
for generic position of the curve $C$, is a set of points. 
The existence of such $3$--$7$ strings in the present context, found first in \cite{Witten:1996qb}, has also been pointed out in \cite{Haghighat:2015ega}, and their importance  in low-dimensional compactifications for the gauge anomalies on the 7-branes was stressed independently in \cite{Schafer-Nameki:2016cfr,Apruzzi:2016iac}. 
The contribution from this sector consists of 
\begin{equation} \label{Ferminumber}
  n_{37} =  8\, c_1(B_2) \cdot C 
\end{equation}
(0,4) half-Fermi multiplets. 
This might seem counterintuitive at first since the number of intersection points of the D3-brane with the discriminant locus is given by $12 \, c_1(B_2) \cdot C$.
However, even though an $SL(2,\mathbb Z)$ transformation allows us to view
each individual point as the intersection of the D3-brane with a (1,0) 7-brane,
globally not all 7-branes can be dualised into the same $(p,q)$-frame. This is
the origin of the reduction of the number of independent zero-modes to
(\ref{Ferminumber}).  We offer three independent derivations for the specific
value (\ref{Ferminumber}): First, by considering the perturbative Sen limit in
section \ref{IIBorientifolds} one realizes that the D7-brane locus is in the
class $8\,K^{-1}_{B_2}$, the remaining $ 4\, c_1(B_2) \cdot C$ points being
associated with the intersection with the O7-plane.  The latter do not host
any additional $3$--$7$ modes.  The extra $3$--$7$ states are also seen,
independently of a deformation to weak coupling, by comparison with the
M5-brane picture in section \ref{sec:M5CY3}, and their number is universally
and uniquely predicted by anomaly cancellation as discussed in section
\ref{sec:anomalies}. The counting (\ref{Ferminumber}), and the arguments
leading to it, are independent of the dimension of $Y_n$ and hence hold true
for D3-branes wrapping curves on any base $B_{n-1}$ in F-theory.

Given the spectrum in table \ref{tbl:CY3fcmultiplets} one can read off the
right- and left-moving central charges of the effective 2d theory as 
\begin{equation}
  c_R = 6 g + 6 c_1(B_2) \cdot C \,,\qquad c_L = 6
  g + 4 c_1(B_2) \cdot C + n_{37} \,,
\end{equation}  
where $n_{37} = 8 c_1(B_2) \cdot C$ is the contribution from the
$3$--$7$ string sector.
It can then be seen that there is a gravitational anomaly
\begin{equation}
  c_L - c_R = 6 c_1(B_2) \cdot C \,,
\end{equation}
which will be cancelled by anomaly inflow from the bulk of the 6d $(1,0)$
theory as discussed in section \ref{sec:anomalies}.


\subsection{Strings in 4d $N=1$} \label{sec:D3CY4}

Let us now consider a D3-brane wrapping a curve $C$ inside a K\"ahler three-fold $B_3$ serving as the base of an elliptic Calabi-Yau 4-fold $Y_4$ on which we compactify F-theory to four dimensions.
This setup gives rise to a string in the extended spacetime $\mathbb R^{1,3}$ and is expected to preserve $(0,2)$ supersymmetry.

\subsubsection{Duality Twist}

The $SU(4)_R$ symmetry of the $N=4$ SYM theory on the D3-brane decomposes, as in (\ref{CY4RSym}), into the rotation group $SO(2)_T$ associated with the two extended dimensions transverse to the string as well as an $SU(2)_R \times U(1)_R$ group. The latter represents the structure group of the normal bundle $N_{C/B_3}$.
Together with the universal decomposition (\ref{eqn:LorentzDecomp}) of $SO(1,3)_L$, this results in the following decomposition of the supercharges of the $N=4$ SYM theory:
\begin{equation}
  \begin{aligned}
  G_{\rm total} \ &\rightarrow \ SU(2)_R \times SO(1,1)_L \times U(1)_C \times U(1)_R \times SO(2)_T \times U(1)_ D\cr 
  ({\bf 2}, {\bf 1}, {\bf \bar 4})_1 \  &\rightarrow \ 
    {\bf 2}_{1;1,0,-1,1} \oplus {\bf 2}_{-1;-1,0,-1,1} \oplus  {\bf 1}_{1;1,-1,1,1} 
    \cr &\qquad\oplus {\bf 1}_{-1;-1,1,1,1} \oplus 
    {\bf 1}_{1;1,-1,1,1} \oplus {\bf 1}_{-1;-1,1,1,1} \cr 
  ({\bf 1}, {\bf 2}, {\bf 4})_{-1} \  &\rightarrow \ 
{\bf 2}_{1;-1,0,1,-1} \oplus {\bf 2}_{-1;1,0,1,-1} \oplus  {\bf
1}_{1;-1,1,-1,-1} \cr &\qquad\oplus {\bf 1}_{-1;1,1,-1,-1} \oplus 
{\bf 1}_{1;-1,-1,-1,-1} \oplus {\bf 1}_{-1;1,-1,-1,-1} \, .
  \end{aligned}
\end{equation}
The topological and duality twist required to preserve the expected two right-moving supersymmetries take the form 
\be\ba
 T_C^{\text{twist}} &= \frac{1}{2}(T_C + T_R) \cr
 T_D^{\text{twist}} &= \frac{1}{2}(T_D + T_R) \,.
\ea\ee
After this twist the supercharges decompose as in
  \begin{align}
  G_{\rm total} \  &\rightarrow \ SU(2)_R \times SO(1,1)_L \times U(1)^{\rm
  twist}_C \times U(1)^{\rm twist}_{D}  \times SO(2)_T \cr 
  ({\bf 2}, {\bf 1}, {\bf \bar 4})_{1}  &\rightarrow \ {\bf 2}_{1;\frac{1}{2},\frac{1}{2},-1} \oplus {\bf 2}_{-1;-\frac{1}{2},\frac{1}{2},-1} \oplus  {\bf 1}_{1;0,0,1} \oplus {\bf 1}_{-1;-1,0,1} \oplus 
{\bf 1}_{1;1,1,1} \oplus {\bf 1}_{-1;0,1,1} \\
  ({\bf 1}, {\bf 2}, {\bf 4})_{-1} &\rightarrow \  {\bf 2}_{1;-\frac{1}{2},-\frac{1}{2},1} \oplus {\bf 2}_{-1;\frac{1}{2},-\frac{1}{2},1} \oplus  {\bf 1}_{1;0,0,-1} \oplus {\bf 1}_{-1;1,0,-1} \oplus 
{\bf 1}_{1;-1,-1,-1} \oplus {\bf 1}_{-1;0,-1,-1} \,, \nonumber 
\end{align}
and one ends up with two positive chirality scalar supercharges
in 2d
\be
  Q_+ \, :\, {\bf 1}_{1;0,0,1}\qquad \qquad  \widetilde{Q}_+ \,:\, {\bf 1}_{1;0,0,-1} \,,
\ee
with $U(1)_R$ charge $\pm 1$. 
To determine the bulk matter on the twisted D3-brane we analyze the decomposition of the 
4d $N=4$ vector multiplet into fields
transforming in the following representations
  \begin{align} \label{N4decompD3inB3}
    &\qquad SU(2)_R \times SO(1,1)_L \times U(1)_C^{\rm twist} \times
    U(1)_D^{\rm twist} \times SO(2)_T \times U(1)_R \cr
    A \,&:\, \quad {\bf 1}_{2,0,*,0,0} \oplus {\bf 1}_{-2,0,*,0,0} \oplus {\bf
    1}_{0,1,*,0,0} \oplus {\bf 1}_{0,-1,*,0,0} \cr
       &  \qquad    = v_+ \oplus v_- \oplus  \bar a \oplus  a \cr
    \phi \,&:\, \quad {\bf 1}_{0,0,0,2,0} \oplus {\bf 1}_{0,0,0,-2,0} \oplus {\bf
    2}_{0,\frac{1}{2},\frac{1}{2},0,1} \oplus {\bf
    2}_{0,-\frac{1}{2},-\frac{1}{2},0,-1} \cr
      &  \qquad =   \bar g \oplus g \oplus  \varphi \oplus \bar
      \varphi \\
    \Psi \,&:\, \quad {\bf 2}_{1,\frac{1}{2},\frac{1}{2},1,0} \oplus {\bf
    1}_{1,1,1,-1,1} \oplus {\bf 1}_{1,0,0,-1,-1} \oplus {\bf
    2}_{-1,-\frac{1}{2},\frac{1}{2},1,0} \oplus {\bf 1}_{-1,0,1,-1,1} \oplus
    {\bf 1}_{-1,-1,0,-1,-1} \cr
    & \qquad = \mu_+ \oplus \psi_+ \oplus \gamma_+ \oplus \rho_- \oplus \lambda_- \oplus \beta_- \cr
    \widetilde{\Psi} \,&:\, \quad {\bf 2}_{1,-\frac{1}{2},-\frac{1}{2},-1,0}
    \oplus {\bf 1}_{1,-1,-1,1,-1} \oplus {\bf 1}_{1,0,0,1,1} \oplus {\bf
    2}_{-1,\frac{1}{2},-\frac{1}{2},-1,0} \oplus {\bf 1}_{-1,0,-1,1,-1} \oplus
    {\bf 1}_{-1,1,0,1,1} \cr
     & \qquad  = \tilde\mu_+ \oplus \tilde\psi_+ \oplus
     \tilde\gamma_+ \oplus \tilde\rho_- \oplus \tilde\lambda_- \oplus
     \tilde\beta_-  \,. \nonumber
  \end{align} 

The fields in the topologically half-twisted theory on $\mathbb R^{1,1} \times
C$ transform as bundle valued differential forms on $C$, which can be
determined in a manner similar to the discussion for the string in 6d in
section \ref{sec:CY3D3}.
 The fundamental representation ${\bf 2}$ of the $SU(2)_R$ symmetry group indicates that the differential forms take value in the normal bundle $N_{C/B_3}$, whose structure group this $SU(2)_R$ is.
The form degree in turn is determined by the topological twist charge $q_C^{\rm twist}$, with our conventions being that a state of topological and duality twist charges $(q_C^{\rm twist}, q_D^{\rm twist}) = (1,0)$ or $(-1,0)$  transforms as an element of $\Omega^{0,1}(C)$ or, respectively $\Omega^{1,0}(C)$. 
In addition sections of the normal bundle $N_{C/B_3}$ carry twist charges. To compute these consider the adjunction formula 
\be \label{adjunctionD3B3}
K_C = K_{B_3}|_C \otimes  \wedge^2 N_{C/B_3} =  {\cal L}^{-1}_D \otimes  \wedge^2 N_{C/B_3} \,.
\ee
Since in our conventions we associate with $K_C$ twist charge $q_C^{\rm twist}=-1$ and with the duality twist bundle 
$\mathcal{L}_D$ twist charge $q_D^{\rm twist} = -1$,
this fixes the charges associated with sections of $N_{C/B_3} $ as 
\be
N_{C/B_3}:   (q_C^{\rm twist}, q_D^{\rm twist}) = \left(-\frac{1}{2}, - \frac{1}{2}\right)\,.
\ee
Following this logic one systematically arrives at the bundle assignments displayed in the last column of  table \ref{tbl:CY4fc}. The results are in agreement with the way the fields enter the topologically twisted action, to which we now turn.

\begin{table}
  \centering
\begin{equation*}
  \begin{array}{|c|c|c|c|c|c|}
    \hline
    (q_C^{\rm twist}, q_D^{\rm twist}) & \multicolumn{2}{c|}{\rm Fermions} & \multicolumn{2}{c|}{\rm Bosons} &
    \text{Form Type}  \cr\hline
    (1/2, 1/2) & {\bf 2}_1 & \mu_+ & {\bf 2}_0 & \varphi & \Omega^{0,1}(C,
    N_{C/B_3} \otimes
    \mathcal{L}_D^{-1}) 
     \cr
  (-1/2, -1/2) & {\bf 2}_1 & \tilde\mu_+& {\bf 2}_0 & \bar \varphi
  &\Omega^{0,0}(C, N_{C/B_3}) 
   \cr\hline
   (1, 1) & {\bf 1}_1 & \psi_+ &{\bf 1}_0 & \bar a & \Omega^{0,1}(C, \mathcal{L}_D^{-1}) 
    \cr
 (-1, -1) & {\bf 1}_1 &\tilde\psi_+   & {\bf 1}_0 & a & \Omega^{0,0}(C,
 \wedge^2 N_{C/B_3}) 
 \cr\hline
  (0, 0) & {\bf 1}_1 & \gamma_+ & {\bf 1}_0 &  g & \Omega^{0,0}(C) 
 \cr   
(0, 0) & {\bf 1}_1  &\tilde \gamma_+  & {\bf 1}_0 & \bar g   & \Omega^{0,0}(C) 
 \cr   \hline
      (-1/2, 1/2) & {\bf 2}_{-1}& \rho_- & & & \Omega^{0,0}(C, N_{C/B_3} \otimes
      \mathcal{L}_D^{-1})
     \cr
        (1/2, -1/2) & {\bf 2}_{-1} & \tilde\rho_-& && \Omega^{0,1}(C, N_{C/B_3}) 
         \cr\hline
  (-1, 0) & {\bf 1}_{-1} & \beta_-& & & \Omega^{0,0}(C, K_C) 
    \cr
       (1, 0) & {\bf 1}_{-1} & \tilde \beta_- &&& \Omega^{0,1}(C)
        \cr\hline
        (0, 1) & {\bf 1}_{-1} &\lambda_- & {\bf 1}_2 & v_+ & \Omega^{0,0}(C,
        \mathcal{L}_D^{-1}) 
    \cr
    (0, -1) & {\bf 1}_{-1}  &\tilde \lambda_-& {\bf 1}_{-2}& v_- &
    \Omega^{0,0}(C, \mathcal{L}_D^{-1})  
     \cr\hline
  \end{array}
  \end{equation*}
  \caption{The bulk field content of the $(0,2)$ field theory on a D3-brane
    along $C \subset B_3$. The fields are the twisted fields from the
    reduction of the $N=4$ SYM spectrum as defined in (\ref{N4decompD3inB3}). 
    We display only the 2d chirality, and the $SU(2)_R$-symmetry
    representation. The form type of each field is computed the action of the
    internal kinetic operators on the fields determined via the equations of
    motion and the BPS equations.}\label{tbl:CY4fc} 
\end{table}

\subsubsection{Action and BPS Equations}

For a D3-brane wrapping a curve in a Calabi-Yau four-fold the gauge
and topological parts of the action are the
same as in (\ref{GaugeAction}) for Calabi-Yau three-folds, as 
they do not depend on the choice of decomposition of the
R-symmetry. The fermion and scalar Lagrangians are
\begin{equation}
  \begin{aligned}
    \mathcal{L}_\text{fermion} = 
    &+ \psi\partial_-\widetilde{\psi} 
    - \psi\partial_\mathcal{A}\widetilde{\lambda} 
    - \beta\bar\partial\widetilde{\gamma}
    + \beta\partial_+\widetilde{\beta}
    + \lambda\bar\partial_\mathcal{A}\widetilde{\psi}
    - \lambda\partial_+\widetilde{\lambda}
    - \gamma\partial_-\widetilde{\gamma}
    + \gamma\partial\widetilde{\beta}
    \cr &+ \rho\partial_+\widetilde{\rho}
    - \rho\bar\partial_\mathcal{A}\widetilde{\mu}
    + \mu\partial_\mathcal{A}\widetilde{\rho}
    - \mu\partial_-\widetilde{\mu}
    + \widetilde{\psi}\partial_-\psi
    + \widetilde{\psi}\bar\partial_\mathcal{A}\lambda 
    - \widetilde{\gamma}\partial_-\gamma
    - \widetilde{\gamma}\bar\partial\beta
    \cr &- \widetilde{\lambda}\partial_\mathcal{A}\psi
    - \widetilde{\lambda}\partial_+\lambda
    + \widetilde{\beta}\partial\gamma
    + \widetilde{\beta}\partial_+\beta
    - \widetilde{\rho}\partial_+\rho
    - \widetilde{\rho}\partial_\mathcal{A}\mu
    + \widetilde{\mu}\bar\partial_\mathcal{A}\rho
    + \widetilde{\mu}\partial_-\mu \,,
  \end{aligned}
\end{equation}
and
\begin{equation}
  \begin{aligned}
    \mathcal{L}_\text{scalar} = 
    &- \partial_-g\partial_+\bar g
    - \partial_-\bar g\partial_+g
    + \bar\partial g\partial\bar g
    + \bar\partial\bar g\partial g
    \cr &+ \partial_+\varphi\partial_-\bar\varphi
    + \partial_+\bar\varphi\partial_-\varphi
    - \partial_\mathcal{A}\varphi\bar\partial_\mathcal{A}\bar\varphi
    - \partial_\mathcal{A}\bar\varphi\bar\partial_\mathcal{A}\varphi \,,
  \end{aligned}
\end{equation}
with the total action given by the combination (\ref{TotalAction}).
The supersymmetry variations, again obtained via dimensional reduction and for 
the fields as identified in appendix \ref{app:D3action},  are 
\be
\ba
    \sqrt{\tau_2}\delta v_- &= -2i(\widetilde{\epsilon}_-\lambda_- +
    \epsilon_-\widetilde{\lambda}_-) \cr
    \sqrt{\tau_2} \delta a &= -2i\epsilon_-\widetilde{\psi}_+ \cr
    \delta g &= -2 i \epsilon_-\gamma_+  \cr
    \delta\varphi_A &= 2i\widetilde{\epsilon}_-\widetilde{\mu}_{+A} 
\ea\qquad \quad 
\ba
    \delta v_+ &= 0 \cr
    \sqrt{\tau_2}\delta\bar a &= 2i\widetilde{\epsilon}_-\psi_+ \cr
    \delta\bar g &= -2i \widetilde{\epsilon}_-\widetilde{\gamma}_+  \cr
    \delta\bar\varphi_A &=   -2i\epsilon_-\mu_{+ A} \,.
\ea
\ee
We are here making manifest the $SU(2)_R$ representations of the fields, with index $A$ referring to the {\bf 2} of $SU(2)_R$.
For the fermions the variations are
\be
\ba
  \delta\lambda_- &= - \epsilon_-(\sqrt{\tau_2}F_{01} +
  \mathcal{F}_\mathcal{A}) 
   \cr
     \delta\beta_- &= -\widetilde\epsilon_-\partial g \cr
    \delta\psi_+ &= \epsilon_-\sqrt{\tau_2}(\bar\partial v_+ - \partial_+\bar a) 
    \cr
    \delta\gamma_+ &= \widetilde\epsilon_-\partial_+ g \cr
    \delta\mu_{+A} &= -\widetilde{\epsilon}\partial_+\varphi_A
   \cr
    \delta\rho_{-A} &= \widetilde{\epsilon}\partial_\mathcal{A}\varphi_A\,.
\ea
\qquad \qquad 
\ba
 \delta\widetilde\lambda &=
     - \widetilde\epsilon_-(\sqrt{\tau_2}F_{01} - \mathcal{F}_\mathcal{A}) 
     \cr
     \delta\widetilde{\beta} &=
    \epsilon_-\bar\partial\bar g \cr
    \delta\widetilde{\psi}_+& = -
        \widetilde{\epsilon}_-\sqrt{\tau_2}(\partial v_+ - \partial_+a)\cr
    \delta\widetilde{\gamma}_+ &=
    \epsilon_-\partial_+\bar g  \cr
    \delta\widetilde{\mu}_{+A} &= 
    \epsilon_-\partial_+\bar\varphi_A \cr
       \delta\widetilde{\rho}_{-A} &=
    \epsilon\bar\partial_\mathcal{A}\bar\varphi_A \,.
\ea
\ee
The BPS equations readily follow
\begin{align}
  F_{01} =\frac{1}{2}( \partial_-v_+ - \partial_+v_-) &= 0  & 
  \mathcal{F}_\mathcal{A} = \frac{1}{2} \sqrt{\tau_2} (\bar\partial a - \partial\bar a) &= 0\cr
  \bar\partial v_+ - \partial_+\bar a &= 0  &
  \partial v_+ - \partial_+ a &= 0  \cr
  \partial_+g &= 0  & \partial_+\bar g &= 0 \cr
  \partial g &= 0 & 
  \bar\partial \bar g &= 0  \cr
  \partial_\mathcal{A}\varphi_{A} &= 0  &
  \bar\partial_\mathcal{A}\bar\varphi_{A} &= 0  \cr
  \partial_+\varphi_{A} &= 0  & \partial_+\bar\varphi_A &= 0 \,.
\end{align}
  As in
section \ref{sec:CY3D3} the internal kinetic operator that acts on the twisted
fields is read off from the BPS equations for the bosonic fields and the
equations of motion for the fermionic fields. 
 The bundle sections listed in table \ref{tbl:CY4fc} which count the zero modes are consistent with this kinetic operator.


\subsubsection{Spectrum}

Upon dimensional reduction on $C$, the zero-modes of the effective 2d (0,2) theory 
are counted by the Dolbeault cohomology groups associated with the bundles in
table \ref{tbl:CY4fc}.
The spectrum consists of
three pairs of chiral plus conjugate-chiral multiplets, two pairs of Fermi plus
conjugate Fermi-multiplets, and the vector multiplet.
As usual, the appearance of $\epsilon_-$ versus $\tilde \epsilon_-$ in the supersymmetry variations fixes the notion of chiral versus conjugate-chiral as well as of Fermi versus conjugate-Fermi superfields. 
In this regard we stick to  the conventions of \cite{Witten:1993yc}.

Note that in table \ref{tbl:CY4fc} the number of zero-modes for multiplets
and their conjugates are equal; this follows from the discussion in appendix
\ref{app:bdlcoho}, which can be used to show that
\begin{equation}
  h^i(C, N_{C/B_3}) = h^{1-i}(C, N_{C/B_3} \otimes \mathcal{L}_D^{-1}) \,.
\end{equation}
Finally, the vector multiplet must be counted in the same way as for a D3-brane on a curve in $B_2$, discussed in section \ref{sec:D3CY3spec}. 
Unless the curve $C$ does not intersect the discriminant locus so that ${\cal
L}_D$ is trivial the vector multiplet is projected out at the massless level.
The bulk spectrum for non-trivial $\mathcal{L}_D$ is given in table \ref{eqn:CY4D3zm}.
The bulk spectrum is completed by $8 c_1(B_3) \cdot C$ Fermi multiplets from
massless $3$--$7$ string excitations.

 \begin{table}
    \centering
  \begin{tabular}{|c|c|c|c|}
  \hline
    \text{Fermions} & \text{Bosons} & \text{(0,2)  Multiplet} &    Zero-mode Cohomology 
    \cr\hline
    $\mu_+$ & $\varphi$ &    \text{ Chiral}   
    &
    \multirow{2}{*}{$h^0(C, N_{C/B_3})$} \cr
   $ \tilde \mu_+$ &$ \bar \varphi $& \text{Conjugate Chiral}  
   & \cr\hline
$ \tilde \psi_+ $ &$  a$ & \text{ Chiral}   
    &
    \multirow{2}{*}{$h^0(C, K_C \otimes \mathcal{L}_D) = g - 1 +
    c_1(B_3) \cdot C$} \cr
   $  \tilde \psi_+$ &$ \bar a $& \text{Conjugate Chiral}   & \cr\hline  
   $  \gamma_+ $ &$  g$ & \text{ Chiral}   
    &
    \multirow{2}{*}{$h^0(C) = 1$} \cr
   $  \tilde \gamma_+$ &$  \bar g $& \text{Conjugate Chiral}   & \cr\hline  

   $ \rho_- $& \text{---} &  \text{ Fermi} 
    &
    \multirow{2}{*}{$h^1(C, N_{C/B_3}) = h^0(C, N_{C/B_3}) -
     c_1(B_3) \cdot C$} \cr
  $ \tilde \rho_- $& \text{---} & \text{ Conjugate Fermi} \ 
    & \cr\hline
    $ \beta_- $& \text{---} &  \text{Fermi} 
    &
    \multirow{2}{*}{$h^1(C) = g$} \cr
  $ \tilde \beta_- $& \text{---} & \text{ Conjguate Fermi} \ 
    & \cr\hline
    $\lambda_-$&$ v_+$ &   \multirow{2}{*}{Vector}  
     & \multirow{2}{*}{0} \cr
 $   \tilde\lambda_- $& $v_-$& & \cr
\hline
  \end{tabular}
\caption{Structure of the 2d $(0,2)$ bulk multiplets in the effective theory on $\mathbb R^{1,1} \times C$ for $C$ inside the base $B_3$ of $Y_4$. \label{eqn:CY4D3zm}}
\end{table}

The spectrum as given in table \ref{eqn:CY4D3zm} allows the computation of the
right- and left-moving central charges, which will now depend on the further
numerical value $h^0(C, N_{C/B_3})$. This extra freedom is not present in the
Calabi-Yau three-fold set-up as the normal bundle is its own determinant
bundle. The total numbers of right- and left-moving degrees of freedom are
\begin{equation}
  n_R = 3( g + c_1(B_3) \cdot C + h^0(C, N_{C/B_3})) \,,\qquad
  n_L = 3( g + h^0(C, N_{C/B_3})) + c_1(B_3) \cdot C + n_{37} \,,
\end{equation}
where the left-moving central charge again includes the contribution 
\begin{equation}
  n_{37} = 8 c_1(B_3) \cdot C 
\end{equation}
from the $3$--$7$ sector.
The gravitational anomaly, which depends only on the line bundle
$\mathcal{L}_D = K^{-1}_{B_3}|_C$, is 
\begin{equation}
  c_L - c_R = 6 c_1(B_3) \cdot C \,.
\end{equation}


\subsection{`Strings' in 2d $(0,2)$ Theories}

In 2d F-theory compactifications tadpole cancellation necessitates the
inclusion of a D3-brane sector, describing additional `strings' filling the
full spacetime.  F-theory compactifications on an elliptic fibration $Y_5$
have been introduced in \cite{Schafer-Nameki:2016cfr,Apruzzi:2016iac} and give
rise to 2d $N=(0,2)$ supersymmetric theories.  For the above reason, the
status of the D3-brane sector is rather different in 2d compared to 6d and
4d F-theory models.  The D3-branes are an integral component of the
definition of the vacuum rather than a defect sector, and their understanding
is imperative to characterize the compactification.

We now describe the theory on a D3-brane wrapping a curve $C$ on the base
$B_4$ of an elliptically fibered Calabi-Yau five-fold $Y_5$, and will return
to the tadpole constraint in the context of the anomaly considerations in section 
\ref{sec:anomalies}.

\subsubsection{Duality Twist}

Since the 2d effective theory on
the D3-branes along $\mathbb R^{1,1}$ is now spacetime-filling,
there is no group of rotations transverse to the D3-brane theory in the
non-compact dimensions. The $SU(4)_R$ symmetry of the $N=4$ SYM theory
on the D3-branes simply decomposes into $SU(3)_R \times U(1)_R$, which is the
structure group of the normal bundle $N_{C/B_4}$ to the curve $C$ inside the
internal K\"ahler space $B_4$.  From the decomposition of the supercharges,
\be\label{CY5Gtotal}
\ba
G_{\rm total} \ & \rightarrow \ SU(3)_R \times SO(1,1)_L \times U(1)_C \times U(1)_R  \times U(1)_ D\cr 
({\bf 2}, {\bf 1}, {\bf \bar 4})_{+1} \  &\rightarrow \ {\bf \bar 3}_{1;1,1,1} \oplus {\bf \bar 3}_{-1;-1,1,1} \oplus {\bf 1}_{1;1,-3,1} \oplus {\bf 1}_{-1;-1,-3,1} \cr
({\bf 1}, {\bf 2}, {\bf 4})_{-1} \  &\rightarrow \ {\bf 3}_{1;-1,-1,-1} \oplus  {\bf 3}_{-1;1,-1,-1} \oplus  {\bf 1}_{1;-1,3,-1} \oplus {\bf 1}_{-1;1,3,-1}  \, ,
\ea
\ee
one infers that the topological and duality twists
\begin{align}\label{CY5Twist}
  T_C^{\text{twist}} &= \frac{1}{6}(3T_C + T_R) \cr
  T_D^{\text{twist}} &= \frac{1}{6}(3T_D + T_R) \,,
\end{align}
result in the expected two right-moving 
 scalar supercharges. These transform under $SO(1,1)_L \times U(1)^{\rm twist}_C \times U(1)^{\rm twist}_D\times U(1)_R$ as 
\begin{align}
  Q_+ \,&:\, {\bf 1}_{1;0,0, 3} & \widetilde{Q}_+ \,&:\, {\bf 1}_{1;0,0, -3} \,.
\end{align}
The $N=4$ fields have the following reduction:
\begin{equation}
  \begin{aligned}    \label{N4decompD3inB4} 
    &\qquad SU(3)_R \times SO(1,1) \times U(1)_C^{\rm twist} \times
    U(1)_D^{\rm twist} \times U(1)_R \cr
    A \,&:\, \quad {\bf 1}_{2,0,*,0} \oplus {\bf 1}_{-2,0,*,0} \oplus {\bf
    1}_{0,1,*,0} \oplus {\bf 1}_{0,-1,*,0} \cr
    &\, \quad = v_+ \oplus v_- \oplus \bar a \oplus a \cr
    \phi \,&:\, \quad {\bf 3}_{0,\frac{1}{3},\frac{1}{3},2} \oplus {\bf
      \overline{3}}_{0,-\frac{1}{3},-\frac{1}{3},-2} \cr 
      & \, \quad = \varphi \oplus \bar\varphi \cr
    \Psi \,&:\, \quad {\bf 3}_{1,\frac{1}{3},\frac{1}{3},-1} \oplus {\bf
    1}_{1,1,1,3} \oplus {\bf 1}_{-1,0,1,3} \oplus {\bf
    3}_{-1,-\frac{2}{3},\frac{1}{3},-1}  \cr
    &\, \quad  = \mu_+ \oplus \psi_{+} \oplus \lambda_- \oplus \rho_{-} \cr
    \widetilde{\Psi} \,&:\, \quad {\bf
      \overline{3}}_{1,-\frac{1}{3},-\frac{1}{3},1}
    \oplus {\bf 1}_{1,-1,-1,-3} \oplus {\bf 1}_{-1,0,-1,-3} \oplus {\bf
      \overline{3}}_{-1,\frac{2}{3},-\frac{1}{3},1} \cr
      &\, \quad  = \tilde\mu_+ \oplus \tilde\psi_{+} \oplus \tilde\lambda_- \oplus \tilde\rho_{-} \,.
  \end{aligned} 
\end{equation}

\begin{table}
  \centering
  \begin{tabular}{|c|c|c|c|c|c|}
    \hline
    $(q_C^\text{twist}, q_D^\text{twist})$ & \multicolumn{2}{c|}{Fermions} &
    \multicolumn{2}{c|}{Bosons} & Form Type \cr\hline
    $(1/3,1/3)$ & ${\bf 3}_1$ &  $\mu_+$ & ${\bf 3}_0$ & $\varphi $& $\Omega^{0,1}(C, \wedge^2N_{C/B_4}
    \otimes \mathcal{L}_D^{-1})$ \cr
    $(-1/3,-1/3)$ & ${\bf \overline{3}}_1$ & $\tilde\mu_+$ & ${\bf \overline{3}}_0$ &  $\bar \varphi$ & $\Omega^{0,0}(C,
     N_{C/B_4})$ \cr\hline
     $(1,1)$ & ${\bf 1}_1$ &$\psi_+$ & ${\bf 1}_0$ & $\bar a$  &
     $\Omega^{0,1}(C, \mathcal{L}_D^{-1})$ \cr
     $(-1,-1)$ & ${\bf 1}_1$& $\tilde\psi_+$ & ${\bf 1}_0$ &$a$& $\Omega^{0,0}(C, K_C \otimes
     \mathcal{L}_D)$ \cr\hline
     $(-2/3,1/3)$ & ${\bf 3}_{-1}$ & $\rho_-$ & - & - & $\Omega^{0,0}(C, \wedge^2N_{C/B_4}
         \otimes \mathcal{L}_D^{-1})$\cr
         $(2/3,-1/3)$ & ${\bf \overline{3}}_{-1}$ & $\tilde \rho_-$& -& - & $\Omega^{0,1}(C, N_{C/B_4})$ \cr\hline
         $(0,1)$ & ${\bf 1}_{-1}$ & $\lambda_-$ &  ${\bf 1}_{2}$ & $v_+$ & $\Omega^{0,0}(C,
         \mathcal{L}^{-1}_D)$ \cr
     $(0,-1)$ & ${\bf 1}_{-1}$ &$\tilde\lambda_-$& ${\bf 1}_{-2}$ & $v_-$ &$\Omega^{0,0}(C,
             \mathcal{L}^{-1}_D)$\cr\hline
  \end{tabular}
  \caption{The bulk field content of the $(0,2)$ field theory for a D3-brane wrapping a curve  $C \subset B_4$.
  We only display the $SO(1,1)_L$ charge and the
 $SU(3)_R$ symmetry representation.}\label{tbl:CY5fc}
\end{table}

The spectrum of the partially  topologically twisted theory on $\mathbb R^{1,1} \times C$  is given in table \ref{tbl:CY5fc}.
The bundles appearing in the last row are determined by the representation of the fields with respect to $SU(3)_R \times U(1)^{\rm twist}_C \times U(1)^{\rm twist}_L$ in 
close analogy to the procedure spelled out in section
\ref{sec:D3CY4}. 
This time, the adjunction formula 
\be
K_C = K_{B_4}|_C \otimes \wedge^3 N_{C/B_4} 
\ee
and the usual assignments of twist charges imply that sections of $K_C$,
$\mathcal{L}_D$ and of $N_{C/B_4} $ carry twist charges
\be
\ba
K_C: \qquad (q_C^{\rm twist},q_D^{\rm twist}) &= (-1,0) \cr 
\mathcal{L}_D: \qquad (q_C^{\rm twist},q_D^{\rm twist}) &= (0,-1) \cr
 \qquad N_{C/B_4}: \qquad (q_C^{\rm twist},q_D^{\rm twist}) &= (-1/3,-1/3) \,.
\ea
\ee

\subsubsection{String Action and BPS Spectrum}

Finally, let us turn to the action and supersymmetry transformations. The total action is (\ref{TotalAction}), where 
the gauge and topological parts of the Lagrangian are  (\ref{GaugeAction}),
while the scalar and fermion parts of the Lagrangian are twist dependent and in the present case given by
\begin{equation}
\ba
    \mathcal{L}_\text{fermion} =& 
    - \lambda_-\partial_+\widetilde\lambda_-
    + \lambda_-\bar\partial_\mathcal{A}\widetilde\psi_+
    - \rho_-\partial_+\widetilde\rho_- 
    + \rho_-\bar\partial_\mathcal{A}\widetilde\mu_+
    + \mu_+\partial_-\widetilde\mu_+ 
    - \mu_+\partial_\mathcal{A}\widetilde\rho_-
     \cr
    &
    + \psi_+\partial_-\widetilde\psi_+
    - \psi_+\partial_\mathcal{A}\widetilde\lambda_-
   - \widetilde\rho_-\partial_+\rho_-
    - \widetilde\rho_-\partial_\mathcal{A}\mu_+
    - \widetilde\lambda_-\partial_+\lambda_-
    - \widetilde\lambda_-\partial_\mathcal{A}\psi_+
     \cr
    &
    + \widetilde\mu_+\partial_-\mu_+
    + \widetilde\mu_+\bar\partial_\mathcal{A}\rho_-
    + \widetilde\psi_+\partial_-\psi_+
    + \widetilde\psi_+\bar\partial_\mathcal{A}\lambda_- \cr 
  \mathcal{L}_\text{scalar} =& 
    - \partial_+\varphi\partial_-\bar\varphi
    - \partial_+\bar\varphi\partial_-\varphi
    +
    \partial_\mathcal{A}\varphi\bar\partial_\mathcal{A}\bar\varphi
    +
    \partial_\mathcal{A}\bar\varphi\bar\partial_\mathcal{A}\varphi
    \,.
    \ea
\end{equation}
The bosonic supersymmetry variations leaving  this theory invariant take the form
\be\ba
    \sqrt{\tau_2}\delta v_- &=
    -2i(\widetilde\epsilon_-\lambda_- + \epsilon_-\widetilde\lambda_-)  \cr
    \sqrt{\tau_2}\delta a &= -2i\epsilon_-\widetilde\psi_+  \cr
    \delta\varphi_\alpha &= 2i\epsilon_-\mu_{+\alpha} 
\ea
\qquad 
\ba
    \delta v_+ &= 0  \cr
    \sqrt{\tau_2}\delta\bar a &= 2i\widetilde\epsilon_-\psi_+  \cr
    \delta\bar\varphi^{\dot{\alpha}} &= -
    2i\widetilde\epsilon_-\widetilde\mu_+^{\dot{\alpha}}  \,.
\ea
\ee
Their fermionic counterparts are found to be
\be
\ba
  \delta\lambda_- &= \epsilon_-(\sqrt{\tau_2}F_{01} + \mathcal{F}_\mathcal{A}) \cr
    \delta\psi_+ &= \sqrt{\tau_2}\epsilon_-(\bar\partial v_+ - \partial_+\bar a)  \cr
    \delta \mu_{+\alpha} &= -\widetilde\epsilon\partial_+\varphi_\alpha  \cr
    \delta\rho_{-\alpha} &=
    \widetilde\epsilon\partial_{\mathcal{A}}\varphi_\alpha 
\ea
\qquad
\ba
  \delta\widetilde\lambda_- &= \widetilde\epsilon_-(\sqrt{\tau_2}F_{01} -
  \mathcal{F}_\mathcal{A}) \cr
    \delta\widetilde\psi_+ &= -\sqrt{\tau_2}\widetilde\epsilon(\partial v_+ -
    \partial_+ a)  \cr
    \delta \widetilde\mu_+^{\dot{\alpha}} &=
    \epsilon\partial_+\bar\varphi^{\dot{\alpha}} \cr
    \delta\widetilde\rho_-^{\dot{\alpha}} &=
    \epsilon\bar\partial_\mathcal{A}\bar\varphi^{\dot{\alpha}} \,.
\ea
\ee
Here the indices $\alpha$ and $\dot \alpha$ refer to the ${\bf 3}$ and ${\bf \bar 3}$ representation of $SU(3)_R$, respectively. More details are given in appendix \ref{app:D3action}. 
The BPS equations are easily read off from the above variations. 

After dimensional reduction to two dimensions, the 2d zero-modes  organize  into two pairs of chiral plus
conjugate-chiral multiplets, a Fermi plus conjugate-Fermi multiplet, and the
vector multiplet. The latter is absent unless ${\cal L}_D$ is trivial,
i.e. unless the curve $C$ does not intersect the discriminant locus of the
elliptic fibration.  The number of zero-modes for each of these fields is
computed by the cohomology groups associated with the bundles appearing in table \ref{tbl:CY5fc}.
Serre duality together with the identity 
\be
\wedge^2 N_{C/B_4} = N_{C/B_4}^\vee \otimes  \wedge^3 N_{C/B_4}\,,
\ee
which follows from (\ref{eqn:dualbdlthm}), guarantee that each multiplet and
its conjugate indeed appear with the same multiplicity. In particular,
\be
h^{i}(C,N_{C/B_4}) = h^{1-i}(C, N_{C/B_4}^\vee \otimes K_C) = h^{1-i}(C,
\wedge^2 N_{C/B_4}^\vee \otimes {\cal L}_D^{-1}|_C)\,.
\ee
The zero-mode spectrum is given in table \ref{tab:CY5D3zm}. 

  
  \begin{table}
    \centering
  \begin{tabular}{|c|c|c|c|}
  \hline
    \text{Fermions} & \text{Bosons} & \text{(0,2)  Multiplet} &    Zero-mode Cohomology 
    \cr\hline
    $\mu_+$ & $\varphi$ &    \text{Chiral}   
    &
    \multirow{2}{*}{$h^0(C, N_{C/B_4})$} \cr
   $ \tilde \mu_+$ &$ \bar\varphi $& \text{Conjugate Chiral}  
   & \cr\hline
$ \tilde \psi_+ $ &$  a$ & \text{ Chiral}   
    &
    \multirow{2}{*}{$h^0(C, K_C \otimes \mathcal{L}_D) = g - 1 + c_1(B_4) \cdot C$} \cr
   $  \psi_+$ &$ \bar a $& \text{Conjugate Chiral}   & \cr\hline  
   $ \rho_- $& \text{---} &  \text{Fermi} 
    &
    \multirow{2}{*}{$h^1(C, N_{C/B_4}) = h^0(C, N_{C/B_4}) + g - 1 -
    c_1(B_4) \cdot C$} \cr
  $ \tilde \rho_- $& \text{---} & \text{Conjugate Fermi} \ 
    & \cr\hline
    $\lambda_-$&$ v_+$ &   \multirow{2}{*}{Vector}  
     & \multirow{2}{*}{0} \cr
 $   \tilde\lambda_- $& $v_-$& & \cr
\hline
  \end{tabular}
\caption{Structure of 2d (0,2) bulk multiplets in effective theory on $\mathbb R^{1,1} \times C$ inside base of $CY_5$. \label{tab:CY5D3zm}}
\end{table}

The bulk spectrum is completed by $8 c_1(B_4) \cdot C$ Fermi multiplets from
the $3$--$7$ sector. From the spectrum the central charges can be computed as
\begin{equation}
\ba
  n_R &= 3( g + c_1(B_4) \cdot C + h^0(C, N_{C/B_4}) - 1)\cr 
    n_L &= 3( g + h^0(C, N_{C/B_4}) - 1) + 9
  c_1(B_4) \cdot C \,,
\ea
\end{equation}
where we have included in the computation of $n_L$ the $n _{37} = 8
c_1(B_4) \cdot C$ modes
from the $3$--$7$ sector. The gravitational anomaly is then
\begin{equation}
  c_L - c_R = 6 \, c_1(B_4) \cdot C \,.
\end{equation}


\subsection{Duality Twist for 2d $N=(0, 6)$ Theories}
\label{sec:Nis6}

So far we have focused on chiral 2d theories arising from duality twisted $N=4$ SYM on a curve with a brane realization in terms of D3-branes wrapping curves in an F-theory compactification, which naturally incorporates the varying coupling. 
In this section we shall study one case that does not have such a brane realization, but nevertheless represents a consistent topological twist of the $N=4$ SYM theory, and upon twisted dimensional reduction results in an $N=(0,6)$ theory. This may in particular be of interest in the light of recent developments on novel supersymmetric theories such as $N=3$ SYM in 4d \cite{Garcia-Etxebarria:2015wns}. 

For this, we consider again the R-symmetry decomposition of $N=4$ SYM which we used for the CY$_5$ twist in (\ref{CY5RSym})
\be
\ba
SU(4)_R &\ \rightarrow\  SU(3)_R \times U(1)_R\cr
    {\bf 4} &\ \rightarrow\  {\bf 1}_3 \oplus {\bf 3}_{-1} \cr
    {\bf 6} &\ \rightarrow\  {\bf 3}_2 \oplus {\bf 3}_{-2} \,,
\ea
\ee
however unlike for the twist discussed in the CY$_5$ brane realization (\ref{CY5Twist}) we could instead consider the twist 
\begin{align}\label{SixTwist}
  T_C^{\text{twist}} &= \frac{1}{2}(T_C - T_R) \cr
  T_D^{\text{twist}} &= \frac{1}{2}(T_D - T_R) \,.
\end{align}
With the same representation theoretic decomposition in (\ref{CY5Gtotal}), but the twist defined as in (\ref{SixTwist}), we end up with the following supercharges under $SU(3) \times SO(1,1) \times U(1)_C^{\rm twist} \times U(1)_D^{\rm twist}$ 
\be
Q: \qquad \bar{\bf 3}_{1; 0,0} \oplus {\bf 3}_{1; 0, 0}  \,,
\ee
thus giving an $N=(0,6)$ supersymmetric theory in 2d. It would be interesting to explore geometric realizations of these theories in more detail in the future. 


\subsection{Hitchin Equations with $\mathcal{L}_D$-twist}
\label{sec:Hitch}

The BPS equations involving the reduction of the 4d gauge field $A_\mu$ are
the same across all dimensions of Calabi-Yau compactifications that we here
consider. 
Of particular interest is the equation
\begin{align} \label{BPSb}
{\cal F}_{\cal A} = \frac{1}{2} \Big( \bar\partial_\mathcal{A}(\sqrt{\tau_2}a) -
  \partial_\mathcal{A}(\sqrt{\tau_2}\bar{a}) \Big)  &= 0 \,.
\end{align}
This equation is expressed in terms of the
$U(1)_D$ eigenstate fields $\sqrt{\tau_2}a$ and $\sqrt{\tau_2}\bar{a}$. These
fields transform as bundle-valued sections,
\begin{align} \label{dualitytwistedafields}
  \sqrt{\tau_2} \bar a &\in \Gamma(\Omega^{0,1}(C, \mathcal{L}_D^{-1})) \cr
  \sqrt{\tau_2} a &\in \Gamma(\Omega^{0,0}(C, K_C \otimes \mathcal{L}_D)) \,,
\end{align}
where the two bundles in questions are dual to each other. The number of sections
of these particular bundles, for a curve $C$ inside a Calabi-Yau $Y_n$ with $n=3, 4, 5$ intersecting the discriminant locus, is
\begin{equation}
  g - 1 + \text{deg}(\mathcal{L}_D)  = g -1 + c_1(B_{n-1})\cdot C\,.
\end{equation}

We see that the equation (\ref{BPSb}) is a $U(1)_D$ duality twisted version of the usual abelian Hitchin equation ${\cal F}= 0$, with ${\cal F}$ the abelian field strength along $C$. Apart from the fact that $\sqrt{\tau_2} a$ and $\sqrt{\tau_2} \bar a$ are sections of the bundles (\ref{dualitytwistedafields}), the expression ${\cal F}_{\cal A}$ involves the $U(1)_D$ covariant derivatives defined in section (\ref{sec:D3CY3spec}). The solutions will be $U(1)_D$ twisted flat connections on $C$. It would be interesting to study the extension of such a duality twisted Hitchin system to the non-abelian case and the underlying mathematics further.

\subsection{Type IIB Limit and Quantum Higgsing in F-theory}\label{IIBorientifolds}

It is very instructive to compare our findings for the spectrum on a D3-brane wrapping a curve in F-theory to the situation in the special case of perturbative 
orientifolds.
If the F-theory setup has a smooth Type IIB orientifold limit, the base space $B_{n-1}$ of the fibration is the quotient of a Calabi-Yau $(n-1)$-fold $X_{n-1}$ by a holomorphic involution $\sigma$. Let us assume that the curve $C \subset B_{n-1}$ wrapped by the D3-brane is not contained in the discriminant locus of the elliptic fibration. 
In Type IIB two different cases are to be considered: Either the double cover $\tilde{C} \subset X_{n-1}$ of the curve $C$ is invariant as a whole under the involution $\sigma$, but not pointwise, or it splits into two irreducible curves $\tilde C = C_+ \cup C_-$ exchanged by the orientifold, $\sigma(C_\pm) = C_\mp$.
In the first case the perturbative gauge group $U(1)$ is broken by the orientifold action to $O(1) = \mathbb Z_2$, and the $U(1)$ gauge potential and gaugino partners are projected out. 
The second case is more subtle: 
If $[C_+] = [C_-]$ in homology, then 
perturbatively there exists an unbroken gauge group 
\be
U(1)_- = U(1)_{C_+} - U(1)_{C_-}
\ee
 with an associated massless gauge field and gaugino partners.
 If $[C_+] \neq [C_-]$ a St\"uckelberg mechanism involving the orientifold-odd Ramond-Ramond two-form potential $C_2$ renders the gauge boson massive and breaks the gauge symmetry to a global $U(1)$ or $\mathbb Z_k$ symmetry.

The possibility of a massless $U(1)_- = U(1)_{C_+} - U(1)_{C_-}$  is to be contrasted with our non-perturbative F-theory result that a massless vector multiplet is found in the effective theory along the string if and only if 
the elliptic surface   $\widehat C \subset Y_{n}$, defined by the restriction of the elliptic fibration to $C$, is trivially fibered, i.e.   $\widehat C = C \times T^2$ globally. 
This situation occurs precisely if $C$ does not intersect the discriminant locus $\Delta$.
In the language of the Type IIB orientifold, this is equivalent to the statement that the curve $\tilde C \subset X_{n-1}$ does not intersect the divisor wrapped by the orientifold plane, whose class we denote by $[O7]$.
Indeed, the 7-brane tadpole cancellation condition relates  $[O7]$ to the class of the divisors wrapped by the D7-branes and their orientifold images, 
$8 [{\rm O7}] = \sum_i ([{\rm D7}]_i + [{\rm D7}]'_i)$. Therefore if the curve $\tilde C$ intersects any of the 7-branes, it necessarily intersects also the O7-plane and vice versa. 
A trivial fibration  $\widehat C = C \times T^2$ then corresponds, in perturbative Type IIB orientifolds, to a D3-brane on $\tilde C = C_+ \cup C_-$ with 
$C_+ \cap O7 = \emptyset = C_- \cap O7$. 

By contrast, if $\tilde C = C_+ \cup C_-$ but $C_+ \cap C_- \neq \emptyset$, then the corresponding $\widehat C$ is non-trivially fibered. Non-perturbatively no massless gauge bosons are found in the effective 2d action despite the fact that the perturbative limit does exhibit a $U(1)_-$ gauge group, at least if $[C_+] =[C_-]$.   

The key to understanding this apparent paradox is that 
at $C_+ \cap C_-$
extra zero-modes localize from strings between the D3-brane on $C_+$ and its image brane on $C_-$. They carry charge $ q_- = 2$
with respect to the perturbative gauge group $U(1)_-$ (in units where the
$3$--$7$ strings carry charge $q_- = 1$). These zero-modes assemble into  chiral multiplets, and are crucial for consistency of the setup: First, since the fermions in the chiral multiplets are of positive chirality, they precisely cancel the $U(1)_-$ anomaly due to the negative chirality $3$--$7$ strings. Indeed, by assumption the $U(1)_-$ is not rendered massive by a St\"uckelberg mechanism and hence must be non-anomalous because massless $U(1)$s do not participate in a Green-Schwarz mechanism in 2d.  
Second,
the chiral scalar fields correspond to the recombination moduli of $\tilde C = C_+ \cup C_-$. In the Type IIB orientifold a non-trivial VEV of the scalar component triggers a recombination of the split curve $\tilde C$ into an irreducible smooth curve.
This recombination higgses $U(1)_- \rightarrow \mathbb Z_2$, consistent with the fact that after the recombination $\tilde C$ is orientifold invariant and carries gauge group $O(1) = \mathbb Z_2$.

Our results on the duality twisted D3-brane theory strongly suggest that {\it generically}\footnote{There is one degenerate exception described below.} F-theory  does not distinguish between the perturbative phase at the origin of the Higgs branch with perturbative gauge group $U(1)_-$ and the phase away from the origin with gauge group $O(1)$. This will indeed be corroborated further from a geometric perspective below.
Our physical interpretation is that
strong coupling effects dynamically higgs the perturbative gauge group in F-theory - an effect which is invisible in the perturbative description via Type IIB orientifolds.
Note that the string coupling $g_s$ indeed enters the non-perturbative regime near the O7-plane, which is where the recombination modes in the $C_+-C_-$ sector are localised.\footnote{More precisely, the O7-plane famously splits in F-theory precisely due to strong coupling effects \cite{Sen:1997gv}. See also \cite{Cvetic:2011gp,Halverson:2016vwx} for recent discussions. } 

It is worth stressing that such a non-perturbatively triggered brane recombination process does not occur for 7-branes intersecting conventional $O7^-$-planes. A transverse intersection locus $D7_i \cap O7$ in F-theory is always 6-dimensional, whereas the locus $D3 \cap O7$ in the configurations considered here is  only 2-dimensional.
The orientifold action projects out the zero-modes in the spectrum of the D7-D7' strings located at the orientifold and thus no non-perturbative recombination is possible. The D7-D7' zero-modes located away from the O7-plane, if any, are not projected out by the orientifold, but since these are not located in the non- region of perturbative $g_s$,  no recombination is triggered in F-theory. Indeed, for 7-branes both the recombined phase and the phase with gauge group $U(1)$ are accessible as described in \cite{Grimm:2010ez}.\footnote{A stack of 7-branes in Type IIB orientifolds with gauge group $U(N)$, on the other hand, does give rise to zero-modes in the anti-symmetric representation on top of the O7-plane, but again for non-abelian gauge groups no non-perturbative Higgsing is observed.}

To understand this effect further, consider the Weierstrass model for an elliptic Calabi-Yau $n$-fold $Y_n$  with base  $B_{n-1} $. We can parametrise the Weierstrass model following Sen as \cite{Sen:1997gv}
\be
\ba \label{WeierSen1}
y^2 &= x^3 + f x  + g  \\
f &= -3 h^2 +   \epsilon   \eta \, \qquad 
g = -2 h^3 +  \epsilon  h  \eta -  \frac{1}{12} \epsilon^2   \chi \,  \\
\Delta & \simeq \epsilon^2 h^2 (\eta^2 - h \chi) + {\cal O}(\epsilon^3) \,,
\ea
\ee
with $f$ and $g$ sections of $K^{-4}_{B_{n-1}}$ and $K^{-6}_{B_{n-1}}$, respectively, and $h$, $\eta$ and $\chi$ generic polynomials of appropriate degree.
The orientifold double-cover is described by the Calabi-Yau space $X_{n-1}$ 
\bea
X_{n-1}:  \,  \xi^2 = h   
\eea
by adding a new coordinate $\xi$ and letting $h$ depend on the coordinates on $B_{n-1}$.
The orientifold action $\sigma: \xi \rightarrow - \xi$ leaves the O7-plane located at $h=0$ invariant. 

Consider now the family of curves 
\bea \label{Cdelta}
C_\delta: h  =  p_1^2 + \delta \, p_2  \, \subset \, B_{n-1}
\eea
with $p_1$ and $p_2$ generic polynomials on $B_{n-1}$ transforming as sections of $K^{-1}_{B_{n-1}}$ and $K^{-2}_{B_{n-1}}$ and a parameter $\delta \in \mathbb R$.    
The double cover of this curve on the Type IIB Calabi-Yau $X_{n-1}$ is given by
\bea
\tilde C_\delta: \,  \xi^2 =  p_1^2 + \delta \,  p_2  \, \subset \, X_{n-1}.
\eea
If $\delta =0$, $\tilde C_\delta$
splits into $\tilde C_0 = C_+ \cup C_-$ with
 \bea
 C_+:  \, \xi = p_1, \qquad C_-: \, \xi = - p_1.
 \eea
The two components $C_+$ and $C_-$ are exchanged by the orientifold action $\xi \rightarrow - \xi$ and intersect at the orientifold locus $\xi=0$.
Hence the parameter $\delta$ parametrizes the Higgs branch for $U(1)_-$ in the Type IIB limit: For $\delta =0$, the perturbative gauge potential of $U(1)_-$ is massless, while for $\delta \neq 0$ it is broken by a Higgs effect in which the $3-3'$ modes acquire a VEV.


The topology of the curve $C_\delta$ in $B_{n-1}$ wrapped by the D3-brane in F-theory does not change as drastically for varying $\delta$.
Rather, the parameter $\delta$ characterizes the intersection of $C_\delta$ with the locus $h=0$, given by
\bea \label{hint1}
\{h=0\} \cap C_\delta:    \quad      \{h=0\} \cap \{p_1 = \pm \sqrt{\delta\, p_2} \} \,.
\eea
 { In the Type IIB limit $\epsilon \rightarrow 0$}, and only in this limit, $\{h=0\}$ describes the location of the O7-plane.
As $\delta \rightarrow 0$, the pairs of intersection points $p_1 = \pm \sqrt{\delta\, p_2}$ with $h=0$ come closer together and finally coalesce as $\delta =0$, corresponding to vanishing Higgs VEV.
Therefore, in the limit  $\epsilon \rightarrow 0$, we can geometrically identify  the VEV of the Higgs fields with the separation of pairs of intersection points of $C_\delta$ with the O7-plane.
This is also intuitive because the Higgs fields correspond to $3-3'$ modes at the intersection of $C_+$ with $C_-$. This is precisely the intersection with the O7-plane in Type IIB, which in fact is a double point from the perspective of $C_\delta$ at $\delta =0$: The two signs in (\ref{hint1}) correspond to the intersection of $C_+$ and $C_-$ with the O7-plane. As these points are separated  for $\delta \neq 0$, the $3-3'$ string modes acquire mass - indicating the Higgsing. 

\begin{figure}
\centering
\includegraphics[width=11cm]{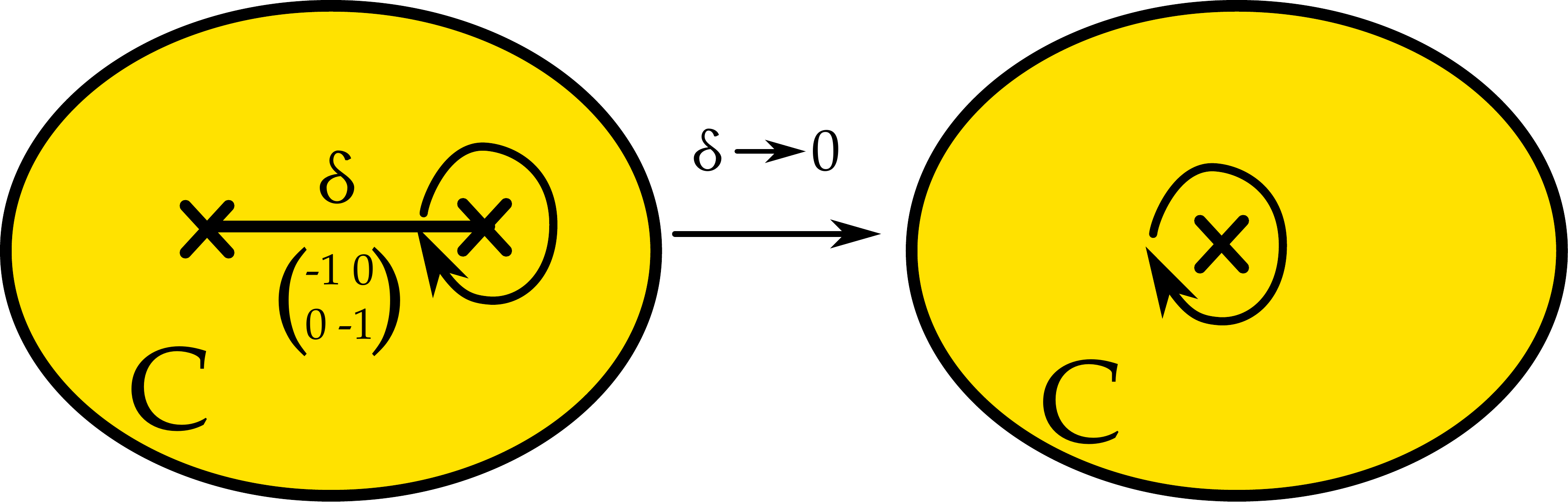}
\caption{$SL(2,\mathbb Z)$  monodromies on the D3-brane in a degenerate $SO(8)$ model. The perturbative restoration of the $U(1)$ gauge symmetry as $\delta \rightarrow0$ corresponds to absence of any monodromies along $C$.  }\label{fig:SO8}
\end{figure}

Finally, we are in a position to address the non-perturbative effects in F-theory. Such effects are parametrized by non-zero values of the parameter $\epsilon$ in (\ref{WeierSen1}).
Famously, as we take the effects of order $\epsilon^3$ into account, the locus $h=0$ in the discriminant $\Delta$ no longer corresponds to the true location of the orientifold plane. Rather the double zeroes of $\Delta$ at $h^2=0$ obtained by ignoring these terms split into two, with their separation being controlled by $\epsilon$. This is Sen's celebrated splitting of the orientifold plane \cite{Sen:1997gv} into a mutually non-local $(p,q)$-7-brane system. If we intersect with $C_\delta$, we see that, independently of the effect from $\delta$, a splitting of the intersection points with the O7-plane system occurs due to strong coupling effects parametrized by $\epsilon$. We claim that this separation must still be viewed as parametrizing the VEV of the Higgs fields. In particular, as $\epsilon \neq 0$, a Higgsing occurs in F-theory irrespective of whether or not $\delta  \neq 0$. This is the non-perturbative Higgsing alluded to above.

The reason why the Higgsing is inevitable in F-theory is because the string coupling always diverges close to the orientifold plane.
The only exception is  a situation in which the 7-brane tadpole is cancelled locally everywhere, corresponding to a configuration of $SO(8)$ gauge group.
Such a background is engineered by taking 
the vanishing orders of $(f,g,\Delta)$  as $(2, 3, 6)$ along $h=0$.
From the perspective of (\ref{WeierSen1}) this is paramount to setting $\epsilon \equiv 0$ in $f$ and $g$. Note that we cannot afford any gauge group different from $SO(8)$ on any other locus if we want to ensure that no strong coupling effects occur. 
The resulting Weierstrass model is degenerate because self-intersections of the $SO(8)$ divisor in codimension-two necessarily lead to non-minimal fibers with vanishing orders ${\rm ord}(f,g, \delta) \geq (4,6,12)$. 
Let us nonetheless analyze the fibration over $C_\delta$, given by the surface 
\bea
\widehat C_\delta: y^2 = x^3  - 3 (p_3^2 + \delta \, p_6)^2 x  - 2  (p_3^2 + \delta \, p_6)^3 \,.
\eea
 Now, the limit  $\delta \rightarrow 0$ differs considerably also in the full F-theory.
 The discriminant locus of  $ \widehat C_\delta$ consists of pairs of points $p_3 = \pm \sqrt{\delta \,  p_6}$. Each of these points describes a defect along the curve $C_\delta$ due to an intersection with the $SO(8)$ locus. Pairs of such defects are connected by a branch-cut associated with a duality wall for the topologically twisted theory, as studied in \cite{Martucci:2014ema} for D3-branes wrapping surfaces. 
 As one encircles an $SO(8)$ defect and crosses a duality wall
 one picks up an 
 $SL(2,\mathbb Z)$ monodromy given by
 \bea
 M_{SO(8)} = \left(\begin{matrix} -1 & 0 \\ 0 & -1 \end{matrix}\right) \,.
 \eea
This monodromy acts as
 \bea
 \left(\begin{matrix} F  \\ F_D  \end{matrix}\right) \rightarrow M_{SO(8)}  \left(\begin{matrix} F  \\ F_D  \end{matrix}\right) \,,
 \eea
with $F$ the $U(1)_-$ gauge field on the D3-brane and $F_D$ its magnetic dual as in (\ref{Fdual1}). 
This is responsible for projecting out the $U(1)_-$ gauge potential. But as $\delta \rightarrow 0$ pairs of $SO(8)$
defects coalesce, and the monodromy disappears, see figure \ref{fig:SO8}. As a result the $U(1)$ gauge potential survives as a massless field. 
To conclude, the $U(1)_-$ is massless if and only if both $\epsilon =0$ and $\delta =0$.


These observations also suggest an interesting connection with the results of \cite{Harvey:2007ab,Cvetic:2011gp} concerning the localisation of some of the bulk D3 modes at the intersection with the O7-plane, as required from the perspective of anomaly inflow. 
As the brane-image brane pair recombines in Type IIB, one linear combination of the two complex recombination scalars remains as a massless modulus of the brane.
Typically we would expect this field to arise as a delocalized bulk field along the cycle. However, it is clear that e.g. for small recombination parameter $\delta $
the avatars of the recombination modes approximately localize around the former intersection locus between the two branes, which lies on top of the O7-plane.
From this perspective it is natural to identify the analogue of the  localised  bulk modes discussed in \cite{Harvey:2007ab,Cvetic:2011gp} as  these recombination modes. It would be interesting to verify or falsify this conjecture.

Let us also note that a similar effect occurs not only for strings from D3-branes wrapped on curves in F-theory, but also for D3-brane instantons associated with Euclidean D3-branes along surfaces. As reviewed e.g. in \cite{Blumenhagen:2008zz} the  D-brane instanton literature differentiates between so-called $U(1)$ instantons and $O(1)$ instantons. The $U(1)$ instantons correspond to split brane-image brane divisors and the $O(1)$ instantons to their recombined phase. The results of this section explain why no such difference must be made in F-theory or M-theory: As long as an instanton intersects the discriminant locus it always behaves like an $O(1)$ instanton due to strong coupling effects.

Finally we turn to  the microscopic counting of the rather mysterious massless Fermi multiplets from D3-7-brane intersections in the Sen limit:
The discriminant takes the form
\bea
\Delta  \simeq \epsilon^2 h^2 (\eta^2 - h \chi) + {\cal O}(\epsilon^3)
\eea
with $\eta^2 - h \chi$ the D7-brane in class $K^{-8}_{B_{n-1}}$ and $h=0$ representing the orientifold plane.
The independent extra modes from the $3$--$7$ sector are located only at the
intersection of the locus $\eta^2 - h \chi$ with the D3-brane, while the D3-O7
sector does not host {\it independent, extra} $3$--$7$ modes.
Hence, the number of extra $3$--$7$ Fermi multiplets is indeed given by $8 c_1(B_{n-1}) \cdot C$. 


\section{Dual Points of View: M5s and M2s} \label{sec:M5M2}

We now complement the analysis in the previous section by an M-theory dual point of view, given the 
 somewhat non-standard twist that we used for the compactification of the D3-branes. We now describe how  using M/F-duality, this system is mapped to M5- or M2-branes wrapping suitable cycles in the elliptic fibration. 
The background has already been summarized in sections \ref{sec:M5} and \ref{sec:M2over}.  
 Apart from giving a complementary point of view, it also allows a generalization of the previous setup.
The duality twist that allowed us to characterize the strings from the D3-brane point of view in F-theory
is  a priori only defined for the abelian theory. It is useful to map the system to M-theory, mapping the D3 to an M5, where the twist becomes a purely geometric topological twist \cite{Assel:2016wcr}, or an M2-brane, where a generalization to non-abelian theories is possible. 
The M5-brane approach also allows for a completely general derivation of the total number of Fermi multiplets due to $3$--$7$ string excitations.

\subsection{Strings and Particles in 5d}

We begin our analysis with strings in 6d $N=(1,0)$ supersymmetric
compactifications of F-theory on elliptic Calabi-Yau three-folds, which by M/F-duality map to 5d compactifications of M-theory. 
The dual (0,4) string obtained from an M5-brane  is precisely the MSW string \cite{Maldacena:1997de,Vafa:1997gr,
Minasian:1999qn}, specialized to an elliptic Calabi-Yau.
For M2-branes, we find an $N=4$
SQM as the dimensional reduction of the
$(0,4)$ theory to one dimension, and we match the spectra accordingly using automorphic duality.

\subsubsection{M5-branes and MSW-Strings}
\label{sec:M5CY3}

Following the general logic discussed in section \ref{sec:M5}, a D3-brane wrapping $C \subset B_2$ is dual to an M5-brane with worldvolume $\mathbb R^{1,1} \times \widehat C$, where the surface $\widehat{C}$ is the restriction of the elliptic fibration $Y_3$ to $C$.
The Lorentz symmetry decomposes under the reduced holonomy of $\mathbb{R}^{1,1} \times \widehat{C}$ as
\be\ba
  SO(1,5)_L &\quad \rightarrow \quad SU(2)_l \times SO(1,1)_L \times U(1)_l \cr
  {\bf 4} &\quad \rightarrow \quad  {\bf 2}_{1,0} \oplus {\bf 1}_{-1,1} \oplus {\bf
    1}_{-1,-1} \cr
  {\bf 15} &\quad \rightarrow \quad  {\bf 1}_{0,0} \oplus {\bf 3}_{0,0} \oplus
    {\bf 1}_{0,2} \oplus {\bf 1}_{0,0} \oplus {\bf 1}_{0,-2} \oplus
    {\bf 2}_{2,1} \oplus {\bf 2}_{2,-1} \oplus {\bf 2}_{-2,1} \oplus {\bf
    2}_{-2,-1} \,.
\ea\ee
Further we  decompose the R-symmetry group $Sp(4)_R$  by the embedding
\be\ba
  Sp(4)_R &\quad \rightarrow \quad SO(3)_T \times U(1)_R \cr
  {\bf 4} &\quad \rightarrow \quad  {\bf 2}_1 \oplus {\bf 2}_{-1} \cr
  {\bf 5} &\quad \rightarrow \quad  {\bf 1}_2 \oplus {\bf 1}_{-2} \oplus {\bf 3}_0 \,.
\ea\ee
The twist of $U(1)_l$ with $U(1)_R$ defined by
\be
U(1)_{\rm twist} = U(1)_l + U(1)_R
\ee
results in the following decomposition of the fields and supercharges:
\begin{align} \label{M5Y3decomp1}
  SO(1,5)_L \times Sp(4)_R &\quad \rightarrow \quad  SU(2)_l \times SO(3)_T \times
  SO(1,1) \times U(1)_\text{twist} \cr
  B_{\mu\nu}: \quad \,\, ({\bf 15}, {\bf 1}) &\quad \rightarrow \quad  {\bf (1,1)}_{0,0} \oplus
  {\bf (3,1)}_{0,0} \oplus {\bf (1,1)}_{0,2} \oplus {\bf (1,1)}_{0,0} \oplus
  {\bf (1,1)}_{0,-2} \cr
  &\qquad\qquad\qquad \oplus {\bf (2,1)}_{2,1} \oplus {\bf (2,1)}_{2,-1}
  \oplus {\bf (2,1)}_{-2,1} \oplus {\bf (2,1)}_{-2,-1} \cr
  H:\quad \,\,  ({\bf \overline{10}, 1}) & \quad \rightarrow \quad ({\bf 3,1})_{-2,0} \oplus ({\bf 1,1})_{2,2} \oplus
  ({\bf 1,1})_{2,0} \oplus ({\bf 1,1})_{2,-2} \oplus ({\bf 2,1})_{0,1}
  \oplus ({\bf 2,1})_{0,-1} \cr 
  \Phi^{ij}: \qquad  ({\bf 1}, {\bf 5}) &\quad \rightarrow \quad  {\bf (1,1)}_{0,2} \oplus
  {\bf (1,1)}_{0,-2} \oplus {\bf (1,3)}_{0,0} \cr
  Q^i,  \rho^{i}: \qquad  ({\bf \overline{4}, 4}) &\quad \rightarrow \quad  {\bf (2,2)}_{-1,1} \oplus
  {\bf (2,2)}_{-1,-1} \oplus {\bf (1,2)}_{1,2} \cr &\qquad\qquad\qquad \oplus {\bf (1,2)}_{1,0}
  \oplus {\bf (1,2)}_{1,0} \oplus {\bf (1,2)}_{1,-2} \,.
\end{align}
One can see that the decomposition of the $Q^i$ gives rise to
four scalar supercharges which are right-handed with respect to the 2d  $SO(1,1)_L$
Lorentz group,
\begin{equation}
 Q^i:  ({\bf \overline{4}, 4}) \supset ({\bf 1, 2})_{1,0} \oplus ({\bf 1,2})_{1,0} \,.
\end{equation}

We are now ready to discuss the zero-modes from each field.
If a state transforms in the ${\bf 2}$ of the internal holonomy group $SU(2)_l$, it must transform as a one-form on the compactification space, with our conventions being that a state ${\bf 2}_{1}$ and ${\bf 2}_{-1}$ with respect to $SU(2)_l \times U(1)_{\rm twist}$ transforms as an element of $\Omega^{0,1}(\widehat C)$ and, respectively $\Omega^{1,0}(\widehat C)$. This implies that a state ${\bf 1}_{2}$ translates into an element of $\Omega^{0,2}(\widehat C)$ and ${\bf 1}_{0}$ to an element of $\Omega^{0,0}(\widehat C)$. 
The zero-modes correspond to the associated cohomology groups. 

From the 6d scalar $\Phi$ there is a $({\bf 1,3})_{0,0}$ field with twist
charge zero. Such a field contributes both three left- and three right-handed real
chiral scalars since $\Phi$ itself has no intrinsic chirality. Similarly
$\Phi$ contains scalars with twist charges $\pm 2$, whose zero-modes are counted by
$h^{2,0}(\widehat{C})$ and $h^{0,2}(\widehat{C})$ respectively, and also
contribute to both right- and left-hand zero-modes. In summary we obtain the
following real bosonic zero-modes from the scalar field
\begin{equation}\label{CY3Phimultis}
\Phi= ({\bf 1}, {\bf 5}): \qquad 
  \begin{array}{|c|c|c|}
  \hline
  SU(2)_l \times  SO(3)_T \times
    SO(1,1)_L \times U(1)_{\text{twist}} &  \hbox{Multiplicity} & \hbox{L/R}\cr \hline
    ({\bf 1},{\bf 3})_{0,0} & h^{0,0}(\widehat{C}) & \hbox{L and R}\cr
    ({\bf 1},{\bf 1})_{0,2} & h^{0,2}(\widehat{C})&  \hbox{L and R} \cr
    ({\bf 1},{\bf 1})_{0,-2} &  h^{2,0}(\widehat{C})&  \hbox{L and R}\cr\hline
  \end{array}
\end{equation}

The self-dual three-form $H$\footnote{Recall that to compute the zero-modes of the potential $B_{\mu 
\nu}$ we can start from the decomposition of  the self-dual field strength $H$. At the
level of the zero-modes, the exterior derivative does not change the representation
content of the internal symmetry groups. Specifically the
$SU(2)_l$ and $U(1)_\text{twist}$ charges are the same for the modes associated with the potential $B_{\mu \nu}$ and those of the field strength. }
contributes 
left-handed real scalar fields 
corresponding to the representation $({\bf 3,1})_{-2,0}$. The
zero-modes of this field are counted by the number $h^{1,1}(\widehat{C}) - 1$ of anti-self-dual two-forms on
$\widehat{C}$. 
To understand this counting directly from the representation under $SU(2)_l \times U(1)_{\rm twist}$ note that elements of $H^{1,1}(\widehat C)$ transform under  as a ${\bf 2}_{1} \otimes {\bf 2}_{-1} \rightarrow {\bf 3}_0 \oplus {\bf 1}_0$. Taking into account the singlet this indeed leads to the counting of $h^{1,1}(\widehat{C}) - 1$ massless modes transforming as a  ${\bf 3}_0$ under $SU(2)_l \times U(1)_{\rm twist}$.
The right-handed scalar fields
arise from the fields strengths with representations $({\bf 1,1})_{2,2}$,
$({\bf 1,1})_{2,0}$, and $({\bf 1,1})_{2,-2}$, which are counted by
$h^{0,2}(\widehat{C})$, $h^{0,0}(\widehat{C})$, and $h^{2,0}(\widehat{C})$
respectively. Indeed the number $2 h^{2,0}(\widehat{C}) + 1$ counts the self-dual two-forms on $\hat C$.
Finally, the modes $({\bf 2,1})_{0, \pm 1}$ of the decomposition of $H$ in (\ref{M5Y3decomp1}) correspond to the fields ${\bf (2,1)}_{\pm 2,\pm 1}$ in $B$. The associated zero-modes would have to be represented by elements of $H^{0,1}(\widehat{C})$. A one-form on the K\"ahler surface has no definite behaviour under Hodge duality and hence the appearance of such states would be in conflict with the fact that the 3-form is self-dual.
This implies that the would-be zero-modes must be discarded, a conclusion which has also been reached e.g. in \cite{Apruzzi:2016nfr}. This is in fact as expected from the dual D3-brane perspective as the modes would give rise to additional vector potentials along the string, which are clearly absent on the D3-brane side.

In summary we find from the self-dual three form the following real scalar modes:  
\begin{equation}
H_{\mu\nu\rho}= ({\bf \overline{10}}, {\bf 1}): \qquad 
  \begin{array}{|c|c|c|c|}
  \hline
  SU(2)_l \times SO(3)_T \times
    SO(1,1)_L \times U(1)_{\text{twist}} &  \hbox{Multiplicity}&\hbox{L/R}\cr \hline
    ({\bf 3},{\bf 1})_{-2,0} & h^{1,1}(\widehat{C})-1 &  \hbox{L}\cr
    ({\bf 1},{\bf 1})_{2,2} & h^{0,2}(\widehat{C}) & \hbox{R} \cr
    ({\bf 1},{\bf 1})_{2,0} & h^{0,0}(\widehat{C})& \hbox{R}\cr
    ({\bf 1},{\bf 1})_{2,-2} &  h^{2,0}(\widehat{C})& \hbox{R}\cr\hline
  \end{array} \label{CY3Hmultis}
\end{equation}

The remaing task is to count the fermions:
The states in the decomposition of the $({\bf \overline 4},{\bf 4})$ in (\ref{M5Y3decomp1}) all come in pairs with opposite $U(1)_{\rm twist}$ charge. Each of these pairs describes a complex Weyl fermion together with its complex conjugate state.
The independent states are the right-handed Weyl fermions $({\bf 1}, {\bf 2})_{1,0}$, $({\bf 1}, {\bf 2})_{1,2}$ counted by 
$h^{0,0} (\widehat{C})$ and, respectively, $h^{0,2} (\widehat{C})$, as well as 
the left-moving Weyl fermions  $({\bf 2}, {\bf 2})_{-1,1}$ counted by 
$h^{0,1}(\widehat{C})$. In summary we obtain from the
fermions
\begin{equation}
\rho= (\overline{\bf 4}, {\bf 4}): \qquad 
  \begin{array}{|c|c|c|c|}
  \hline
  SU(2)_l \times SO(3)_T \times
    SO(1,1)_L \times U(1)_{\text{twist}} &  \hbox{Multiplicity}&\hbox{L/R}\cr \hline
    ({\bf 1},{\bf 2})_{1,0} + c.c. &  h^{0,0}(\widehat{C}) &  \hbox{R}\cr
    ({\bf 1},{\bf 2})_{1,2} + c.c. & h^{0,2}(\widehat{C}) &  \hbox{R}\cr
    ({\bf 2},{\bf 2})_{-1,1} + c.c. & h^{0,1}(\widehat{C}) &  \hbox{L}\cr
  \hline
  \end{array}  \label{CY3rhomultis}
\end{equation}

These states assemble into 2d $(0,4)$ multiplets as we shall now describe.  
It is shown in appendix \ref{app:ellipticsurfaces}, that the mutiplicities appearing in (\ref{CY3Hmultis}) satisfy the relation
\be \label{h11identitya}
h^{1,1}(\widehat C) - 2  h^{0,2}(\widehat C) - 2 = 8 \, {\rm deg}({\cal
L}_{D}) \geq 0 \,.
\ee
Here we recall briefly that ${\cal L}_D=K_{B_2}^{-1}|_C$ is the line bundle describing the elliptic fibration of the
surface $\widehat C$, which is also the duality line bundle associated to the
$U(1)_D$ bonus symmetry of the D3-brane theory.
This line bundle is always non-negative because it is the restriction of the line bundle describing the elliptic fibration of the Calabi-Yau 3-fold restricted to $\widehat C$. It is trivial if and only if $\widehat C$ is globally a direct product $C \times  T^2$. 
In particular, 
$2 h^{0,2}(\widehat C) + 1 \leq h^{1,1}(\widehat C) -1$, and 
thus $2 h^{0,2}(\widehat C) + 1$ of the left- and right-moving scalar  modes from the self-dual 2-form can combine into 
$2 h^{0,2}(\widehat C) + 1$ real non-chiral massless scalar fields. 
The remaining $h^{1,1}(\widehat C) - 2  h^{0,2}(\widehat C) - 2$ left-moving real scalar fields can be dualised into left-handed complex Weyl fermions. 
 In this way we arrive at the 2d
 (0,4) multiplet structure displayed in table \ref{M5CY3multiplets}. 

 \begin{table}
   \centering
  \begin{equation*} 
  \begin{array}{|c|c|c|c|c| }
  \hline
    \hbox{Multiplicity}                                & (0,4) \,  \hbox{Multiplet}  &   \hbox{Complex scalars} & \hbox{R-Weyl}                   & \hbox{L-Weyl}   \cr \hline\hline
   h^{0,0}(\widehat C) = 1                              & \hbox{Hyper}                   &  2                                      &  2                                       & -     \cr \hline
   h^{0,1}(\widehat C) = g                             & \hbox{Fermi}                   & -                                       & -                                        & 2    \cr \hline
   h^{0,2}(\widehat C) = g - 1 + c_1(B_2) \cdot C
   & \hbox{Hyper}                    &  2                                     &  2                                       & -     \cr  \hline
   h^{1,1}(\widehat C)- 2 h^{0,2}(\widehat C) - 2 = 8 \, c_1(B_2) \cdot C
      & \hbox{half-Fermi}        & -                                    & -                                          & 1 \cr \hline
     \end{array}
\end{equation*}
\caption{The $(0,4)$ mutiplets and their multiplicity in a compactification of
  an M5-brane wrapping an elliptic surface inside of an elliptic Calabi-Yau
three-fold.} \label{M5CY3multiplets}
\end{table}

In view of (\ref{h11identitya}) the half-Fermi multiplets in the last line are absent precisely if $\widehat C = C \times  T^2$.
These Fermi multiplets are the M5-brane incarnation of the zero-modes due
to $3$--$7$ strings in the dual F-theory picture. As promised, their counting as $8 \, {\rm deg}({\cal L}_D) = 8 \, c_1(B_2) \cdot C$ is completely universal.
In particular it remains valid for arbitrary configurations of 7-branes,
corresponding to fiber degenerations of the elliptic fibration in codimension
one. Once we work on the resolution of the elliptic fibration, which have been
studied explicitly in
\cite{Esole:2011sm,MS,Krause:2011xj,Lawrie:2012gg,Hayashi:2013lra,Hayashi:2014kca},
these
degenerations translate into fibral curves over the 7-brane locus, which in
turn restrict to fibral curves of the M5-brane surface $\widehat C$
intersecting the 7-brane locus. The net number of Fermi multiplets as given by
$8 \, {\rm deg}({\cal L}_D)$ is already taking these fibral curves into
account.

The numerical values of 
the Hodge numbers appearing in (\ref{CY3Phimultis}), (\ref{CY3Hmultis}) and (\ref{CY3rhomultis}) are computed via a Leray spectral sequence in appendix \ref{app:ellipticsurfaces}.
In particular this allows us to compare with the D3-spectrum in
 table \ref{tbl:CY3fcmultiplets}.
 If we include in the
latter also the $3$--$7$ string zero-modes, perfect agreement is found.


\subsubsection{M2-branes and $N=4$ SQM}

Another dual description of our string theories is in terms of an M2-brane wrapping the same curve $C \subset B_2$ as the D3-brane, as already summarized in section \ref{sec:M2over}.  In this case the effective theory is 1d $N=4$ SQM. 
To deduce the effective SQM we first note that the $SO(8)_R$-symmetry splits into the rotation group $SO(4)_T$ in the four extended directions
transverse to the brane, and another $SO(4)_R$ in the four
directions internal to the Calabi-Yau normal to the brane. K\"ahlerity of
$Y_3$ specifies a further reduction of the structure group of the normal bundle
$N_{C/Y_3}$ from $SO(4)_R$ to $U(2)_R$. The decompositions of the relevant
representations of $SO(8)_R$ are then
\be
\ba
  SO(8)_R &\quad\rightarrow\quad SO(4)_T \times SU(2)_R \times U(1)_R \cr
  {\bf 8_v}  &\quad\rightarrow\quad({\bf 2,2,1})_0 \oplus ({\bf 1,1,2})_1 \oplus ({\bf
  1,1,2})_{-1} \cr
  {\bf 8_c}  &\quad\rightarrow\quad({\bf 1,2,2})_0 \oplus ({\bf 2,1,1})_1 \oplus ({\bf
  2,1,1})_{-1} \cr
  {\bf 8_s}  &\quad\rightarrow\quad({\bf 1,2,1})_1 \oplus ({\bf 1,2,1})_{-1} \oplus
  ({\bf 2,1,2})_0 \,.
\ea
\ee
Combined with the reduction of $SO(1,2)_L \rightarrow U(1)_L$ along the M2-brane
this implies that the correct 
 twist to create scalar supercharges along $C$ is 
\begin{equation}
  T_{\text{twist}} = \frac{1}{2}(T_L + T_R) \,,
\end{equation}  
which is normalised appropriately. Indeed with this twist the
supersymmetry parameters and fields of the 3d $N=8$ theory on the M2-brane reduce as 
\begin{align}
  SO(1,2)_L \times SO(8)_R &\quad\rightarrow\quad  SO(4)_T \times SU(2)_R \times
      U(1)_{\text{twist}} \cr
        \epsilon \,:\  ({\bf 2, 8_s}) &\quad\rightarrow\quad ({\bf 1,2,1})_1 \oplus 2 \times ({\bf 1,2,1})_0\oplus ({\bf 1,2,1})_{-1} \oplus ({\bf 2,1,2})_{\frac{1}{2}}
    \oplus ({\bf 2,1,2})_{-\frac{1}{2}}\cr 
  \phi \,:\, ({\bf 1, 8_v})&\quad\rightarrow\quad  ({\bf 2,2,1})_0 \oplus ({\bf
  1,1,2})_{\frac{1}{2}} \oplus ({\bf 1,1,2})_{-\frac{1}{2}} \\
  \rho \,:\, ({\bf 2, 8_c}) &\quad\rightarrow\quad  ({\bf 1,2,2})_{\frac{1}{2}} \oplus
  ({\bf 1,2,2})_{-\frac{1}{2}} \oplus ({\bf 2,1,1})_1 \oplus ({\bf 2,1,1})_{-1}
  \oplus  2 \times ({\bf 2,1,1})_0  \,. \nonumber
\end{align}
The first line, containing the supersymmetry parameters, implies that there are $N=4$ scalar supercharges in 1d. 
The states and the zero-mode counting are summarised in table
\ref{tbl:M2CY3fc2}. The cohomology groups are determined as follows.
By adjunction, $K_C = \Lambda^2 N_{C/Y_3}$ and hence by the usual rationale zero-modes of fields forming a representation ${\bf 2}_{-1/2}$ and ${\bf 2}_{1/2}$ of $SU(2)_R \times U(1)_{\rm twist}$ transform as elements of $H^0(C,N_{C/Y_3})$ and $H^1(C,N_{C/Y_3})$, respectively. Indeed, we assign twist charge $-1$ to $K_C$ and therefore $-1/2$ to $N_{C/Y_3}$.
Note that the dimensions of these groups are in fact equal, as required by the multiplet structure of $N=4$ SQM. This follows from
the theorem on complex vector bundles (\ref{eqn:dualbdlthm}),
which in this instance implies that
\begin{equation}
  N_{C/Y_3} = N_{C/Y_3}^\vee \otimes K_C \,.
\end{equation}
Together with Serre duality this can be used to show that
\begin{equation}
  H^1(C, N_{C/Y_3}) = [H^0(C,K_C \otimes N_{C/Y_3}^\vee)]^\vee = [H^0(C,
  N_{C/Y_3})]^\vee \,.
\end{equation}

\begin{table}
  \centering
\begin{tabular}{|c|c|c|c||c|}
  \hline
  \multicolumn{2}{|c|}{Fermions} & \multicolumn{2}{c||}{Bosons} & \multirow{2}{*}{$N=4$ Multiplet} \cr\cline{1-4}
  State & Zero-modes & State & Zero-modes & \cr\hline\hline
  $({\bf 1,2,2})_{\frac{1}{2}}$ & \multirow{2}{*}{$h^{1}(C,N_{C/Y_3})$} &
  $({\bf 1,1,2})_{\frac{1}{2}}$ & \multirow{2}{*}{$h^{1}(C,N_{C/Y_3})$}& \multirow{2}{*}{$(2,4,2)$ Chiral} \cr
    $({\bf 1,2,2})_{-\frac{1}{2}}$ & & $({\bf 1,1,2})_{-\frac{1}{2}}$ &  &\cr\hline
  $({\bf 2,1,1})_0$ $(\times 2)$ & $h^0(C)$ & $({\bf 2,2,1})_0$ & $h^0(C)$ & $(4,4,0) $ Hyper
  \cr\hline
  $({\bf 2,1,1})_{ 1}$ & $h^{1,0}(C)$ & \multirow{2}{*}{---} &\multirow{2}{*}{---}& \multirow{2}{*}{$(0,4,4)$ Fermi} \cr
  $({\bf 2,1,1})_{- 1}$ & $h^{0,1}(C)$ & & & \cr\hline
\end{tabular}
\caption{The field content from a single M2-brane wrapping a curve inside a
Calabi-Yau three-fold. The representations are in terms of the remnant symmetry groups $SO(4)_T \times SU(2)_R \times U(1)_{\rm twist}$. \label{tbl:M2CY3fc2}}
\end{table}

It is necessary for the sake of comparison to associate the bundles
$N_{C/Y_3}$ to the bundles $N_{C/B_2}$ that appear in the spectrum
computations from the D3 branes. The long exact sequence in cohomology
associated to the short exact sequence of normal bundles in appendix
\ref{app:bdlcoho} yields the result that
\begin{equation}
  h^0(C, N_{C/Y_3}) = h^0(C, N_{C/B_2}) = g - 1 + \text{deg}(\mathcal{L}_D)  \,.
\end{equation}
The zero-modes of the states in table \ref{tbl:M2CY3fc2} can then be counted
explicity and compared to the D3-brane spectrum in table \ref{tbl:CY3fcmultiplets}. 
The
fermionic content of the spectrum readily matches the total fermion content of the
D3-brane analysis summarized in table \ref{tbl:CY3fcmultiplets}. 
To understand the relation of the bosonic spectra, we must carefully 
compare the multiplet structure of the 2d $(0,4)$ spectrum in table
\ref{tbl:CY3fcmultiplets} and the structure in table \ref{tbl:M2CY3fc2}. We
denote the 1d $N=4$ SQM multiplets by $(n, N, N-n)$ with $n$ the number of
bosonic degrees of freedom, $N$ the number of fermions and supersymmetries,
and $N-n$ the number of auxiliary fields, as explained in appendix \ref{app:SQM}. These are to be compared to the circle reduction of the 2d $(0,4)$ multiplets from the D3-brane analysis.
The 2d twisted hypermultiplet of table \ref{tbl:CY3fcmultiplets} straightforwardly reduces to the $(4,4,0)$ hypermultiplet in 1d, and similarly the Fermi multiplets in both tables map onto each other. More subtle is the $(2,4,2)$ multiplet.
The fermions in the $(2,4,2)$ multiplet correspond to the $\psi$ and $\tilde\psi$ modes of table \ref{tbl:CY3fcmultiplets}, and the two real scalar fields are identified with the modes $\sigma$ and $\tilde{\sigma}$. Both of these types of states are counted by 
\be
h^{1}(C, N_{C/Y_3}) = h^{0}(C, K_C \otimes \mathcal{L}_D)\,,
\ee
which matches the counting in table \ref{tbl:CY3fcmultiplets}. The crucial point concerns the Wilson line modes $a, \bar{a}$ of table \ref{tbl:CY3fcmultiplets}.
As discusssed in section \ref{sec:M2over}, after reduction to 1d these can be dualised into the two real auxiliary fields in the $(2,4,2)$ multiplet. 
Automorphic duality can be applied to the Wilson line scalars because these enjoy a shift symmetry. 
In this way we find perfect match between the D3-brane and the M2-brane spectrum.  

\subsection{Strings and Particles in 3d}

\subsubsection{M5-branes and $(0,2)$-Strings}
\label{sec:M5CY4}

Let us now consider an M5-brane along $\mathbb R^{1,1} \times \widehat{C}$ with  $\widehat{C}$ an elliptic surface inside a
Calabi-Yau four-fold $Y_4$.  There is only a single
transverse non-compact direction, which does not admit any continuous rotation group. Thus the reduction of the R-symmetry takes the form
\begin{equation}
  \begin{aligned}
    SO(5)_R &\rightarrow SO(4)_R \rightarrow SU(2)_R \times U(1)_R \cr
    {\bf 4} &\rightarrow {\bf 2}_0 \oplus {\bf 1}_1 \oplus {\bf 1}_{-1} \cr
    {\bf 5} &\rightarrow {\bf 2}_1 \oplus {\bf 2}_{-1} \oplus {\bf 1}_0 \,,
  \end{aligned}
\end{equation}
with $SU(2)_R \times U(1)_R$ being the structure group associated with the
normal bundle $N_{\widehat{C}/Y_4}$ to $\widehat{C}$ inside $Y_4$.
We recall that the Lorentz symmetry along the M5-brane decomposes into
\begin{equation}
  SO(1,5)_L \rightarrow SO(1,1)_L \times SU(2)_l \times U(1)_l \,.
\end{equation}
The partial topological twist 
\begin{equation} \label{twistM5onY4}
  T_{\text{twist}} = T_l + T_R 
\end{equation}
gives rise to exactly two positive chirality supercharges transforming as 
scalars along $\widehat{C}$ as required for $(0,2)$  supersymmetry along $\mathbb R^{1,1}$.
With this twist the field content and the supercharges decompose as
  \begin{align} \label{M5decomprho1}
    & \qquad SU(2)_l \times SU(2)_R \times SO(1,1)_L \times
    U(1)_{\text{twist}} \times U(1)_R
    \cr
    \rho\,,\, Q \,&:\, \quad ({\bf 2,2})_{-1,0,0} \oplus ({\bf 2,1})_{-1,1,1}
    \oplus ({\bf 2,1})_{-1,-1,-1} \oplus ({\bf 1,2})_{1,-1,0} \oplus ({\bf
    1,1})_{1,0,1} \cr &\qquad \oplus ({\bf 1,1})_{1,-2,-1} \oplus ({\bf
    1,2})_{1,1,0}
    \oplus ({\bf 1,1})_{1,2,1} \oplus ({\bf 1,1})_{1,0,-1} \\
    \Phi \,&:\, \quad ({\bf 1,2})_{0,1,1} \oplus ({\bf 1,2})_{0,-1,-1} \oplus ({\bf
    1,1})_{0,0,0} \cr
    H \,&:\, \quad ({\bf 3,1})_{-2,0,0} \oplus ({\bf 1,1})_{2,2,0} \oplus ({\bf
    1,1})_{2,0,0} \oplus ({\bf 1,1})_{2,-2,0} \oplus ({\bf 2,1})_{0,1,0} \oplus
    ({\bf 2,1})_{0,-1,0} \,. \nonumber
  \end{align}

The representations of the various fields with respect to $SU(2)_l \times SU(2)_R \times U(1)_{\text{twist}} $ determine the cohomology groups which count the massless modes as follows:
States transforming as a ${\bf 2}$ of $SU(2)_l$ correspond to one-forms on $\widehat C$, while $SU(2)_l$ singlets can a priori be zero-forms or two-forms.
Similarly, states transforming as a ${\bf 2}$ of $SU(2)_R$ must take values in the normal bundle $N_{\widehat{C}/Y_4}$.
The remaining ambiguities are fixed by the $U(1)_{\rm twist}$ charge. A
$(0,q)$-form contributes $+q$ to the total twist charge. Importantly, the
normal bundle also carries twist charge because of (\ref{twistM5onY4}).
This can also be seen as follows: First, by adjunction
\be
K_{\widehat{C}} = \wedge^2 N_{\widehat{C}/Y_4} \,,
\ee
and combined with the fact that $H^0(\widehat{C}, K_{\widehat{C}}) = H^{2,0}(\widehat{C})$, we assign the following twist charges to  $K_{\widehat{C}}$  and $N_{\widehat{C}/Y_4}$
\be
q^{\rm twist}(K_{\widehat{C}}) = -2, \qquad \quad q^{\rm
twist}(N_{\widehat{C}/Y_4}) = -1. 
\ee
 Thus, for instance the zero-mode of a state transforming as a $({\bf 1},{\bf
 2 })_{-1}$ under $SU(2)_l \times SU(2)_R  \times
    U(1)_{\text{twist}}$ is associated with an element of
 $H^0(\widehat{C},N_{\widehat{C}/Y_4})$, while  massless states of the form
 $({\bf 2},{\bf 2 })_{0}$ are associated to $H^1(\widehat{C},N_{\widehat{C}/Y_4})$.
This puts us in a position to assemble the zero-modes. From the decomposition of the scalars we find
\begin{equation}
\Phi= ({\bf 1}, {\bf 5}): \qquad 
  \begin{array}{|c|c|c|}
  \hline
  SU(2)_l \times SU(2)_R \times SO(1,1)_L \times
    U(1)^{\text{twist}} &  \hbox{Multiplicity} & \hbox{L/R}\cr \hline
    ({\bf 1},{\bf 1})_{0,0} & 1 & \hbox{L and R}\cr
    ({\bf 1},{\bf 2})_{0,1} &   h^{2}(\widehat{C}, N_{\widehat C/Y_4})  &  \hbox{L and R} \cr
    ({\bf 1},{\bf 2})_{0,-1} &  h^{0}(\widehat{C}, N_{\widehat C/Y_4})&  \hbox{L and R}\cr\hline
  \end{array}  \label{CY4Phimultis}
\end{equation}
The 3-form modes lead to the same types of left- and right-moving real scalar fields as for an M5-brane on a Calabi-Yau 3-fold, discussed in detail in section \ref{sec:M5CY3},
\begin{equation}
H_{\mu\nu\rho}= ({\bf \overline{10}}, {\bf 1}): \qquad 
  \begin{array}{|c|c|c|c|}
  \hline
  SU(2)_l \times SU(2)_R \times
    SO(1,1)_L \times U(1)_{\text{twist}} &  \hbox{Multiplicity}&\hbox{L/R}\cr \hline
    ({\bf 3},{\bf 1})_{-2,0} & h^{1,1}(\widehat{C})-1 &  \hbox{L}\cr
    ({\bf 1},{\bf 1})_{2,2} & h^{0,2}(\widehat{C}) & \hbox{R} \cr
    ({\bf 1},{\bf 1})_{2,0} & h^{0,0}(\widehat{C})& \hbox{R}\cr
    ({\bf 1},{\bf 1})_{2,-2} &  h^{2,0}(\widehat{C})& \hbox{R}\cr\hline
  \end{array} \label{CY4Hmultis}
\end{equation}
The component $({\bf 3,1})_{-2,0,0}$  of $H$ results in $h^{1,1}(\widehat{C}) -1$ left-moving  real scalar zero-modes.
Of these, $h^{0,2}(\widehat{C}) + h^{2,0}(\widehat{C}) + 1$ combine with the
right-moving scalar zero-modes into non-chiral real scalars. The remaining
$h^{1,1}(\widehat{C}) - 2 h^{0,2}(\widehat{C})  -2  = 8  \, {\rm deg}({\cal L}_D)$
left-moving real scalars can be dualised into left-moving complex fermion zero-modes.  The states $({\bf 2,1})_{0,\pm1,0}$ from $H$ do not give rise to any zero-modes for the reasons discussed in section \ref{sec:M5CY3}.

From the decomposition of the fermions we obtain
\begin{equation}
\rho= ({\bf \bar 4}, {\bf 4}): \qquad 
  \begin{array}{|c|c|c|c|}
  \hline
  SU(2)_l \times SU(2)_R \times
    SO(1,1)_L \times U(1)_{\text{twist}} &  \hbox{Multiplicity}&\hbox{L/R}\cr \hline
  
  ({\bf 2,2})_{-1,0}  & h^{1}(\widehat{C}, N_{\widehat C/Y_4}) &  \hbox{L}\cr 
  ({\bf 2,1})_{-1,1} + c.c. &  h^{0,1}(\widehat{C}) &  \hbox{L}\cr 
  ({\bf 1,2})_{1,1}  + c.c.  &  h^{2}(\widehat{C}, N_{\widehat C/Y_4}) &  \hbox{R}\cr 
   ({\bf 1,1})_{1,0}   + c.c.  &  h^{0}(\widehat{C}) &  \hbox{R}\cr 
   ({\bf 1,1})_{1,2} + c.c. &  h^{0,2}(\widehat{C}) &  \hbox{R}\cr \hline
       \end{array} \label{CY4rhomultis}
\end{equation}
We interpret pairs of states with opposite twist charge in the decomposition of the $({{\bf \overline 4}, {\bf 4}})$    in (\ref{M5decomprho1}) as complex conjugate Weyl fermions, the independent ones being as listed in lines $2-5$ of (\ref{CY4rhomultis}). The state  $({\bf 2,2})_{-1,0}$ is special to the extent that it forms its own complex conjugate and must therefore be counted as real.

The spectrum now organizes into 2d $(0,2)$ multiplets with the multiplicities
as shown in table \ref{M5CY4multiplets}. Note in particular that we have assembled the 
  $h^{1}(\widehat{C}, N_{\widehat C/Y_4})$ left-moving Majorana-Weyl fermions from (\ref{CY4rhomultis}) into   $\frac{1}{2}  h^{1}(\widehat{C}, N_{\widehat C/Y_4})  $ Fermi multiplets, each of which contains one (complex) left-moving Weyl fermion. 
  This is possible  because $  h^{1}(\widehat{C}, N_{\widehat C/Y_4})  $ is in fact an even integer.
  To see this and to compare the spectrum to the results obtained from the D3-brane on $C$,  we use the Leray spectral sequence discussed in appendix
\ref{app:ellipticsurfaces} together with the results of appendix
\ref{app:bdlcoho}. This relates the cohomology groups on
$\widehat{C}$ to cohomology groups on $C$. 
In particular one can see from the second line of (\ref{eqn:M5Chatdecomp})
that $h^{1}(\widehat{C}, N_{\widehat C/Y_4})  \in 2 \mathbb Z$.
Altogether we find perfect match with the D3-brane spectrum of table
\ref{eqn:CY4D3zm} once the $3$--$7$ Fermi zero-modes are taken into consideration.


\begin{table}
  \centering
 \begin{equation*}
  \begin{array}{|c|c|c|c|c| }
  \hline
    \hbox{Multiplicity}                                & (0,2)   &   \hbox{ complex scalars} & \hbox{R-Weyl}                   & \hbox{L-Weyl}   \cr \hline
   h^{0,0}(\widehat C) = 1                             & \hbox{Chiral}                   &  1                                      &  1                                       & -     \cr \hline
   h^{0,1}(\widehat C) = g                            & \hbox{Fermi}                   &   -                                       & -                                        & 1    \cr \hline
   h^{0,2}(\widehat C) = g - 1 +  c_1(B_3) \cdot C                             & \hbox{Chiral}                    &  1                                     &  1                                       & -     \cr  \hline \hline
   h^{0}(\widehat{C}, N_{\widehat C/Y_4}) = h^0(C, N_{C/B_3})  & \hbox{Chiral}                    &  1                                     &  1                                       & -     \cr  \hline
   \frac{1}{2}  h^{1}(\widehat{C}, N_{\widehat C/Y_4}) = h^0(C, N_{C/B_3}) -
    c_1(B_3) \cdot C   & \hbox{Fermi}                   &   -                                       & -                                        & 1    \cr \hline \hline
   h^{1,1}(\widehat C)- 2 h^{0,2}(\widehat C) - 2 = 8\,
    c_1(B_3) \cdot C   & \hbox{Fermi}        & -                                    & -                                          & 1 \cr \hline
   \end{array}
\end{equation*}
\caption{The $(0,2)$ multiplets of the 2d theory arising when an M5-brane
wraps an elliptic surface inside an elliptic Calabi-Yau four-fold.}\label{M5CY4multiplets}
\end{table}


\subsubsection{M2-branes and $N=2$ SQM}

In the dual M2-brane setup, compactification of the M2-brane on the curve $C \subset Y_4$ makes it necessary to 
reduce the Lorentz symmetry as $SO(1,2)_L \rightarrow U(1)_L$, which $U(1)_L$ the holonomy group on $C$. The R-symmetry $SO(8)_R$ decomposes,  as appropriate for a curve in a Calabi-Yau four-fold, as
\be
\ba
SO(8)_R  &\quad  \rightarrow \quad  SU(3)_R \times U(1)_R  \times SO(2)_T \cr 
{\bf 8}_{\bf v} & \quad \rightarrow \quad {\bf 3}_{2,0} \oplus \bar{\bf 3}_{-2, 0} \oplus {\bf 1}_{0,2} \oplus {\bf 1}_{0, -2} \cr 
{\bf 8}_{\bf s} & \quad \rightarrow \quad {\bf 3}_{-1, 1} \oplus \bar{\bf 3}_{1,-1} \oplus {\bf 1}_{3, 1}  \oplus {\bf 1}_{-3, -1} \cr 
{\bf 8}_{\bf c} & \quad \rightarrow \quad {\bf 3}_{-1, -1} \oplus \bar{\bf 3}_{1,1} \oplus {\bf 1}_{3, -1}  \oplus {\bf 1}_{-3, 1}  \,.
\ea
\ee
Geometrically we can think of $SU(3)_R \times U(1)_R$ as the structure group of the
normal bundle $N_{C/Y_4}$ of $C$ in $Y_4$. 
The supersymmetry parameters transform in $({\bf 2}, {\bf 8}_{\bf s})$ under $SO(1,2)_L\times SO(8)_R$. 
Twisting with
\be
T_{\rm twist} = {1\over 6} (3 T_L + T_R) \,,
\ee
where $T$ denotes the generators of the respective abelian groups, the supersymmetry variation parameters contain the modes
\be
\epsilon \quad \supset \quad {\bf 1}_{-1; 0} \oplus {\bf 1}_{+1; 0} \,, 
\ee
under $U(1)_L \times U(1)_{\rm twist}$ generating the $N=2$ supersymmetry. The matter fields decompose and twist into the following fields 
\be
\ba
SO(3)_L\times SO(8)_R  &\quad  \rightarrow \quad    SU(3)_R \times SO(2)_T \times U(1)_{\rm twist} \cr 
\rho: ({\bf 2}, {\bf 8}_{\bf c}) & \quad \rightarrow \quad   
{\bf 3}_{-1;{1\over 3}} \oplus {\bf 3}_{-1; - {2\over 3}} \oplus \bar{\bf 3}_{1;{2\over 3}}  \oplus \bar{\bf 3}_{1; -{1\over 3}} \oplus 
{\bf 1}_{-1; 1} \oplus {\bf 1}_{-1;0} \oplus {\bf 1}_{1; 0} \oplus {\bf 1}_{1; -1} \cr 
\phi: ({\bf 1}, {\bf 8}_{\bf v}) & \quad \rightarrow \quad 
{\bf 3}_{  0; {1\over 3}  }  \oplus  \bar{\bf 3}_{0; -{1\over 3} }  \oplus {\bf 1}_{2; 0} \oplus {\bf 1}_{-2, 0} \,.
\ea
\ee
In our conventions, the canonical bundle $K_C$ carries twist charge $-1$. In view of the adjunction formula
\be
K_C = \Lambda^3 N_{C/Y_4} \,,
\ee
it is hence appropriate to assign 
the normal bundle $N_{C/Y_4}$ twist charge $-1/3$ such that a state in $\bar{\bf 3}_{1; - {1\over 3}}$ gives rise to $h^0(C, N_{C/Y_4})$ zero-modes.\footnote{The identification of the ${\bf \overline 3}$, as opposed to the ${\bf 3}$, of $SU(3)_R$ with a section of the normal bundle with structure group  $SU(3)_R$ is of course a pure matter of convention. The choice here is made to keep our convention for the twist charge of $K_C$.}
Furthermore the zero-modes associated with the state $\bar{\bf 3}_{1; {2\over 3}}$      are correctly counted by $h^1(C, N_{C/Y_4})$.
Serre duality guarantees that this is the same number as the number of zero-modes of ${\bf 3}_{-1; -{2\over 3}}$, which a priori is counted by $h^0(C, N^\vee_{C/Y_4} \otimes K_C)$. The short exact sequence (\ref{app:SESM2}) and the resulting relation (\ref{app:SESM2b})
in turn imply that these states receive two types of contributions, one from $h^{1}(C, N_{C/B_3})$, and another from $h^{0}(C, K_C \otimes \mathcal{L}_D)$. 
Then the twisted component fields fall into 1d 2B supermultiplets, and are summarized in table \ref{CY4M2Spec}.
We recall the notation $(n,N,N-n)$ for an SQM multiplet (in a theory with $N$ supercharges) containing $n$ scalar, $N$ fermionic and $N-n$ auxiliary field real degrees of freedom.

\begin{table}
$$
  \begin{array}{|c|c|c|c|c|}\hline
    \text{Fermions} & \text{Bosons} & \text{2B Multiplet} &  \text{Zero-modes} & \text{Fields} \cr\hline\hline
    \bar{\bf 3}_{1; - {1\over 3}} \oplus {\bf 3}_{-1; {1\over 3}} &  
    \bar{\bf 3}_{0; - {1\over 3}} \oplus {\bf 3}_{0; {1\over 3}}  & (2,2,0)\, {\rm Chiral} & h^0(C, N_{C/B_3}) & \mu, \tilde{\mu}; \varphi, \bar\varphi\cr  \hline
 {\bf 1}_{1; 0} \oplus {\bf 1}_{-1; 0} &  
  {\bf 1}_{2; 0} \oplus {\bf 1}_{-2; 0} & (2,2,0) \, {\rm Chiral} & h^0(C) & \gamma, \tilde{\gamma}; g, \bar{g}\cr\hline
    {\bf 1}_{1; -1} \oplus {\bf 1}_{-1; 1} & \text{---}& (0,2,2)\, {\rm Fermi}  & h^{1}(C) & \beta, \tilde\beta\cr \hline
 \multirow{2}{*}{$\bar{\bf 3}_{1; {2\over 3}}\oplus{\bf 3}_{-1; -{2\over 3}}$} &   \multirow{2}{*}{---}& 
  \multirow{2}{*}{(0,2,2) \, {\rm Fermi} } & h^{1}(C, N_{C/B_3}) & \rho, \tilde{\rho}  \cr 
  && &h^{0}(C, K_C \otimes \mathcal{L}_D) & \psi, \tilde{\psi} \cr\hline 
    \end{array}
    $$
\caption{Spectrum of an M2-brane on $C\times \mathbb{R}$ with $C\subset Y_4$ in terms of the 1d $N=2$ SQM.\label{CY4M2Spec}}
\end{table}
Using the decomposition of 2d $(0,2)$ chiral/Fermi multiplets into 1d 2B chiral/Fermi multiplets outlined in appendix \ref{app:SQM}, the above spectrum matches the one determined from the F-theory computation of D3-branes with duality twist in table \ref{eqn:CY4D3zm}. 
Following the logic of section \ref{sec:M2over}, circle reduction of the Wilson line scalars $a, \bar{a}$ and fermion partners $\psi, \tilde{\psi}$ a priori gives rise to  $(2,2,0)$ chiral multiplets, which by the automorphic duality, discussed in appendix \ref{app:SUSYS}, are dual to the 2B Fermi multiplets counted by $h^{0}(C, K_C \otimes \mathcal{L}_D)$ in table \ref{CY4M2Spec}. In this map the Wilson line modes  are mapped to 
the auxiliary fields of the Fermi multiplets.


\subsection{M2-branes and SQM from M-theory on CY5}

We now come to the M-theory dual description of F-theory compactified on a Calabi-Yau $Y_5$. The latter gives rise to a 2d (0,2) theory \cite{Schafer-Nameki:2016cfr,Apruzzi:2016iac}.
Since, in the F-theory description, the strings from D3-branes on a curve $C \subset B_4$ are already spacetime-filling, there are no 
transverse non-compact dimensions along which we can T-dualise such as to end up with a description of the system in terms of an M5-brane. The only M-theory dual configuration to a D3-brane in F-theory on $Y_5$ is hence an M2-brane, obtained by T-dualizing along the D3-brane worldvolume to a D2-brane in IIA, and subsequent uplift to M-theory.  An M2-brane wrapping a curve in the base of $Y_5$ results in an SQM with $N=2$ supercharges. 

The first step of our analysis is to note that the  R-symmetry is reduced as 
\be
\ba
SO(8)_R \quad &\rightarrow \quad SU(4)_R \times U(1)_R \cr 
{\bf 8}_{\bf v}  \quad &\rightarrow \quad {\bf 4}_{1} \oplus \bar{\bf 4}_{-1 } \cr 
{\bf 8}_{\bf c}  \quad &\rightarrow \quad {\bf 4}_{-1} \oplus \bar{\bf 4}_{1 } \cr 
{\bf 8}_{\bf s}  \quad &\rightarrow \quad {\bf 6}_{0} \oplus {\bf 1}_{2} \oplus {\bf 1}_{-2} \,.
\ea
\ee
Twisting the $U(1)_L$ from $SO(1,2)_L \rightarrow U(1)_L$ with $U(1)_R$ by 
\be
T_{\rm twist} = {1\over 4} (2T_{L} + T_R) 
\ee
results in the following spectrum of the M2-brane theory:
\begin{equation}
  \begin{aligned}
SO(3)_L\times SO(8)_R  &\quad \rightarrow \quad   SU(4)_R \times U(1)_{\rm twist} \cr 
\epsilon: ({\bf 2}, {\bf 8}_{\bf s}) &\quad \rightarrow \quad {\bf 6}_{1\over 2} \oplus {\bf 6}_{- {1\over 2}} 
\oplus {\bf 1}_{1} \oplus {\bf 1}_{0}\oplus {\bf 1}_{-1} \oplus {\bf 1}_{0}\cr 
\phi:  ({\bf 2}, {\bf 8}_{\bf c}) &\quad \rightarrow \quad {\bf 4}_{{1\over 4}} \oplus  {\bf 4}_{-{3\over 4}} \oplus 
\bar{\bf 4}_{-{1\over 4}} \oplus  \bar{\bf 4}_{{3\over 4}}   \cr 
\rho: ({\bf 1}, {\bf 8}_{\bf v}) &\quad \rightarrow \quad  {\bf 4}_{1\over 4} \oplus \bar{\bf 4}_{-{1\over 4}} \,.
 \end{aligned}
\end{equation}
The zero-modes can be determined by recalling our usual convention that we assign twist charge $-1$ to $K_C$. Adjunction,
\begin{equation}
  K_C = \wedge^4 N_{C/Y_5} \,,
\end{equation}
hence implies that the normal bundle $N_{C/Y_5}$ with structure group $SU(4)_R \times U(1)_R$ must have twist charge $-1/4$.
For instance the state ${\bf
  \overline{4}}_{-1/4}$ leads to $h^0(C, N_{C/Y_5})$ zero-modes in the SQM.    
As for the zero-modes from ${\bf 4}_{1/4}$, we identify the ${\bf 4}$ as the triple-antisymmetric tensor product of the ${\bf \bar 4}$ and conclude that the zero-modes are counted by $h^1(C, \wedge^3 N_{C/Y_5})$. This is in perfect agreement with their twist charge - recalling that elements of $H^1(C) = H^0(C,K_C^{-1})$ carry an additional  $q_{\rm twist} = 1$.
Then Serre duality and 
 the vector bundle
formula (\ref{eqn:dualbdlthm}),
\begin{equation}
  N_{C/Y_5} = \wedge^3 N_{C/Y_5}^\vee \otimes K_C \,,
\end{equation}
guarantee that $h^1(C, \wedge^3 N_{C/Y_5}) = h^0(C, N_{C/Y_5})$. This number is in fact equal to  $h^0(C, N_{C/B_4})$ as explained in appendix  \ref{app:bdlcoho} around equation (\ref{app:SESM2b}).
Analogous reasoning for the remaining fields leads to the zero-modes in table \ref{tab:CY5M2zm}, where we are invoking once more equation (\ref{app:SESM2b})
to express the modes counted by $h^1(C, N_{C/Y_5})$ in terms of cohomology groups on the base only.

Comparing the modes here with those in table \ref{tab:CY5D3zm} we observe  the same number of fermionic degrees of freedom.
The only difference at first sight is from the 
 $h^0(C, K_C \otimes \mathcal{L}_D)= g - 1 + c_1(B_4) \cdot C$ Fermi multiplets. By automorphic duality these 
get mapped to chiral multiplets associated to the Wilson line scalars $a, \bar{a}$ and their fermionic partners, which completes the match of both approaches.

\begin{table}
$$
  \begin{array}{|c|c|c|c|c|}\hline
    \text{Fermions} & \text{Bosons} & \text{2B Multiplet} &  \text{Zero-modes} & \text{Fields} \cr\hline\hline
    \bar{\bf 4}_{ - {1\over 4}} \oplus {\bf 4}_{{1\over 4}} &  
    \bar{\bf 4}_{ - {1\over 4}} \oplus {\bf 4}_{{1\over 4}}  & (2,2,0)\, \,
    {\rm Chiral} & h^0(C, N_{C/B_4}) & \mu, \tilde{\mu}; \varphi, \bar\varphi\cr  \hline
 \multirow{2}{*}{$\bar{\bf 4}_{{3\over 4}}\oplus{\bf 4}_{-{3\over 4}}$} &   \multirow{2}{*}{---}& 
  \multirow{2}{*}{(0,2,2) \, {\rm Fermi}} & h^{1}(C, N_{C/B_4}) & \rho, \tilde{\rho}  \cr 
  && &h^{0}(C, K_C \otimes \mathcal{L}_D) & \psi, \tilde{\psi} \cr\hline 
    \end{array}
    $$
\caption{Spectrum of an M2-brane on $C\times \mathbb{R}$ where $C\subset Y_5$, in terms of multiplets of the 1d $N=2B$ SQM. \label{tab:CY5M2zm}}
\end{table}


\section{Anomalies for Strings in Various Dimensions}\label{sec:anomalies}

The chiral string theories  analyzed so far exhibit a rich pattern of gravitational and global anomalies.
In this section we investigate the cancellation of the field theoretic
anomalies by anomaly inflow from the ambient space to the string. 
After a general discussion of the anomaly inflow mechanism applied to the
strings under consideration in section \ref{sec:anomgen}, we explicitly match
the field theoretic anomalies with the anomaly inflow for the strings in $d=8,
6, 4$, and $2$ dimensions. As we will see, the structure of anomalies has
interesting implications, in particular for the rather mysterious spectrum of
$3$--$7$ strings.

\subsection{Anomaly Inflow for Strings in $\mathbb R^{1,d-1}$ }\label{sec:anomgen}

The contribution of a positive chirality complex Weyl fermion in representation $R$ to the field theoretic gauge and gravitational anomalies on the string  is characterized by the anomaly 4-form
\bea \label{eqn:anomalyfromferm}
I_{4,R} =  \hat A(T) \,  {\rm tr}_R \, e^{i F} |_{4-{\rm form}}= - \frac{1}{2}
{\rm tr}_R F^2 - \frac{1}{24} p_1(T) \,.
\eea
In terms of this 4-form, the chiral anomaly contribution of a complex Weyl fermion takes the form
\bea
{\cal A}|_{R} = - 2 \pi \int_{\rm string} I_{2,R}^{(1)} \,,
\eea
with 
\bea
{I}_{4,R} = d {I}^{(0)}_{3,R} \,, \qquad   \delta{I}^{(0)}_{3,R} = d
{I}_{2,R}^{(1)} \,.
\eea
For consistency these field theoretic anomalies must be cancelled by the standard mechanism of anomaly inflow from the dimensions transverse to the string.
For a string along 
 $\mathbb R^{1,1} \subset \mathbb R^{1,d-1}$ with $d >2$ the anomaly inflow can be formulated as an inflow from the extended transverse spacetime dimensions.\footnote{The case $d=2$ requires a special treatment and will be discussed separately in section \ref{2danomalies}.}
In the sequel we briefly summarize this standard anomaly inflow to fix our
conventions and notation in a simplified version, which is valid in this form
except in $d=6$, where the string is self-dual and special care must be
applied.

Let $c_2$ denote the 2-form sourced electrically by the string in $\mathbb R^{1,d-1}$. The field strength associated with its magnetically dual $(d-4)$-form $c_{\rm d-4}$ is subject to the modified Bianchi identity
\bea \label{source1}
d F_{\rm d-3} = \delta^{\rm (d-2)}(x_{\rm T}) \,. 
\eea 
Here $\delta^{\rm (d-2)}(x_{\rm T})$ denotes the Poincar\'e dual to the string at $x_{\rm T} = 0$.
The magnetic field strength $F_{\rm d-3}$ enjoys the Chern-Simons-type couplings
\bea \label{CScouplings1}
S_{\rm CS} =   - 2 \pi \int_{\mathbb R^{1,d-1}} F_{\rm d-3} \wedge  {\cal
I}_3^{(0)} \,.
\eea 
The standard descent relations 
\bea
{\cal I}_4 = d {\cal I}^{(0)}_3, \qquad   \delta{\cal I}^{(0)}_3 = d {\cal I}_2^{(1)},
\eea
with $d {\cal I}_4 = 0$, imply that 
in the presence of the source term (\ref{source1}), the CS action (\ref{CScouplings1})  picks up a non-trivial gauge variation of the form
\bea
\delta S_{\rm CS}  &=& - 2 \pi \int_{\mathbb R^{1,d-1}}  F_{\rm d-3} \wedge \delta{\cal I}^{(0)}_3 = -  2 \pi\int_{\mathbb R^{1,d-1}} d F_{\rm d-3} \wedge {\cal I}^{(1)}_2 \\
&=&  - 2 \pi \int_{\rm string}  {\cal I}^{(1)}_2.
\eea
This classical gauge variance cancels the field theoretic anomalies along the
string: 
\bea
 {\cal I}_4 + \sum_{R} \chi_R  \, I_{4,R} = 0 \,.
\eea
Here $\chi_R = n_{R,+} - n_{R,-}$,  and   $n_{R,\pm}$ denotes the number of complex Weyl fermions in representation $R$, with the upper and lower sign referring to positive and negative chirality, respectively.

In our case the string arises from a D3-brane wrapping a curve on an internal compactification space. 
The 2-form $c_2$ coupling electrically to the string in $\mathbb R^{1,d-1}$ is then a linear combination of the modes obtained by reducing the Ramond-Ramond 4-form along $H^{1,1}_{B_n}$ with $B_n$ the base of the F-theory elliptic fibration and $n=5-d/2$. 
The Chern-Simons couplings (\ref{CScouplings1}) can be determined by dimensional reduction of the CS couplings for the 7-branes in the perturbative limit. 
The latter are expressed in terms of the A-roof and Hirzebruch L-roof genera as \cite{Morales:1998ux} 
\be
\ba \label{braneactions1}
S_{\rm D7} &= 2\pi  \int_{\rm D7} C_4 \wedge  \frac{1}{2 }\Tr e^{i F} \sqrt{\widehat{A} (T)} \cr 
S_{\rm O7} &= - 8 \pi \int_{\rm O7} C_4 \wedge \sqrt{\widehat{L}\left({1\over
4} T\right)} \,,
\ea
\ee
in units where the string scale $\ell_s=1$ and $T$ denotes the respective tangent bundles.\footnote{Note that the relative factor of $-4$ in the normalization of $S_{\rm D7}$ and $S_{\rm O7} $ is the one relevant for computations downstairs on the F-theory base as opposed to upstairs, prior to orientifolding, on the Calabi-Yau double cover.}
Furthermore, the A-roof and L-roof genera are expanded into the Pontrjagin classes as 
\be
\ba
\sqrt{\widehat{A} (T) }&=  1 - {1\over 48} \, p_1(R) + \cdots \cr 
\sqrt{\widehat{L}\left({1\over 4} T\right)} &= 1+ {1\over 96} \, p_1 (R) + \cdots \, ,
\ea
\ee
where
\be
p_1 (R) = - {1\over 2} \tr R\wedge R \,.
\ee
Here and in the sequel, the symbol $\tr$ refers to the trace in the fundamental representation of the structure group of the relevant bundle.
As for the trace over the gauge bundle in $S_{\rm D7}$, 
$\Tr F^2$  is related to the trace in the fundamental via $\frac{1}{2} {\rm Tr} F^2 = {\rm tr} F^2$ for the perturbative gauge group $SU(n)$.
Its normalization for other gauge groups will be discussed in more detail in  section \ref{subsec_Flavoranomaly}.

To compute the CS couplings for a string on $\mathbb R^{1,d-1}$ from compactification of F-theory with base $B_n$,
we  
introduce a basis $\omega^{\rm (2n-2)}_\alpha$ of ${(2n-2)}$-forms on $B_n$ and expand
\bea
C_4 = c^\alpha_{\rm d-4}  \wedge \omega^{\rm (2n-2)}_\alpha \,,
\eea
which is possible for $d=8,6,4$ as considered in this section. 
We then plug this ansatz into (\ref{braneactions1}) and use the tadpole relation 
\bea
\sum_a { D}_a  = 4 \, {O7} =  8 \, c_1(B_n) 
\eea
for the divisor classes $D_a$ wrapped by the 7-branes and the orientifold-plane class $O7$ in the perturbative Type IIB limit.
Summing up the gravitational couplings of all 7-branes and the O7-plane gives
\be
\ba \label{CSactionexansion}
(S_{\rm D7} + S_{\rm O7})|_{\rm grav} &= 2\pi \int_{\mathbb R^{1,d-1}} c^\alpha_{\rm d-4}  \wedge p_1(R) \left(\int_{B_n}   \omega^{\rm (2n-2)}_\alpha  \wedge (- \frac{1}{48}\sum_a D_a - \frac{4}{96} O7  )     \right) \cr      
&= {2\pi} \int_{\mathbb R^{1,d-1}} c^\alpha_{\rm d-4}  \wedge
\left(-\frac{1}{4}\right) p_1(R)    \left(\int_{B_n}   \omega^{\rm
(2n-2)}_\alpha  \wedge c_1(B_n) \right)\,. 
\ea
\ee
A similar contribution describes the CS couplings involving the field strengths $F_a$ on each individual brane on divisor $D_a$. From the perspective of the 
string, the gauge symmetry along the 7-branes is related to the flavor
symmetry for the $3$--$7$ modes. 

Let us now come back to the string from a D3-brane wrapped on the curve $C$ in $B_n$.
The CS coupling of the form $c_{\rm d-4}$ which couples magnetically to it 
 is given by
\bea
S_{\rm CS} = 2 \pi \int_{\mathbb R^{1,d-1}} c_{\rm d-4} \wedge \left(  p_1(R)  \left(-  \frac{1}{4}  \, c_1(B_n) \cdot C\right) - \sum_a  \frac{1}{4}\, {\rm Tr} F_a^2    \Big(D_a \cdot C \Big)  \right) \, .
\eea
This identifies
\bea \label{calI41}
 {\cal I}_4 =  p_1(R)  \left(-  \frac{1}{4}  \, c_1(B_n) \cdot C\right) - \sum_a
 \frac{1}{4} \, {\rm Tr} F_a^2   \,  \Big(D_a \cdot C \Big)\,.
\eea
Finally, one can decompose $p_1(R)$ with respect to the tangent bundle along the string and the normal bundle in the transverse $(d-2)$ dimensions as
\bea
p_1(R) = p_1(T) + p_1(N)\,. 
\eea

In the special case $d=6$ extra contributions to the anomaly polynomial (\ref{calI41}) arise due to the self-dual nature of the string, which will modify the normal bundle anomaly inflow. Furthermore one can complete (\ref{calI41}) to include also inflow of  R-symmetry anomalies as will be discussed in section \ref{sec:anomaly6d}. 


\subsection{Anomalies for Strings in 8d and 4d}

These considerations can be most directly applied to strings in 8d and 4d.
We begin with F-theory compactified on an elliptic K3 with base $B_1=
\mathbb{P}^1$. As discussed in appendix \ref{app:K3}, 
a wrapped D3-brane on $C = B_1$ gives rise to 
a string in $\mathbb R^{1,7}$, which couples magnetically to the Type IIB 4-form $C_4$ along the extended spacetime directions.
 Hence in (\ref{CSactionexansion}) no expansion in terms of internal forms is required (i.e. $n=1)$.
The gravitational part of the anomaly inflow (\ref{calI41})  therefore
provides a contribution
\begin{equation}\label{eqn:curvinflow}
{\cal I}_{4,\rm grav + N}   = \Big( - \frac{1}{4}  c_1(B_1) \cdot C \Big) \,
\left(p_1(T) + p_1(N)\right) =   - \frac{1}{2} \left(p_1(T) + p_1(N)\right)\,.
\end{equation}
In the last equation we are using 
 $c_1(B_1) = c_1(\mathbb P^1)  = 2$ and $C=B_1$\,.

From the spectrum in equation (\ref{eqn:D3K3}) one can compute the contribution
of each complex Weyl fermion to the anomaly following
(\ref{eqn:anomalyfromferm}), which leads to the gravitational anomaly
\begin{equation}
{I}_{4,{\rm grav}} =   - \frac{1}{24}p_1(T)(4 - 16) = {1 \over 2}p_1(T) \,.
\end{equation}
This is accompanied by an anomaly in the normal bundle $N$ associated with the uncompactified directions
transverse to the string. The positive chirality fermions in the $(0,8)$
hypermultiplet transform in the ${\bf 4}$ spinor representation of  its structure group
$SO(6)$. The induced normal bundle anomaly is therefore
\begin{equation}
{I}_{4,{\rm N}}   =  - {1 \over 2} \text{tr}_{\bf 4} F^2 = - {1 \over 4}
  \text{tr}  F^2 = {1 \over 2} p_1(N) \,,
\end{equation}
where we have related the trace over the spinor representation to the trace
over the fundamental representation using \cite{Erler:1993zy}. We can see
that these two terms are
cancelled by the anomaly inflow (\ref{eqn:curvinflow}). 
The cancellation of the  flavor group anomalies will be discussed in more
detail in section \ref{subsec_Flavoranomaly}.

A similar pattern occurs for strings along $\mathbb R^{1,3}$ obtained from 
F-theory compactified on an elliptically fibered four-fold $Y_4$.
The
spectrum of the string from the pure D3-brane sector is listed in table
\ref{eqn:CY4D3zm} and  is accompanied by a further
$  8 c_1(B_3) \cdot C $
Fermi multiplets from the $3$--$7$ string sector. 
The gravitational anomaly contribution from these sectors is  
\begin{equation}
  \begin{aligned}
    I_{4,{\rm grav}} &= - {1 \over 24}p_1(T)(2 c_1(B_3) \cdot C) 
  + \frac{1}{24}p_1(T)(8c_1(B_3) \cdot C) \cr
  &= {1 \over 4}p_1(T)   \Big(c_1(B_3) \cdot C\Big) \,.
  \end{aligned}
\end{equation}
Further, the $SO(2)_T$ normal bundle anomaly  takes the form
\begin{equation}
  \begin{aligned}
    I_{4, {\rm N}} = 2 (c_1(B_3) \cdot C) \left( - {1 \over 2} \text{tr}_{\text{spin}}
    F^2 \right)
    = {1 \over 4} \Big(c_1(B_3) \cdot C\Big) p_1(N) \,,
  \end{aligned}
\end{equation}
where we have used that for $SO(2)$
\begin{equation}
  \text{tr}_{\text{spin}} F^2 = {1 \over 8} \text{tr} F^2 \,.
\end{equation}
Thus the gravitational and normal bundle anomaly from the spectrum of the $(0,2)$
theory on the string are both cancelled by the anomaly inflow (\ref{calI41}).


\subsection{Anomalies for Self-dual Strings in 6d}\label{sec:anomaly6d}

The anomaly of strings in 6d theories is on a slightly different footing
compared to the other dimensions, as the string couples to a potential with
self-dual field strength. The anomaly of self-dual strings in 6d were
initially discussed in \cite{Berman:2004ew}. More specifically for F-theory
compactifications on elliptic Calabi-Yau three-folds this has been discussed
in \cite{Shimizu:2016lbw} (see also \cite{Kim:2016foj}) for setups where the
D3 is wrapped on a $\mathbb{P}^1$ with normal bundle degree $-n$, $n\geq 3$.
We provide a generalization of this analysis for any curve $C$ in the base
$B_2$ not contained in the discriminant of the fibration.


Part of the anomaly polynomial derives from the expression (\ref{calI41}) present in all dimensions. The structure group of the normal bundle $N$ to the string in $\mathbb R^{1,5}$ is $SO(4)_T = SU(2)_{T,1} \times SU(2)_{T,2}$, for which one can write
\bea
p_1(N) =  -\frac{1}{2} {\rm tr}F_{SO(4)}^2  =  - \tr F_{T,1}^2 - \tr F_{T,2}^2   \,.   
\eea
There exist, however, two more contributions to the anomaly polynomial special to $d=6$: The first is an extra contribution to the normal bundle anomaly 
which is rooted in the self-dual nature of the string and which is derived in
\cite{Shimizu:2016lbw,Kim:2016foj} to take the form
\bea
- \frac{1}{2} \, (C \cdot C)   \,     \chi_4(N) =   - \frac{1}{2} \, (C \cdot
C) \,  \Big( \frac{1}{2} \tr F_{T,2}^2  -  \frac{1}{2} \tr F_{T,1}^2        \Big)\,.
\eea
Furthermore, there is a contribution to the field theoretic $SU(2)_I$
R-symmetry, which can be inferred on purely field theoretic grounds from the
Green-Schwarz terms of the 6d $(1,0)$ supergravity \cite{Ohmori:2014kda}.  In our conventions and notation, this yields an extra term of the form $ \frac{1}{2} \tr F_I^2$   
 in ${\cal I}_4$. 
Altogether the anomaly inflow polynomial is hence
\be
\ba \label{I4inflow6d}
{\cal I}_4 &= - \frac{1}{4} \Big( c_1(B_2) \cdot C \Big)  p_1(T) - \frac{1}{4} \sum_a {\rm Tr} F_a^2 [D_a] \cdot C  + \frac{1}{2} \tr F_I^2       \cr
&  + \frac{1}{2} \tr F_{T,1}^2     \left( \frac{1}{2} C \cdot C + \frac{1}{2} c_1(B_2) \cdot C  \right)    + 
\frac{1}{2} \tr F_{T,2}^2   \left( - \frac{1}{2} C \cdot C + \frac{1}{2} c_1(B_2) \cdot C  \right)  \,.
\ea
\ee

To compare this to the field theoretic anomaly induced by the spectrum in table \ref{tbl:CY3fcmultiplets}, we first note that all bulk fields transform in real or pseudo-real 
representations and should therefore be counted as Majorana-Weyl fermions. 
This gives rise to an additional factor of $\frac{1}{2}$ compared to (\ref{eqn:anomalyfromferm}). 
The field theoretic anomaly polynomial then takes the form
\be\ba
I_4 &= - \frac{1}{24} p_1(T)   \, \Big(     (2 \times 2 \times \frac{1}{2})  \, ( g-1 + c_1(B_2) \cdot C + (1-g))  -    ( 1 \times 1 \times 1)\,    8 c_1(B_2)\cdot C      \Big) \\
&+ \frac{1}{4} \sum_a {\rm Tr} F_a^2 \,  \Big(D_a \cdot C \Big) -  \frac{1}{2} \tr F^2_I    \times \frac{1}{2} \times 2 \\
& - \frac{1}{2} \tr F^2_{T,1}   \,  \Big(g-1 + c_1(B_2) \cdot C\Big)   \times  \frac{1}{2}  \times 2  
 - \frac{1}{2} \tr F^2_{T,2}      \times  \Big(\frac{1}{2}  \times 2    -g  \times  \frac{1}{2} \times 2 \Big)  \\
&=\frac{1}{4} \Big({c_1(B_2) \cdot C} \Big) \,  p_1(T)  +  \frac{1}{4} \sum_a {\rm Tr} F_a^2  \Big( D_a \cdot C \Big)  - \frac{1}{2} \tr F^2_I  \\
&  - \frac{1}{2} \tr F^2_{T,1}  \Big(g-1 + c_1(B_2) \cdot C\Big) + \frac{1}{2} \tr F^2_{T,2}  \Big(g -1\Big) \,.
\ea
\ee
This precisely cancels the anomaly inflow (\ref{I4inflow6d}) because, by
adjunction,
\bea
C \cdot (C - c_1(B_2)) = 2g -2 \,.
\eea
Note that in the first line we have also included the contribution of the $ 8
c_1(B_2) \cdot C$ complex $(0,4)$ half-Fermi-multiplets from the $3$--$7$ string sector.


\subsection{Anomalies in 2d}\label{2danomalies}

Since the string in two dimensions is spacetime-filling,
there is no room for anomaly inflow from non-compact dimensions transverse to the string.
To study the cancellation of anomalies one has to consider instead 
the standard anomaly inflow to the string from the bulk of the 7-branes in 10
dimensions, where the string is viewed as a defect along the eight-dimensional
7-brane worldvolume. Consider a 7-brane along $\mathbb R^{1,1} \times D_a$ with $D_a$ a complex 3-cycle on the base $B_4$. Viewed as a defect $\mathbb R^{1,1} \times C$ inside this worldvolume the string couples magnetically to the Type IIB 4-form $C_4$ inside  $\mathbb R^{1,1} \times D_a$ such that the Bianchi identity gets modified to
\bea
d F_5 = \delta^{(6)}(x_T) \,,
\eea  
with $\delta^{(6)}(x_T)$ the Poincar\'e dual to the string inside the 7-brane.
The Chern-Simons couplings of $C_4$ along each brane can be summed up, as before, to
\bea
 S_{CS} =  2\pi \int_{\mathbb R^{1,1} \times B_4}   C_4 \wedge
 \Big(-\frac{1}{4} p_1(R) \wedge  c_1(B_4) - \frac{1}{4} \sum_a {\rm Tr} F_a
 \wedge F_a \wedge D_a  \Big)\,,
\eea
which by descent contributes to the gauge variance as
\bea
\delta S_{CS} = 2 \pi \int_{\mathbb R^{1,1} \times B_4}   {{\cal
I}^{(1)}_{2,R}} \wedge C \wedge (-\frac{1}{4} c_1(B_4))  +  \sum_a {{\cal
I}^{(1)}_{2,a}}   \wedge C \wedge D_a\,.  
\eea
Here ${{\cal I}^{(1)}_{2,R}}$ and ${{\cal I}^{(1)}_{2,a}}$ are related by descent to $-\frac{1}{2} {\rm tr} R \wedge R$ and  $-\frac{1}{4} {\rm Tr} F_a \wedge F_a$ respectively.
Of course this leads to the same anomaly form  (\ref{calI41}) cancelling the gauge and gravitational anomalies on the string. 

Consistently the gravitational contribution to the anomaly polynomial from the $3$--$3$ and
$3$--$7$ sectors, which we read off from the spectrum in table
\ref{tab:CY5D3zm}, is
\begin{equation}\label{eqn:D3contrib}
  \begin{aligned}
    I_{4,T} &= - {1 \over 24} p_1(T) (2 c_1(B_4) \cdot C - 8 c_1(B_4) \cdot C)
    \cr
    &= + {1 \over 4} p_1(T) c_1(B_4) \cdot C \,.
  \end{aligned}
\end{equation}
This time there is no normal bundle anomaly.

In fact, much more can be said: Since the D3-brane `strings' fill all of spacetime, the curve class $C$ is fixed by a tadpole condition \cite{Schafer-Nameki:2016cfr,Apruzzi:2016iac} 
as $C = \frac{1}{24} c_4(Y_5)|_{\rm base}$. Tadpole cancellation implies that the sum of all anomaly inflow terms vanishes. Hence the total field theoretic anomalies from all sectors of the 2d theory must sum up to zero. These sectors contain the D3-brane theory as discussed here, but furthermore the moduli sector from the 2d (0,2) supergravity, the actual gravitational sector and the 7-brane sector.
The anomalies due the moduli sector and the gravitational sector are derived in detail in the companion paper \cite{MINI}, and found to be
\begin{equation} \label{I4SUGRA}
I_{4,{\rm SUGRA}} =   - {1\over 24} p_1(T) \left( - \tau(B_4) + \chi_1(Y_5) - 2 \chi_1(B_4) + 24\right) \,.
\end{equation}
Here $\tau$ is the signature of the manifold and $\chi_i$ are the arithmetic genera. 
The anomaly contribution from the 7-branes has been derived in \cite{Schafer-Nameki:2016cfr,Apruzzi:2016iac}; 
in the simplest case of   
an elliptic fibration $Y_5$ which is a smooth Weierstrass
model the 7-branes only contribute moduli fields which are already part of the anomaly (\ref{I4SUGRA}).
In \cite{MINI}
we furthermore determine that when one adds the D3-brane contribution
(\ref{eqn:D3contrib}), augmented with $C$ such that the tadpole is solved, to
the supergravity sector then the total anomaly can be written as
\begin{equation}\label{eqn:D3SUGRA}
  -{1\over 24} p_1(T) \left(-24\chi_0(B_4) + 24\right) \,.
\end{equation}
In terms of Hodge numbers, $\chi_0(B_4)$ is defined as
\begin{equation}
  \chi_0(B_4) = h^{0,0}(B_4) - h^{1,0}(B_4) + h^{2,0}(B_4) -
  h^{3,0}(B_4) +
  h^{0,4}(B_4) \,,
\end{equation}
where the K\"ahler structure of $B_4$ forces all but the first and last terms to
be zero, and $h^{0,0} = 1$. Further it is required that $B_4$ is compatible
with being the base space of a Calabi-Yau elliptic fibration, $Y_5$, and the
existence of any $(0,4)$-forms on $B_4$ would, under the uplift to $Y_5$, ruin
the Calabi-Yau property of the total space. In conclusion, it is necessary for
such a $B_4$ to have $\chi_0(B_4) = 1$ and thus the total anomaly
(\ref{eqn:D3SUGRA}) always vanishes.



\subsection{Consequences for Flavor Symmetry from $3$--$7$ Strings} \label{subsec_Flavoranomaly}

To this juncture we have been principally concerned with the gravitational
sectors of the anomaly inflow. In this section we shall determine constraints on the flavour sector from the anomaly inflow. The flavor group along the string corresponds to the gauge group on the 7-branes. Recall first expression
(\ref{eqn:anomalyfromferm}) describing the contribution to the anomaly polynomial from a
complex positive chirality Weyl fermion in a representation $R$.
This is to be contrasted with the relevant anomaly inflow term
\begin{equation} \label{flavorinflow}
  - \sum_a {1 \over 4} \text{Tr} F_a^2 (D_a \cdot C) \,,
\end{equation}
where $\text{Tr}$ is a normalised trace fixed by requiring that the
smallest topological charge of the embedded $SU(2)$ instanton is $1$. Such a
normalised trace can be related to the trace over the fundamental
representation
through
\begin{equation}
  \text{tr} F^2 = s_G \text{Tr} F^2 \,.
\end{equation}
The $s_G$ are numerical factors depending only on the gauge group $G$ and we refer in particular to appendix A of \cite{Ohmori:2014kda} for details.
For gauge group $G=  SU(k)$, $s_G=\frac{1}{2}$ and hence the expression (\ref{flavorinflow}) agrees with the perturbatively expected form  $ - \sum_a {1 \over 2} \text{tr} F_a^2 (D_a \cdot C)$ obtained by reducing the D7-brane CS action, which is typically written in terms of the trace over the fundamental representation. For non-perturbative gauge groups, however, in particular the exceptional series, $s_G \neq \frac{1}{2}$. In this case a direct derivation via the CS action of perturbative Type IIB theory is of course not possible, but the crucial point is that the expression  (\ref{flavorinflow}) appears in the anomaly polynomial of the 6d $N=(1,0)$ supergravity theory obtained by F-theory compactifications on a Calabi-Yau three-fold \cite{Sadov:1996zm}. Anomaly cancellation in 6d alone fixes the normalization of the traces as above \cite{Grassi:2000we,Ohmori:2014kda}.

This raises the question how the anomaly from the spectrum may be
cancelled off by the anomaly inflow. Let us consider only a single simple
flavour group $G$. The total flavour contribution from the spectrum is of the 
form
\begin{equation} \label{flavoranomaly}
  \left(-{1 \over 2}\text{tr}_R F^2 \right)(n_{R,+} - n_{R,-}) \,,
\end{equation}
where $n_{R,+/-}$ is the number of positive/negative chirality complex Weyl fermions in representation $R$. If $R$ is a real representation of $G$ then this
contribution comes in addition with an overall factor of $1/2$ to account for the Majorana-Weyl nature of the fermions. 
Since the flavour symmetry has its origin in the gauge group on the $7$-branes, it is
only the $3$--$7$ string modes that are charged under it. For these
$n_{R,+}=0$, $n_{R,-} = C \cdot D_G$ with $D_G$ the divisor wrapped by the 7-brane with gauge group $G$.

The anomaly inflow which should cancel off the anomaly (\ref{flavoranomaly}) is
\begin{equation} \label{inflowtermflavor}
  \left(- {1 \over 4} \text{Tr} F^2 \right) (D_G \cdot C) \,. 
\end{equation}
If the $3$--$7$ strings transform in the fundamental representation $R$, then
it is immediately apparent that for such a cancellation it would be necessary
to have $G$ such that either $s_G = 1/2$ and a complex fundmental
representation, or $s_G = 1$ and a real fundamental representation. One can
see from the appendix of \cite{Ohmori:2014kda} that these are the cases where $G = SU(k)$ or $USp(k)$,
and, respectively, $G = SO(k)$  or $G_2$. For any other flavour group $G$, and for any  representation other than the fundamental, the contribution from (\ref{flavoranomaly}) overshoots the inflow term (\ref{inflowtermflavor}).

From this we are lead to the inescapable conclusion that the anomalies
explicitly forbid any other flavour groups than the above. One can see this
feature already when considering the anomaly of the E-string theory, in 6d,
which was written down in \cite{Shimizu:2016lbw}. The E-string theory has a
global $E_8$ flavor symmetry {\it at the infrared fixed-point}, corresponding to strong coupling. One can see this symmetry by considering the
E-string theory as the theory on the worldvolume of an  M2-brane interpolating between an
M5-brane and an end-of-the-world M9-brane. The endpoint on the M9-brane gives
rise to the $E_8$ symmetry. 

Such a setup has a perturbative description in type IIA, see
\cite{Kim:2014dza} for more details, where the E-string theory in turn has a
weakly coupled description. In this understanding of the theory there is not
any $E_8$ symmetry any longer {\it in the UV}, it is removed as the M9-brane becomes a stack
of $8$ D8-branes and an O8-plane in the limit. Such a brane stack has an
$SO(16)$ symmetry, which is then the flavor symmetry in the UV of the
E-string theory. Calculating the anomaly from the spectrum here yields a
successful cancellation \cite{Shimizu:2016lbw}, as the $3$--$7$ string multiplets now transform in a
real representation of an $SO$ group, which as we discussed above, is
sufficient for the cancellation between the inflow and the spectrum.

From these results we conjecture that whenever a D3-brane
wraps a curve intersecting a 7-brane with gauge group $G$ which is not of
$SU$, $USp$, $SO$, or $G_2$ type, then $G$ will not survive intact as a flavour group into the UV description of
the 2d theory.



\subsection*{Acknowledgements}

We thank Benjamin Assel, Chris Couzens, Michele del Zotto, Arthur Hebecker, Neil Lambert, Ling Lin, Dario Martelli, Christoph Mayrhofer, Eran Palti, Christian Reichelt, Fabian Ruehle, Jenny Wong, and Fengjun Xu for discussions. 
SSN acknowledges support by the ERC Consolidator Grant 682608 ``Higgs bundles: Supersymmetric Gauge Theories and Geometry (HIGGSBNDL)".
The work of TW and CL was partially supported by DFG under Grant TR33 'The Dark Universe' and under GK 1940 'Particle Phyiscs Beyond the Standard Model'. TW and CL thank the Fields Institute Toronto for hospitality during important stages of this work.


\appendix

\section{Supersymmetry in 2d and 1d}
\label{app:SUSYS}
\subsection{2d $(0,2)$ and $(0,4)$ Supersymmetry}\label{app:2dmults}

In this appendix we summarize the structure of the supermultiplets in
 2d $(0,2)$ and $(0,4)$ theories. 
 $(0,2)$ theories in two dimensions have an R-symmetry group $U(1)_R$  and the
following multiplets \cite{Witten:1993yc}:
\begin{enumerate}
  \item Vector multiplet: contains a gauge field, $A_\mu$, and an adjoint
    valued left-moving fermion, $\lambda_-$.
  \item Chiral multiplet: contains a complex scalar, $\phi$, and a
    right-moving fermion, $\psi_+$, both valued in a representation $R$ of
    the gauge group. If $\mathcal{R}[-]$ denotes the R-symmetry charge of the given
    field then the fields in the chiral multiplet must satisfy $\mathcal{R}[\psi_+] =
    \mathcal{R}[\phi] - 1$.
  \item Fermi multiplet: contains a single left-moving fermion, $\psi_-$,
    transforming in a representation $R$ of $G$.
\end{enumerate}

The 2d $(0,4)$ multiplets can be built out of the $(0,2)$ multiplets.
Such theories have an $SO(4)_R \cong SU(2)_R \times SU(2)_I$ R-symmetry,
which, in the IR SCFT is broken to a single $SU(2)_R$ R-symmetry
\cite{Witten:1994tz}. The supercharges transform in the $({\bf 2,2})_+$
representation, where the subscript denotes the Lorentz $SO(1,1)$ chirality.
The possible multiplets compatible with supersymmetry are then given as
follows \cite{Tong:2014yna,Putrov:2015jpa}:
\begin{enumerate}
  \item Vector multiplet: contains a $(0,2)$ vector multiplet and an
    additional adjoint valued $(0,2)$ Fermi multiplet. It then contains a
    gauge field, $A_\mu$, and a pair of left-moving complex fermions,
    $\lambda^a_-$, transforming in the $({\bf 2,2})_-$.
  \item Hypermultiplet: contains a pair of $(0,2)$ chiral multiplets in
    conjugate representations of $G$. The pair of complex scalars transforms
    as $({\bf 2,1})$ under the R-symmetry, while the pair of right-moving
    fermions transforms in $({\bf 1,2})$.
  \item Twisted hypermultiplet: much like the regular hypermultiplet it
    consists of a pair of $(0,2)$ chiral multiplets in conjugate
    representations, however the pair of complex scalars transforms in the
    $({\bf 1,2})$, and the right-moving fermions as $({\bf 2,1})$.
  \item Fermi multiplet: contains a pair of $(0,2)$ Fermi multiplets in
    conjugate representations. The left-handed fermions transform in the
    $({\bf 1,1})$ representation of the R-symmetry.
  \item $1/2$--Fermi multiplet: a single $(0,2)$ Fermi multiplet which is a
    singlet under the $SO(4)_R$ symmetry is also consistent with the
    enhanced $(0,4)$ supersymmetry. We shall call this a $1/2$--Fermi to
    contrast it with the ``true'' $(0,4)$ Fermi above.
\end{enumerate}

For the $(0,4)$ theories we are interested in how each multiplet contributes
to the central charges of the theory. The right-moving central charge and the
gravitational anomaly are 
\begin{equation}\label{eqn:contri}
  c_R = 3 \text{Tr}_{\text{Weyl ferm.}} \gamma^3 Q_R^2 \quad \,,\, \quad c_R -
  c_L = \text{Tr}_{\text{Weyl ferm.}} \gamma^3 \,,
\end{equation}
where $Q_R$ is the R-charge of the fermion, and $\gamma^3$ is the chirality. 
Since the $(0,4)$ superconformal algebra
fixes that the fermions of each (twisted) hypermultiplet have R-charge $1$,
and of each
(half-)Fermi multiplet have R-charge $0$, it is easily seen that the
contribution from each multiplet is
\begin{equation}
  c_R = 3 n_R \quad \,,\, \quad c_L = 2n_R + n_L \,,
\end{equation}
where $n_{R/L}$ is the number of right/left Weyl fermions in the multiplet.
The contributions, excluding the vector multiplet,
are
\begin{equation}
  \begin{array}{c|c|c|c}
    & \text{(Twisted) Hyper} & \text{Fermi} & 1/2\text{--Fermi} \cr\hline
    (c_R, c_L) & (6,4) & (0,2) & (0,1)
  \end{array} \,.
\end{equation}
It is not possible to give such a conclusion for the $(0,2)$ theories, as the
R-charge is not determined by the superconformal algebra, one must use the
expressions (\ref{eqn:contri}) for the appropriate $Q_R$ in each case.

\subsection{1d SQM}
\label{app:SQM}

We briefly summarize now the 1d SQM with extended supersymmetry and its multiplet structure, which we obtain from the M2-brane reduction and which will be matched with the dimensional reduction of the 2d $(0,2)$ theories. A nice review of these 1d SQM theories can be found in \cite{Okazaki:2015pfa}. 

An important fact to remember in 1d is that physical scalars and auxiliary
scalars  can be exchanged into each other, by the so-called automorphic
duality \cite{Gates:2002bc} (for an in-depth exposition see
\cite{Bellucci:2005xn}), which corresponds here to exchanging time-derivatives
of physical bosons with auxiliary fields. {This map can be performed as long
as the theory is free or the fields take values in a target whose metric does
not depend on the dualized fields}. Multiplets are therefore denoted by $(n,
N, N-n)$ where $n$ is the number of real bosonic degrees of freedom, $N$ the
number of supersymmetries and $N-n$ the number of auxiliary fields. 
 
The $N=2$ SQM has two types of representations, either $2A$, which is the dimensional reduction of $N=(1,1)$ in 2d, or, more relevant for us, $2B$, which is the reduction of 2d $N=(0,2)$. The $N=2B$ SQM has a $U(1)_R$ symmetry, which rotates the two supercharges into each other, and descends from the R-symmetry of the 2d theory. 
There are chiral and Fermi multiplets of $2B$ SQM:
\begin{enumerate}
\item $2B$ chiral multiplet $(2,2,0)$: this has a complex scalar and complex fermion and no auxiliary field, 
\be
\Phi = \varphi + \theta \psi + {i\over 2 }\theta \bar\theta \dot{\varphi} \,,
\ee 
with 
\be
\begin{aligned}
\delta \varphi &= i\epsilon \psi \,, && \delta \bar\varphi  = i \bar\epsilon \bar\psi \\
\delta \psi & = \bar\epsilon \dot{\varphi} \,,&& \delta \bar\psi = \epsilon \dot{\bar\varphi} \,.
\end{aligned}
\ee
The $2B$ chiral multiplet is related, upon reduction from 2d to 1d, to a
2d $(0,2)$ chiral multiplet with one complex scalar, $\phi$, and fermion,
$\psi_+$, and supersymmetry transformations (see for conventions \cite{Schafer-Nameki:2016cfr})
\be
\begin{aligned}
\delta \phi &= - \sqrt{2} \epsilon_- \psi_+  \,, & &  \delta \bar\phi = \sqrt{2} \bar\epsilon_- \bar{\psi}_+ \cr 
\delta \psi_+& = i \sqrt{2} \partial_+ \phi \bar\epsilon_- \,,& & \delta \bar\psi_+ = - i \sqrt{2} \partial_+ \bar\phi \epsilon_- \,.
\end{aligned}
\ee
Indeed, these transformations reduce (up to rescaling of the scalars by $i/\sqrt{2}$) to the $2B$ multiplet. In a similar fashion one can see that the 2d $(1,1)$ multiplets reduce to $2A$ multiplets, which have real scalars and auxiliary fields. 

\item $2B$ Fermi multiplet $(0,2,2)$: the lowest component is a complex fermion $\rho$, and includes a complex auxiliary scalar $h$ such that 
\be
\Lambda = \rho + \theta h + {i\over 2} \theta \bar\theta \dot\rho \,,
\ee
with the supersymmetry transformations under the two supercharges 
\be
\begin{aligned}
\delta \rho&= i\epsilon h \,, && \delta \bar\rho = -i \bar\epsilon \bar{h} \\
\delta h& = \bar\epsilon \dot{\rho} \,,&& \delta \bar{h}= -\epsilon \dot{\bar\rho} \,.
\end{aligned}
\ee
As in the chiral multiplet case, the Fermi multiplet of 2d $(0,2)$
dimensionally reduces to a 1d 2B Fermi multiplet. 

\end{enumerate}


\section{Duality Twisted Dimensional Reduction of $N=4$ SYM } \label{app:D3action}

\subsection{Conventions and Field Identifications}

In this appendix we shall give some more details about the reduction of the
$N=4$ SYM action and supersymmetry transformations in terms of the duality twisted
fields and the scalar supersymmetries.
Let us begin by summarizing our conventions to help us to identify the field content of the D3-brane theory on $\mathbb R^{1,1} \times C$.
  For all three situations, $C \subset Y_n$ with $n=3,4,5$, we
introduce the coordinates $x^0$, $x^1$ along $\mathbb{R}^{1,1}$ and $x^8$,
$x^9$ along $C$, which we shall usually write in the combinations $z =
\frac{1}{2}(x^8 - i x^9)$ and $\bar{z} = \frac{1}{2}(x^8 + i
x^9)$.\footnote{Note that this slightly unconventional definition of
holomorphic versus anti-holomorphic coordinates matches our assignment that
after the twist a field of topological twist charge $q_C = +1$ corresponds
to a $(0,1)$ form.} We furthermore define the derivatives $\partial_{\pm}
= \partial_0 \pm \partial_1$ as well as $\partial \equiv \partial_z =
\partial_8 + i \partial_9$, $\bar\partial \equiv \partial_{\bar z} =
\partial_8 - i \partial_9$. In these conventions the K\"ahler  form on $C$
becomes
\begin{equation}
  J = 2 dx^8 \wedge dx^9 = - 4 i d z \wedge d \bar z,  \qquad \quad  \ast_C 1
  = J, \qquad \ast_C J = 1 \,. 
\end{equation}
The fields appearing in the decomposition of the 4d vector field $A_\mu$ as
in (\ref{DecompFieldsCY3}),  (\ref{N4decompD3inB3}) and (\ref{N4decompD3inB4}) and  can then be identified, by studying their
transformation properties under $SO(1,1)$ and $U(1)_C$:
The scalar fields $a$ and $\bar a$ represent the components of the $U(1)$
gauge field along the curve $C$, 
\begin{equation}
  a \equiv a_{z} = A_8 + i A_9, \qquad \bar{a} \equiv \bar a_{\bar z} =  A_8 - i A_9
  \,,
\end{equation}
where the subscripts $z$, $\bar{z}$ anticipate the form bi-degree $(1,0)$
and $(0,1)$ respectively of the fields on the curve $C$, and the 
external components of the gauge field organize into 
\begin{equation}
  v_{\pm}  = A_0 \pm A_1 \,.
\end{equation}
In particular, 
\be
i F_{89} = i (\partial_8 A_9 - \partial_9 A_8) =\frac{1}{2} \left(
\bar\partial a_{ z}  - \partial \bar a_{\bar z} \right) \,.
\ee

The treatment of the normal space to the D3 brane differs for compactifications on $Y_3$, $Y_4$ and $Y_5$:

\paragraph*{D3 on $Y_3$}
On the normal space to the D3-brane we introduce coordinates $x^2,\ldots,x^5$ for the external normal space $\mathbb R^4 \subset \mathbb R^{1,5}$ as well as local coordinates $x^6,x^7$ for the normal space of $C$ inside the base $B_2$. The scalar fields $\phi$ transforming in the ${\bf 6}$ of $SU(4)_R$ of the $N=4$ SYM on the D3-brane can then be identified with the normal fluctuations in the transverse directions and written in components as $\phi_m$, $m=2, \ldots,7$. 
The normal fluctuations of the D3-brane inside the Calabi-Yau are given by 
\be
\sigma_{z}  = \phi_6 + i \phi_7, \qquad \bar\sigma_{\bar z} = \phi_6 - i \phi_7,
\ee
while the external normal fluctuations $\phi_2, \ldots \phi_5$ organize as an $SO(4)_T$ bispinor $\varphi_{A \dot B}$ defined as 
\be
\varphi_{A \dot B}  = \left( \begin{array}{cc}   - \phi_2 + i \phi_3 &  \phi_4 - i \phi_5 \cr  \phi_4 +  i \phi_5 &  \phi_2 + i \phi_3   \end{array} \right).
\ee
Finally, chiral $SO(4)_T$ spinors transforming as a $({\bf 2,1})$ of $SO(4)_T \simeq SU(2)_{T,1} \times SU(2)_{T,2}$ are denoted as 
 objects $\psi_A$, while anti-chiral $SO(4)_T$ spinors in the $({\bf 1,2})$ are of the form $\tau^{\dot B}$.
 We will encounter contractions  among chiral spinors defined as $\psi \chi = \psi^A \chi_A = \chi \psi$ (and similarly for anti-chiral spinors as  $\tau \rho = \tau_{\dot A} \rho^{\dot A} = \rho \tau$).\footnote{Spinor indices are raised and lowered by $\varepsilon^{AB} = \left(\begin{array}{cc} 0 & 1 \cr -1 & 0 \end{array}\right)=\varepsilon^{\dot A \dot B}$ and $\varepsilon_{AB} = \left(\begin{array}{cc} 0 & -1 \cr 1 & 0 \end{array}\right)=\varepsilon_{\dot A \dot B}$. 
 }

\paragraph*{D3 on $Y_4$}

Let us denote by 
$x^6, x^7$ the two coordinates transverse to the D3-brane in $\mathbb R^{1,3}$ and by $x^2,x^3,x^4,x^5$ the coordinates normal to $C$ inside $B_3$. 
In terms of  the scalar field $\phi_2, \ldots \phi_7$ in the ${\bf 6}$ of $SU(4)_R$  the remaining scalar fields in the decomposition (\ref{N4decompD3inB3}) are given by the combinations
\begin{align}
&g = \phi_6 + i \phi_7, \qquad   & \bar g = \phi_6 - i \phi_7 \\
&\varphi_A = \left( \begin{array}{c} \phi_4 - i \phi_5 \\ \phi_2 + i \phi_3 \end{array}  \right), \qquad & \bar \varphi_A = \left( \begin{array}{c} -\phi_2 + i \phi_3 \\ \phi_4 + i \phi_5  \end{array} \right).
\end{align}
The index $A$ refers to the ${\bf 2}$ representation of $SU(2)_R$,  the non-abelian part of the structure group of the normal bundle $N_{C/B_3}$.

\paragraph*{D3 on $Y_5$}

Here the normal coordinates $x^2, \ldots, x^7$ lie entirely inside the base
$B_4$. We introduce the index $\alpha$ and $\dot \alpha$ for the ${\bf 3}$ and
${\bf \bar 3}$ of the $SU(3)_R$ structure group of the  normal bundle $N_{C/B_4}$ 
and define the fields  
\bea
\varphi_\alpha = \left(\begin{array}{c}  \phi_2 - i \phi_3 \\-\phi_4 + i
  \phi_5 \\\phi_6 - i \phi_7\end{array} \right)\,, \qquad
\bar\varphi^{\dot\alpha} = \left(\begin{array}{c}  \phi_2 + i \phi_3 \\-\phi_4
  - i \phi_5 \\\phi_6 + i \phi_7\end{array} \right) \,.
\eea
In performing the reduction we follow closely the conventions specified in 
appendix A of \cite{Schafer-Nameki:2016cfr}, to which we refer for more details.

\subsection{Action and Supersymmetry}

We now verify that the dimensionally reduced effective action for the D3-brane on $\mathbb R^{1,1} \times C$
is indeed invariant under the twisted supersymmetry variations stated in the
main text. For brevity we describe this here explicitly only for the (0,4)
theory obtained for a curve $C \subset Y_3$. 

The action of $N=4$ SYM has four components, covering the gauge, topological, fermion,
and scalar sectors. We shall consider each of these terms separately, and fix
the relative factors between the four terms by requiring that the action be
invariant under the twisted supersymmetry. The gauge and topological actions
can be rewritten in terms of the twisted fields as
\begin{align}
  {\cal L}_{\rm g} = F_{\mu \nu} F^{\mu \nu} &= 
  -\frac{1}{2} (\partial \bar a - \bar \partial a)^2  - \frac{1}{2}  (\partial_-
v_+ - \partial_+ v_-)^2  \cr &\quad\, - \partial_- a \, \partial_+ \bar a - \partial_- \bar a
\,  \partial_+ a  - \partial v_+ \, \bar \partial v_- - \partial v_- \,
\bar\partial v_+ \cr &\quad\, + \bar \partial v_+ \, \partial_- a + \partial v_+ \,
\partial_- \bar a + \partial v_-\,  \partial_+ \bar a + \bar \partial v_- \,
\partial_+ a \,, \\
  {\cal L}_{\rm top} =\epsilon^{\mu \nu \alpha \beta} F_{\mu \nu} F_{\alpha
\beta} &= 
  \frac{3i}{2} \Big(  
    \partial v_+ \, \bar\partial v_- - \partial v_- \bar\partial v_+ 
    + \partial_- \bar a \, \partial_+ a - \partial_- a \, \partial_+ \bar a  \cr &\quad\quad\,  
    + \bar\partial v_+ \, \partial_- a - \partial v_+ \, \partial_- \bar a + \partial
\bar a \, \partial_- v_+ - \bar \partial a \, \partial_- v_+  \cr &\quad\quad\,
 -   \bar\partial v_- \, \partial_+ a + \partial v_- \,
\partial_+ \bar a - \partial \bar a \, \partial_+ v_- + \bar \partial a \,
\partial_+ v_- \Big) \,.
\end{align}
They are agnostic towards the R-symmetry part of the $N=4$ theory as
they depend only on the 4d gauge field, which is a singlet under the $SU(4)_R$
symmetry. These terms and fields will then be the same regardless of the
dimension of the K\"ahler base in which the wrapped curve is located. The
fermion and scalar terms will depend on the dimension of the compactification,
and here is where we shall focus on the $(0,4)$ content from the Calabi-Yau
three-fold compactification. Now that the coupling $\tau$ varies over the curve
$C$ there are fields in the decomposition of the scalars and fermions which
transform non-trivially under the $U(1)_D$ duality group. Any derivatives
along $C$ of these fields are, following \cite{Martucci:2014ema}, promoted to $U(1)_D$
covariant derivatives, which were described in (\ref{eqn:covderivs}). The
scalar and fermion actions from the $N=4$ Lagrangian for 
compactification on Calabi-Yau three-fold are then
\begin{align}
  {\cal L}_{\rm sc} =\partial_\mu \phi_i \partial^\mu \phi^i &= 
  - \partial_- \sigma \,  \partial_+ \bar \sigma  - \partial_- \bar\sigma \,
  \partial_+ \sigma + \bar\partial_{\mathcal{A}} \sigma \, \partial_{\mathcal{A}} \bar \sigma + \bar \partial_\mathcal{A}
  \bar \sigma \,  \partial_\mathcal{A} \sigma  \cr &\quad\, + \partial_+ \varphi_{A \dot B} \,
  \partial_- \varphi^{A\dot B} - \partial \varphi_{A \dot B} \, \bar\partial
  \varphi^{A \dot B}  \,, \\
  {\cal L}_{\rm f} = \bar\Psi  \Gamma^\mu \partial_\mu \Psi &= 
  \lambda_- \partial_+ \tilde\lambda_- 
  + \psi_+ \partial_\mathcal{A} \tilde\lambda_- 
  - \tilde\lambda_- \partial_+ \lambda_- 
  - \tilde\lambda_- \partial_\mathcal{A}\psi_+  \cr &\quad\, 
  + \mu_+ \partial_- \tilde\mu_+ - \tilde\mu_+ \partial_-\mu_+ 
  + \rho_- \bar\partial \tilde\mu_+ 
  - \tilde\mu_+ \bar\partial \rho_- \cr &\quad\, 
  - \lambda_- \bar\partial_\mathcal{A} \tilde\psi_+ 
  - \psi_+\partial_- \tilde\psi_+ 
  + \tilde\psi_+ \bar\partial_\mathcal{A} \lambda_- 
  + \tilde\psi_+ \partial_- \psi_+ \cr &\quad\, 
  - \mu_+ \partial \tilde\rho_-
  - \rho_-\partial_+ \tilde\rho_- 
  + \tilde\rho_-\partial\mu_+ 
  + \tilde\rho_- \partial_+ \rho_- \,.
\end{align}
The total action for the twisted theory on $\mathbb R^{1,1} \times C$ can be written as
\bea\label{eqn:CY3actionAPP}
\int_{\mathbb R^{1,1} \times C} {d^4 x \sqrt{|g|}} \left(   c_1 \,  \tau_2 \,  {\cal L}_{\rm g} +   c_2 \, \tau_1 \,  {\cal L}_{\rm top} + c_3\,  {\cal L}_{\rm sc} + c_4 \, {\cal L}_{\rm f}   \right) \,,
\eea
where the constants will be fixed, up to an overall factor, by twisted supersymmetry invariance.

\paragraph*{Supersymmetry}  The supersymmetry variations of the $N=4$ SYM 
theory can be reduced to variations with respect to the surviving supersymmetries and written in terms of the twisted fields. 
The variation of the bosonic fields becomes
\begin{align}
&\sqrt{\tau_2 } \delta a = 2 i \epsilon_- \tilde \psi_{+}\,, \quad
&\sqrt{\tau_2 }  \delta \bar a =   2 i \tilde \epsilon_-  \psi_{+}\,, \label{eq-deltaaCY3} \\
& \delta \sigma = -  2 i \tilde \epsilon_- \tilde \psi_{+}\,, \qquad    & \delta
\bar\sigma =  2  i \epsilon_-  \psi_{+}\,,\label{eq-deltasigmaCY3} \\
& \delta \varphi_{A \dot B} = - 2 i  (\epsilon_{-A} \mu_{+\dot B} + \tilde
\epsilon_{-A} \tilde \mu_{+\dot B})\,,  \label{eq-deltavarphiCY3}\\
& \sqrt{\tau_2}  \, \delta v_- = 2 i (\lambda_- \tilde \epsilon_- + \tilde
\lambda_-  \epsilon_-)\,, \qquad & \delta v_+ = 0 \,.
\end{align}
For the fermionic variations one finds 
\begin{align}
  &\delta \psi_{+} = \epsilon_- \sqrt{\tau_2}  (- \partial_+ \bar a  +  \,
  \bar\partial v_+ ) +  \tilde \epsilon_- \partial_+ \bar \sigma \,,    
  & \delta \tilde \psi_{+} =  \tilde \epsilon_- \sqrt{\tau_2}  (\partial_+ a -
  \partial v_+) +  \epsilon_- \partial_+ \sigma\,,  \label{eq-deltapsiCY3}    \\
  &\delta \mu_+^{\dot B} =  -  \tilde  \epsilon_{- A} \partial_+  \varphi^{A
  \dot B} \,,  
  & \delta \tilde \mu_+^{\dot B} =   +    \epsilon_{ - A}
  \partial_+  \varphi^{A \dot B}\,,  \label{eq-deltamuCY3}\\
  &\delta \rho_-^{\dot B} =  \tilde \epsilon_{ - A}  \partial_{\cal A}
  \varphi^{A \dot B}\,, \quad    
  & \delta {\tilde \rho}_-^{\dot B} =  \epsilon_{- A} \bar\partial_{\cal A}
  \varphi^{A \dot B}\,,\label{eq-deltarhoCY3} \\
  &\delta \lambda_- = -  \epsilon_-  (\sqrt{\tau_2}  F_{01}  +  {\cal F}_{\cal
  A})   - \tilde \epsilon_- \ast_C  \partial_{\cal A} \bar \sigma\,,  
  & \delta \tilde \lambda_- = -  \tilde \epsilon_- (\sqrt{\tau_2}   F_{01}  -
  {\cal F}_{\cal A} )+  \epsilon_- \ast_C \bar\partial_{\cal A} \sigma
  \label{eq-deltalambdaCY3} \,,
\end{align}
where
\bea
F_{01} = \frac{1}{2} \left( \partial_- v_+ - \partial_+  v_-\right)\,, \qquad
{\cal F}_{\cal A} = \frac{1}{2} \sqrt{\tau_2} (\bar\partial a- \partial \bar a) \,.
\eea
It can be verified explicity that the action (\ref{eqn:CY3actionAPP}) is
indeed invariant under these supersymmetry variations, without use of
equations of motion and as long as one ignores boundary
terms, if we fix
\begin{equation}
  4 i c_3 - 2 c_4 = 0 \,, \quad
  c_4 - 2 i c_1 = 0 \,, \quad
  3 c_2 - 2c_1 = 0 \,.
\end{equation} 
It is vital for supersymmetry that the coupling $\tau$ varies holomorphically
along $C$,
\begin{equation}
  \bar \partial\tau = 0 \,,
\end{equation}
as this is required for a cancellation of the variations between
the gauge and the topological sectors. The topological terms would usually be
annihilated directly by the SUSY variation, however, there are terms remaining
in this variation that are proportional to the derivatives of $\tau_1$.
Holomorphicity of $\tau$ allows one to relate
\begin{equation} \label{partialtau1tau2}
  \partial \tau_1 = i \partial \tau_2 \quad \,, \quad \overline{\partial}
  \tau_1 = -i \overline{\partial}\tau_2 \,,
\end{equation}
and one can then notice that these variations are cancelled by the variations
proportional to the derivatives of $\tau_2$ from the gauge action.
Furthermore, the boundary terms are expected to cancel against the variation of the defect action describing the 7-brane insertions along $C$.     This would be
in analogy to a Euclidean D3-brane wrapping a K\"ahler surface as was studied in \cite{Martucci:2014ema}.   

\paragraph*{$U(1)_D$ invariance}
It is observed that the righthand side of (\ref{eq-deltaaCY3}) has a definite
$U(1)_D$ charge, as it is composed to a fermion whose $U(1)_D$ charged is
fixed by the well-definedness of the 4d fermions with respect to the
$U(1)_D$ symmetry. The lefthandside is then an object of the same $U(1)_D$
charge, and as the variation is constructed to be a singlet under $U(1)_D$
it can be determined that the objects $ \sqrt{\tau_2} a$ and $ \sqrt{\tau_2}
\bar a$ have unambiguous $U(1)_D$ charge, despite their origin in the 4d
vector field, which has no definite charge. 

\paragraph*{Equations of motion for the scalars}
\be
\begin{aligned}
\partial \left(\tau_2 ( \partial \bar a  - \bar\partial a) \right)   + \tau_2 \left(2 \partial_+ \partial_- a - \partial_- \partial v_+ - \partial_+ \partial v_-\right) &= 0 \\
\bar\partial \left(\tau_2 (\bar\partial  a  - \partial \bar a)\right) + \tau_2 \left(2 \partial_+ \partial_- \bar a - \partial_- \bar\partial v_+ - \partial_+ \bar\partial v_-\right) &= 0  \\
 \partial (\tau_2 \bar\partial v_- - \tau_2 \partial_- \bar a)  +  \bar\partial (\tau_2 \partial v_- - \tau_2 \partial_-  a)      + \tau_2  \partial_- (\partial_- v_+ -  \partial_+ v_-) &=0 \\
  \partial (\tau_2 \bar\partial v_+ - \tau_2 \partial_+ \bar a)  +
  \bar\partial (\tau_2 \partial v_+ - \tau_2 \partial_+ a)        + \tau_2
  \partial_+(\partial_+ v_- -  \partial_- v_+) &=0 \,. \\
 \end{aligned} 
 \ee

The first two and the last equations vanish identically once the BPS equations
$\partial_+ v_- - \partial_- v_+ = 0$ and $\bar \partial a - \partial \bar a =
0$   as well as $\partial v_+ = \partial_+ a$ and $\partial_+ \bar a = \bar
\partial v_+$ are enforced. The third equation involving $v_-$ does not vanish
identically by means of the BPS equations. This is consistent with the chiral
$(0,4)$ nature of the supersymmetry and in particular the fact that $v_-$ does
not appear independently in the BPS equations. 

%
%
%
%
%



\section{Cohomology Computations in Elliptic Fibrations}

\subsection{Invariants of Elliptic Surfaces}

\label{app:ellipticsurfaces}

In this appendix we collect some results on the topological properties of
elliptic surfaces, following \cite{MR1078016,MR2732092}, that will be useful
in the computation of the zero-modes for the D3 and M5 compactifications. 

A Weierstrass fibration, $S$, over a projective curve $C$ can be defined by
the triple $(\mathcal{L}, f, g)$, where $\mathcal{L}$ is a line bundle on $C$
and $f$ ($g$) is a section of $\mathcal{L}^4$ ($\mathcal{L}^6$) such that
$\Delta = 4f^3 + 27g^2$ is a non-identically-zero section of
$\mathcal{L}^{12}$. The total space of the fibration is not required to be
smooth but can have rational double points. When the elliptic surface $S$ is
chosen as the restriction of an elliptically fibered Calabi-Yau $n$-fold $Y_n$ to a curve $C$, with  
 $\mathcal{L}  \equiv {\cal L}_D \cong K_{B_{n-1}}^{-1}|_C$, then we shall call the resulting
surface $\widehat{C}$, as is frequently used in the main text. However in this
appendix we shall consider a general $\mathcal{L}$.

We shall first consider the Leray spectral sequence relating the
vector-bundle valued cohomology groups on $S$ to those on $C$. For an elliptic
surface the spectral sequence degenerates to
\begin{equation}
  \label{Leray1}
  \vcenter{\xymatrix@R=10pt@M=4pt@H+=22pt{
         0 \ar[r] & H^0(C,\pi_\star V)  \ar[r] &
       H^0(S, V)  \ar[r] &
      0  
      \ar`[rd]^<>(0.5){}`[l]`[dlll]`[d][dll] & 
      \\
        & H^1(C,\pi_\star V)  \ar[r] &
       H^1(S, V)  \ar[r] &
      H^0(C, R^1 \pi_\star V)  
      \ar`[rd]^<>(0.5){}`[l]`[dlll]`[d][dll] & 
      \\
      & 0  \ar[r]  &
       H^2(S, V)   \ar[r] &
      H^1(C, R^1 \pi_\star V) \ar[r] &
      0,
      \,
    }}
\end{equation}
where $V$ is a vector bundle on $S$.
In the cases of interest to us, the right derived images can be obtained with the help of the projection formula
\bea
R^q\pi_\star( M \otimes \pi^\star{\cal N} ) = R^q\pi_\star( M) \otimes {\cal N}, 
\eea
where ${\cal N}$ is a vector bundle on $C$.
In particular, for any pullback bundle we obtain
\bea
R^0\pi_\star( \pi^\star{\cal N} ) &=&  \pi_\star( \pi^\star{\cal N} )   =  {\cal N},  \\
R^1\pi_\star( \pi^\star{\cal N} ) &=&  R^1\pi_\star( {\cal O}_{S} )  \otimes \pi^\star{\cal N}    =  {\cal L}^\vee \otimes {\cal N}.
\eea
In the last equality we have used that 
\bea
R^1\pi_\star( {\cal O}_{S})   ={\cal L}^{\vee}.
\eea

For instance, the cohomology groups counting the massless spectrum on an M5-brane wrapping
$\widehat C$, the elliptic surface over $C$ in $Y_n$ with $\mathcal{L} = \mathcal{L}_D \cong K_{B_{n-1}}^{-1}|_C$, will
involve 
\begin{equation}
  V = N_{\widehat{C}/Y_n} = \pi^*N_{C/B_{n-1}} \, .
\end{equation}
As will be discussed in the next section, this
particular bundle $V$ together with the fixed form of $\mathcal{L}$ will
allow the computation, via adjunction, of the dimensions of the cohomology
groups explicitly as used in the main text in sections \ref{sec:M5CY3} and \ref{sec:M5CY4}.

Now let us summarise some of the results for a general elliptic surface, which
follow from the above spectral sequence, and compute the numerical invariants
of the surface.
The fibration $S$ has the form of a product of $C$ with a smooth elliptic
curve if and only if  $\mathcal{L} \cong \mathcal{O}_C$. The global invariants, as just discussed, can be computed via the
Leray spectral sequence for the projection $\pi : S \rightarrow C$. One finds
that
\begin{equation}
  q = h^{0,1}(S) = \begin{cases}
    g \,\,\,\,\,\qquad S \text{ is not a product} \cr
    g + 1 \quad S \text{ is a product}
  \end{cases} \,,
\end{equation}
with $g$ the genus of $C$. Thus when $S$ carries a non-trivial elliptic fibration all one-forms on
$S$ are pullbacks from $C$. Similarly one can compute the geometric genus
\begin{equation}
  p_g = h^{0,2}(S) = \begin{cases}
    g + \text{deg}(\mathcal{L}) - 1 \quad S \text{ is not a product} \cr
    g + \text{deg}(\mathcal{L}) \,\,\,\,\qquad S \text{ is a product}
  \end{cases} \,.
\end{equation}
The Euler characteristic can be computed as the alternating sum of the above
Hodge numbers and in both the trivial and non-trivial cases is
\begin{equation}
  \chi(S) = \text{deg}(\mathcal{L}) \,.
\end{equation}
Using the Noether formula and that $K_S^2 = 0$ determines the Euler number
\begin{equation}
  e(S) = 12\text{deg}(\mathcal{L}) \,,
\end{equation}
which can be used to determine the final unknown entry in the Hodge diamond of
$S$: 
\begin{equation}
  h^{1,1}(S)  = e(S) + 4h^{0,1} - 2h^{0,2} - 2 = \begin{cases}
    10 \, \text{deg}(\mathcal{L}) + 2g \,\,\,\,\,\qquad  S \text{ is not a product} \cr
    10 \, \text{deg}(\mathcal{L}) + 2g + 2 \quad  S \text{ is a product} 
  \end{cases} \,.
\end{equation}
Finally we know that if $(\mathcal{L}, f, g)$ forms Weierstrass data over a
projective curve $C$ then $\text{deg}(\mathcal{L}) \geq 0$, since
$\mathcal{L}^{12}$ must have a section that is not identically zero for the
discriminant $\Delta$ to not be identically zero. This can be used to make a
statement purely about the cohomology on the base curve $C$; if $(\mathcal{L},
f, g)$ provides the Weierstrass data for a Weierstrass fibration over $C$ then
\begin{equation}
  h^0(C, \mathcal{L}^{-1}) = \begin{cases}
    0 \quad  S \text{ is not a product} \cr
    1 \quad  S \text{ is a product}
  \end{cases} \,.
\end{equation}


\subsection{Bundle Cohomology Computations}
\label{app:bdlcoho}

Consider $\pi: Y_n \rightarrow B_{n-1}$ an elliptically fibered Calabi-Yau $n$-fold
with curve $C \subset B_{n-1}$ such that $\widehat{C} = \pi^{-1}(C)$ is a
non-trivially fibered elliptic surface. As noted, in this case ${\cal L} = {\cal L}_D = K^{-1}_{B_{n-1}}|_C$. To determine the number of zero-modes
in each of the compactifications in the main text we need to be able to relate the following
Hodge numbers
\begin{equation}
  \begin{aligned} \label{app:cohoms1}
    D3 \,&:\, & &h^0(C, K_C \otimes \mathcal{L}_D) \,,\quad h^0(C, N_{C/B_{n-1}})
    \,,\quad h^1(C, N_{C/B_{n-1}}) \cr
    M2 \,&:\, & &h^0(C, N_{C/Y_{n}}) \,,\quad h^1(C, N_{C/Y_n})  \cr
    M5 \,&:\, & &h^0(\widehat{C}, N_{\widehat{C}/Y_n}) \,,\quad h^1(\widehat{C},
    N_{\widehat{C}/Y_n}) \,,\quad h^2(\widehat{C}, N_{\widehat{C}/Y_n}) \,.
  \end{aligned}
\end{equation}
One can see that all of the multiplets in
\cref{tbl:CY3fcmultiplets,eqn:CY4D3zm,tab:CY5D3zm,M5CY3multiplets,tbl:M2CY3fc2,M5CY4multiplets,CY4M2Spec,tab:CY5M2zm} 
are counted by one of these Hodge numbers (or just by a Hodge number of $C$ or $\widehat C$). It may be necessary to utilize Serre duality or
other bundle isomorphisms to write it in the above form.

If $V$ is an arbitrary vector bundle over $C$ then the Hirzebruch-Riemann-Roch
theorem reads
\begin{equation}
  \chi(C, V) = \int_C \big(c_1(V) - \frac{1}{2}\text{rk}(V)c_1(K_C)\big) \,.
 \end{equation}
Since $\mathcal{L}_D$ is an ample line bundle 
\begin{equation}
  h^1(C, K_C \otimes \mathcal{L}_D) = h^0(C, \mathcal{L}_D^\vee) = 0 \,,
\end{equation}
by the Kodaira vanishing theorem, and thus
\begin{equation}
  h^0(C, K_C \otimes \mathcal{L}_D) = g - 1 + \text{deg}(\mathcal{L}_D) \,.
\end{equation}
Furthermore the adjunction theorem in this notation reads that
\begin{equation}
  \text{det}(N_{C/B_{n-1}}) = K_C \otimes \mathcal{L}_D \,.
\end{equation}
For $Y_3$ the determinant acts trivially and thus the dimensions of $H^i(C,
N_{C/B_{2}})$ translate into the expressions in (\ref{app:cohoms1}). Recalling that the first Chern class is
blind to the determinant, and that  ${\rm rk}(N_{C/B_{n-1}}) = n-2$ one can compute the Euler characteristic to determine that
\begin{equation}\label{eqn:eulerD3}
  h^0(C, N_{C/B_{n-1}}) - h^1(C, N_{C/B_{n-1}}) = \text{deg}(\mathcal{L}_D) + (g - 1)(4 -
  n) \,.
\end{equation} 
As such one can write all of the degrees of freedom in terms of the three
quantities
\begin{equation}
  \begin{aligned}
    &g \,, & &\text{deg}(\mathcal{L}_D) \,, & &h^0(C, N_{C/B_{n-1}}) \,.
  \end{aligned}
\end{equation}

For the M2-brane quantities one can use adjunction on $Y_n$, $K_{B_{n-1}} =
N_{B_{n-1}/Y_n}$, to write the short exact sequence for normal bundles as
\begin{equation} \label{app:SESM2}
  0 \rightarrow N_{C/B_{n-1}} \rightarrow N_{C/Y_{n}} \rightarrow \mathcal{L}_D^\vee
  \rightarrow 0 \,.
\end{equation}
The associated long exact sequence in cohomology yields the equivalences
\begin{equation} \label{app:SESM2b}
  \begin{aligned}
    h^0(C, N_{C/Y_n}) &= h^0(C, N_{C/B_{n-1}}) \cr
    h^1(C, N_{C/Y_n}) &= h^1(C, N_{C/B_{n-1}}) + h^0(C, K_C \otimes \mathcal{L}_D) \,.
  \end{aligned}
\end{equation}
The last expression can be simplified using (\ref{eqn:eulerD3}) to
\begin{equation}
  h^1(C, N_{C/Y_n}) = h^0(C, N_{C/B_{n-1}}) + (g - 1)(n - 3) \,.
\end{equation}

For the M5-brane the bundle of interest is
\begin{equation}
  N_{\widehat{C}/Y_{n}} = \pi^*N_{C/B_{n-1}} \,,
\end{equation}
whenceforth one can use the Leray spectral sequence for elliptic surfaces
(\ref{Leray1}) to determine that
\begin{equation}\label{eqn:protoM5rels}
  \begin{aligned}
    h^0(\widehat{C}, N_{\widehat{C}/Y_n}) &= h^0(C, N_{C/B_{n-1}}) \cr
    h^1(\widehat{C}, N_{\widehat{C}/Y_n}) &= h^0(C, N_{C/B_{n-1}} \otimes
        \mathcal{L}_D^\vee) + h^1(C, N_{C/B_{n-1}}) \cr
    h^2(\widehat{C}, N_{\widehat{C}/Y_n}) &= h^1(C, N_{C/B_{n-1}} \otimes
    \mathcal{L}_D^\vee) \,.
  \end{aligned}
\end{equation}
At this point we note that the only situation where these particular
cohomology groups are of interest is when $Y_4$ is a four-fold; as such we may
restrict now to this case without consequence. For a general complex vector
bundle $\mathcal{G}$ of rank $r$ it follows \cite{MR2093043} that
\begin{equation}\label{eqn:dualbdlthm}
  \wedge^n \mathcal{G} = \wedge^{r - n} \mathcal{G}^\vee \otimes \wedge^r
    \mathcal{G} \,.
\end{equation}
By combining this relation with Serre duality one can compute that for
Calabi-Yau four-folds there exists the relationship
\begin{equation}
  h^i(C, N_{C/B_3} \otimes \mathcal{L}_D^\vee) = h^{1-i}(C, N_{C/B_3}) \,.
\end{equation}
One can thus rewrite the relations (\ref{eqn:protoM5rels}) as
\begin{equation}\label{eqn:M5Chatdecomp}
  \begin{aligned}
    h^0(\widehat{C}, N_{\widehat{C}/Y_4}) &= h^0(C, N_{C/B_3}) \cr
    h^1(\widehat{C}, N_{\widehat{C}/Y_4}) &= 2(h^0(C, N_{C/B_3}) -
    \text{deg}(\mathcal{L}_D)) \cr
    h^2(\widehat{C}, N_{\widehat{C}/Y_4}) &= h^0(C, N_{C/B_3}) \,.
  \end{aligned}
\end{equation}



\section{Strings in 8d...}\label{app:K3}

This section is devoted to strings in 8d compactifications of F-theory on an elliptically fibered Calabi-Yau twofold, i.e. 
an elliptic K3 surface with base
 isomorphic to a smooth rational curve.
An M5-brane wrapping a K3 surface is well-known to be the heterotic string \cite{Witten:1995ex} and the
low-energy 2d theory on the worldvolume of the M5-brane has $(0,8)$
supersymmetry. In contradistinction to the situation studied in the previous section, it is of course
not necessary to twist here as the elliptic surface wrapped by the
M5-brane is itself Calabi-Yau.

This $(0,8)$ theory has an F-theory description in terms of a D3-brane
wrapping $C$, the smooth rational curve forming the base of the elliptic
fibration. 
As we will see, the treatment of this theory deviates in an interesting way from its counterparts investigated in section 
\ref{sec:StringsfromD3}. 
Finally we will consider the $N=8$ SQM that arises from an
M2-brane on the base $C$ of $K3$.

\subsection{...from D3-branes}

If one tries to compactify the D3-brane theory, $N=4$
SYM theory, on a $2$-sphere one
might be be tempted to conclude that there is no obvious mechanism with which to twist the theory, as the
entire $SO(6)_R$ symmetry group is used up to describe rotations in the transverse
non-compact space to the brane. Naively one would be lead to believe that in
this setup all supersymmetry on the worldvolume of the D3-brane is broken. The resolution to this puzzle
is that in F-theory on $K3$ the D3-brane intersects 24 7-branes at various points in its
worldvolume, and the resulting variation of the axio-dilaton
provides an additional duality twist which 
ensures that 8 supercharges are preserved.

Indeed, the topological twist giving rise to chiral supersymmetry along the string
twists the $U(1)_C$ holonomy group along curve $C$ by the $N=4$ bonus symmetry $U(1)_D$ such that 
\be \label{TwistK3}
T_{\rm twist} = {1\over 2} \left(T_{C} - T_{D} \right) \,.
\ee
Since the D3-brane wraps the entire base $C = B_1 = \mathbb P^1$, there is no $U(1)_R$ symmetry available which would be associated with the transverse directions of $C$ within the base. The twist (\ref{TwistK3}) can hence be viewed as the {\it difference} of the would-be individual twists of $U(1)_C$ and $U(1)_D$ by $U(1)_R$, which is independent of $U(1)_R$. This is the only option in absence of a well-defined $U(1)_R$. 
The appearance of a single twist is clear also from the M5-brane point of view, where the M5 wraps the full K3 and thus, as pointed out already, does not require any topological twist at all. The twist (\ref{TwistK3}) is the remnant of the non-triviality of the elliptic fibration in the M5-brane picture. 

We obtain the following decomposition of the supercharges under the symmetry groups 
\be
\ba
SO(1,3)_L \times SU(4)_R \times U(1)_D \quad &\rightarrow \quad SO(1,1) \times U(1)_C \times U(1)_D \cr 
({\bf 2}, {\bf 1}, \bar{\bf 4})_1 \quad & \rightarrow \quad \bar{\bf 4}_{1, 1, 1} \oplus \bar{\bf 4}_{-1, -1, 1} \cr 
({\bf 1}, {\bf 2}, {\bf 4})_{-1} \quad & \rightarrow \quad {\bf 4}_{1, -1, -1} \oplus {\bf 4}_{-1, 1, -1}  \,.
\ea
\ee
The twist (\ref{TwistK3})
results in two positive chirality scalar supercharges which transform as 
\be
SU(4)_R \times SO(1,1)_L \times U(1)_{\rm twist}: \qquad {\bf 4}_{1, 0} \oplus \bar{\bf 4}_{1, 0} \,.
\ee
The resulting theory is an $N=(0,8)$ supersymmetric theory with the following massless spectrum: The scalars remain in a ${\bf 6}$ of the R-symmetry $SU(4)_R$ and are trivially charged under $U(1)_{\rm twist}$, whereby their multiplicity is counted by $h^{0}(C)=1$. The fermions decompose as 
\be
\ba
SO(1,3)_L \times SU(4)_R \times U(1)_D \quad &\rightarrow \quad SO(1,1) \times U(1)_{\rm twist} \cr 
\Psi: \quad ({\bf 2}, {\bf 1}, {\bf 4})_{1} \quad & \rightarrow \quad {\bf 4}_{1, 0} \oplus {\bf 4}_{-1, -1} \cr 
\widetilde\Psi:\quad ({\bf 1}, {\bf 2}, \bar{\bf 4})_{-1} \quad & \rightarrow \quad \bar{\bf 4}_{1, 0} \oplus \bar{\bf 4}_{-1, 1}  \,. \cr 
\ea
\ee
These are counted by $h^0(C) = 1$ for $q_{\rm twist} =0$, and by $h^{0,1}(C)=0$ and $h^{1,0}(C) = 0$ for $q_{\rm twist} = \pm 1$. 
Finally, the gauge field decomposes as in (\ref{eqn:4dvecdecomp}). This gives rise to scalar fields $a_z$ and $\bar a_{\bar z}$ of twist charge $q_{\rm twist} = 0$, i.e. transforming as ${\bf 1}_{0}$, of multiplicity 1. The twist charge can either be inferred directly from the supersymmetry variation, or, more heuristically perhaps, by noting that 
$U(1)_{\rm twist}$ corresponds to the difference of the individual $U(1)^{\rm twist}_{C}$ and  $U(1)^{\rm twist}_{D}$ which would arise in presence of a $U(1)_R$ symmetry on higher dimensional base spaces; since the Wilson line scalars $a_z$ and $\bar a_{\bar z}$ have equal charge under these two twisted $U(1)$ symmetries in the higher dimensional examples, this explains $q_{\rm twist}=0$. 
The remaining gauge field components $v_{\pm}$ have zero multiplicity, again  by supersymmetry as in the lower-dimensional examples. 
Finally, the $3$--$7$ sector gives rise to the remaining $8 c_1(B_1) \cdot C=16$ complex Weyl fermions of negative chirality, where we are using that $C=B_1 = \mathbb P^1$. 
The complete spectrum is thus:
\begin{equation}\label{eqn:D3K3}
  \begin{array}{c|c|c}
 \text{Bosons} & \text{Fermions} & \text{Zero-modes} \cr\hline
 {\bf 6}_0 , \  2 \times {\bf 1}_{0} & {\bf 4}_{1}  \oplus \overline{\bf 4}_{1} & h^{0}({C}) = 1 \cr
&  {\bf 4}_{1-}  \oplus \overline{\bf 4}_{-1}  & h^{1}({C}) = 0 \cr 
 & {\bf 1}_{-1} & 8 c_1(B_1) \cdot C = 16
  \end{array}
\end{equation}
The first row provides the field content of a  2d $(0,8)$ hypermultiplet and the $16$ complex Weyl fermions from the $3$--$7$ sector represent purely fermionic $N=(0,8)$ multiplets.

\subsection{...from M5-branes}

When considering an M5-brane wrapping a K3 surface there are five transverse
non-compact directions, which are rotated by the $SO(5)$ R-symmetry of the
$(0,2)$ theory on the worldvolume of the M5. It is only the 6d Lorentz
symmetry along the M5-brane which decomposes since the 6d spacetime has the form
$\mathbb{R}^{1,1} \times K3$,
\begin{equation}
  \begin{aligned}
    SO(1,5)_L &\ \rightarrow \ SO(1,1)_L \times SU(2)_l \cr
    {\bf 4} &\ \rightarrow\  {\bf 2}_1 \oplus 2 \times {\bf 1}_{-1} \cr
    {\bf 10} &\ \rightarrow\  {\bf 3}_2 \oplus 3 \times {\bf 1}_{-2} \oplus 2 \times
    {\bf 2}_0 \,.
  \end{aligned}
\end{equation}
The full field content of the abelian tensor multiplet can then be decomposed
under this holonomy reduction, and one finds:
\begin{equation}
  \begin{aligned}
    SO(1,5) \times Sp(4) &\ \rightarrow \ SO(1,1)_L \times SU(2)_l \times Sp(4)
    \cr
H:    ({\bf \overline{10}, 1}) &\ \rightarrow \ ({\bf 3, 1})_{-2} \oplus 3 \times ({\bf
    1,1})_2 \oplus 2 \times ({\bf 2,1})_0 \cr
 \Phi:    ({\bf 1, 5}) &\ \rightarrow\  ({\bf 1, 5})_0 \cr
 Q, \rho:   ({\bf \overline{4}, 4}) &\ \rightarrow\  ({\bf 2,4})_{-1} \oplus 2 \times ({\bf
    1,4})_1 \,.
  \end{aligned}
\end{equation}
The theory has eight right-moving supercharges which are scalars under
the internal $SU(2)_l$ holonomy. 

The counting of zero-modes is determined entirely by the internal $SU(2)_l$ representation under which
the twisted fields transform:
\begin{equation}\label{eqn:M5K3}
  \begin{array}{c|c|c|c}
    SU(2)_l & \text{Bosons} & \text{Fermions} & \text{Zero-modes} \cr\hline
    {\bf 1} & 3 \times ({\bf 1,1})_2 \,\,,\, ({\bf 1, 5})_0 & 2 \times ({\bf
    1,4})_1 & h^{0,0}(\widehat{C}) = 1 \cr
    {\bf 2} & 2 \times ({\bf 2,1})_0 & ({\bf 2,4})_{-1} & h^{1,0}(\widehat{C})
    = 0 \cr
    {\bf 3} & ({\bf 3,1})_{-2} & & h^{1,1}(\widehat{C}) - 1 = 19 
  \end{array}
\end{equation}
Note that the three $({\bf 1,1})_2$ scalar bosons arising from the decomposition
of the self-dual two-form are purely right-moving because of the chirality of
$B$. To complete the top row of (\ref{eqn:M5K3}) into a full $(0,8)$
hypermultiplet, where all of the eight real scalar degrees of freedom are both
left- and right-moving, it is necessary to combine them with three of the
left-moving bosonic degrees of freedom from the $({\bf 3,1})_{-2}$ zero-modes.
In addition to the single $(0,8)$ hypermultiplet this leaves us, after
fermionization, with $16$ left-moving complex Weyl fermions in the 2d theory,
which is consistent with the heterotic string \cite{Witten:1995ex}.

From the field content described in table (\ref{eqn:M5K3}) one can compute the
left- and right-moving central charges. A complete $(0,8)$ hypermultiplet
contains two $(0,4)$ hypermultiplets, whose central charge we know from appendix
\ref{app:SUSYS}. Each hypermultiplet thus contributes $(c_R, c_L) =
(12, 8)$. The
$19-3$ remaining left-moving Weyl fermions each contribute $+1$
to $c_L$. As such the central charges are 
\begin{equation}
  c_L = 24 \,, \qquad c_R = 12 \,.
\end{equation}
The resulting gravitational anomaly 
\begin{equation}
  c_L - c_R = 12 
\end{equation}
along the string will be discussed from the point of view of anomaly inflow in the next section.


\subsection{...from M2-branes}

For an M2-brane wrapping the base inside the K3 surface\footnote{M2-branes on general K3-surfaces have been discussed in \cite{Okazaki:2014sga,Okazaki:2015pfa}.}  there are six
non-compact transverse directions, and two real directions related to how the
curve can move inside the K3.  As such the R-symmetry of the M2-brane theory
is broken as
\begin{equation}
  \begin{aligned}
    SO(8)_R &\ \rightarrow\  SO(6)_T \times U(1)_R \cr
  {\bf 8}_v &\ \rightarrow\  {\bf 6}_0 \oplus {\bf 1}_{\pm 2} \cr
  {\bf 8}_c &\ \rightarrow\ {\bf 4}_1 \oplus {\bf \overline{4}}_{-1} \cr
  {\bf 8}_s &\ \rightarrow\ {\bf 4}_{-1} \oplus {\bf \overline{4}}_1 \,.
  \end{aligned}
\end{equation}
Combining this decomposition with the field content of the $3d$ theory one
finds
\begin{equation}
  \begin{aligned}
    SO(1,2) \times SO(8) &\ \rightarrow\  SO(6)_T \times U(1)_L \times U(1)_R \cr
  \epsilon:    ({\bf 2, 8}_s) &\ \rightarrow\  {\bf 4}_{1,-1} \oplus {\bf 4}_{-1,-1} \oplus {\bf
      \overline{4}}_{1,1} \oplus {\bf \overline{4}}_{-1,1} \cr
  \rho:     ({\bf 2, 8}_c) &\ \rightarrow\  {\bf 4}_{1,1} \oplus {\bf 4}_{-1,1} \oplus {\bf
      \overline{4}}_{1,-1} \oplus {\bf \overline{4}}_{-1,-1} \cr
   \Phi:    ({\bf 1, 8}_v) &\ \rightarrow\  {\bf 6}_{0,0} \oplus {\bf 1}_{0,\pm 2} \,.
  \end{aligned}
\end{equation}
With the obvious additive twist
\be
T_{\rm twist} = \frac{1}{2} (T_L + T_R)
\ee
between the two $U(1)$s one can see that the
supersymmetry parameters in the $({\bf 2, 8}_s)$ contain eight scalar modes with respect to $U(1)_{\rm twist}$, as required for an $N=8$ SQM. The cohomology groups counting the zero-modes
follow from the $U(1)_\text{twist}$ charges of the fields:

\begin{equation}
  \begin{array}{c|c|c|c}
    U(1)_{\text{twist}} & \text{Bosons} & \text{Fermions} & 
    \text{Zero-modes} \cr\hline
    0 & {\bf 6}_0 & {\bf 4}_0 \,\,,\, {\bf \overline{4}}_0 & h^{0,0}(C) = 1 
    \cr
    +1 & {\bf 1}_{1} & {\bf 4}_1 & h^{1,0}(C) = 0   \cr
    -1 & {\bf 1}_{-1} & {\bf \overline{4}}_{-1} & h^{0,1}(C) = 0 
  \end{array}
\end{equation}

The scalars and fermions in the first line assemble into a $(6,8,2)$ multiplet of $N=8$ SQM.
This multiplet is matched with the 2d $N=(0,8)$ hypermultiplet along the string obtained from the dual D3-brane configuration by automorphic duality:
After circle reducing the 2d theory the two Wilson line degrees of freedom, $a_z$ and $\bar a_{\bar z}$, from decomposition of the $N=4$ SYM gauge field can be dualised into the auxiliary fields of the    $(6,8,2)$ multiplet.



\begin{thebibliography}{10}

\bibitem{Bershadsky:1995qy}
M.~Bershadsky, C.~Vafa and V.~Sadov, \emph{{D-branes and topological field
  theories}}, \href{http://dx.doi.org/10.1016/0550-3213(96)00026-0}{\emph{Nucl.
  Phys.} {\bf B463} (1996) 420--434},
  [\href{https://arxiv.org/abs/hep-th/9511222}{{\tt hep-th/9511222}}].

\bibitem{Benini:2013cda}
F.~Benini and N.~Bobev, \emph{{Two-dimensional SCFTs from wrapped branes and
  c-extremization}},
  \href{http://dx.doi.org/10.1007/JHEP06(2013)005}{\emph{JHEP} {\bf 06} (2013)
  005}, [\href{https://arxiv.org/abs/1302.4451}{{\tt 1302.4451}}].

\bibitem{Benini:2015bwz}
F.~Benini, N.~Bobev and P.~M. Crichigno, \emph{{Two-dimensional SCFTs from
  D3-branes}}, \href{http://dx.doi.org/10.1007/JHEP07(2016)020}{\emph{JHEP}
  {\bf 07} (2016) 020}, [\href{https://arxiv.org/abs/1511.09462}{{\tt
  1511.09462}}].

\bibitem{Vafa:1994tf}
C.~Vafa and E.~Witten, \emph{{A Strong coupling test of S duality}},
  \href{http://dx.doi.org/10.1016/0550-3213(94)90097-3}{\emph{Nucl. Phys.} {\bf
  B431} (1994) 3--77}, [\href{https://arxiv.org/abs/hep-th/9408074}{{\tt
  hep-th/9408074}}].

\bibitem{Kapustin:2006pk}
A.~Kapustin and E.~Witten, \emph{{Electric-Magnetic Duality And The Geometric
  Langlands Program}},
  \href{http://dx.doi.org/10.4310/CNTP.2007.v1.n1.a1}{\emph{Commun. Num. Theor.
  Phys.} {\bf 1} (2007) 1--236},
  [\href{https://arxiv.org/abs/hep-th/0604151}{{\tt hep-th/0604151}}].

\bibitem{Yamron:1988qc}
J.~P. Yamron, \emph{{Topological Actions From Twisted Supersymmetric
  Theories}}, \href{http://dx.doi.org/10.1016/0370-2693(88)91769-8}{\emph{Phys.
  Lett.} {\bf B213} (1988) 325--330}.

\bibitem{Martucci:2014ema}
L.~Martucci, \emph{{Topological duality twist and brane instantons in
  F-theory}}, \href{http://dx.doi.org/10.1007/JHEP06(2014)180}{\emph{JHEP} {\bf
  06} (2014) 180}, [\href{https://arxiv.org/abs/1403.2530}{{\tt 1403.2530}}].

\bibitem{Assel:2016wcr}
B.~Assel and S.~Schafer-Nameki, \emph{{Six-dimensional Origin of
  $\mathcal{N}=4$ SYM with Duality Defects}},
  \href{https://arxiv.org/abs/1610.03663}{{\tt 1610.03663}}.

\bibitem{Intriligator:1998ig}
K.~A. Intriligator, \emph{{Bonus symmetries of N=4 superYang-Mills correlation
  functions via AdS duality}},
  \href{http://dx.doi.org/10.1016/S0550-3213(99)00242-4}{\emph{Nucl. Phys.}
  {\bf B551} (1999) 575--600},
  [\href{https://arxiv.org/abs/hep-th/9811047}{{\tt hep-th/9811047}}].

\bibitem{Intriligator:1999ff}
K.~A. Intriligator and W.~Skiba, \emph{{Bonus symmetry and the operator product
  expansion of N=4 SuperYang-Mills}},
  \href{http://dx.doi.org/10.1016/S0550-3213(99)00430-7}{\emph{Nucl. Phys.}
  {\bf B559} (1999) 165--183},
  [\href{https://arxiv.org/abs/hep-th/9905020}{{\tt hep-th/9905020}}].

\bibitem{Vafa:1996xn}
C.~Vafa, \emph{{Evidence for F theory}},
  \href{http://dx.doi.org/10.1016/0550-3213(96)00172-1}{\emph{Nucl. Phys.} {\bf
  B469} (1996) 403--418}, [\href{https://arxiv.org/abs/hep-th/9602022}{{\tt
  hep-th/9602022}}].

\bibitem{Morrison:1996na}
D.~R. Morrison and C.~Vafa, \emph{{Compactifications of F theory on Calabi-Yau
  threefolds. 1}},
  \href{http://dx.doi.org/10.1016/0550-3213(96)00242-8}{\emph{Nucl. Phys.} {\bf
  B473} (1996) 74--92}, [\href{https://arxiv.org/abs/hep-th/9602114}{{\tt
  hep-th/9602114}}].

\bibitem{Morrison:1996pp}
D.~R. Morrison and C.~Vafa, \emph{{Compactifications of F theory on Calabi-Yau
  threefolds. 2.}},
  \href{http://dx.doi.org/10.1016/0550-3213(96)00369-0}{\emph{Nucl. Phys.} {\bf
  B476} (1996) 437--469}, [\href{https://arxiv.org/abs/hep-th/9603161}{{\tt
  hep-th/9603161}}].

\bibitem{Franco:2015tna}
S.~Franco, D.~Ghim, S.~Lee, R.-K. Seong and D.~Yokoyama, \emph{{2d (0,2) Quiver
  Gauge Theories and D-Branes}},
  \href{http://dx.doi.org/10.1007/JHEP09(2015)072}{\emph{JHEP} {\bf 09} (2015)
  072}, [\href{https://arxiv.org/abs/1506.03818}{{\tt 1506.03818}}].

\bibitem{Franco:2015tya}
S.~Franco, S.~Lee and R.-K. Seong, \emph{{Brane Brick Models, Toric Calabi-Yau
  4-Folds and 2d (0,2) Quivers}},
  \href{http://dx.doi.org/10.1007/JHEP02(2016)047}{\emph{JHEP} {\bf 02} (2016)
  047}, [\href{https://arxiv.org/abs/1510.01744}{{\tt 1510.01744}}].

\bibitem{Schafer-Nameki:2016cfr}
S.~Schafer-Nameki and T.~Weigand, \emph{{F-theory and 2d $(0, 2)$ theories}},
  \href{http://dx.doi.org/10.1007/JHEP05(2016)059}{\emph{JHEP} {\bf 05} (2016)
  059}, [\href{https://arxiv.org/abs/1601.02015}{{\tt 1601.02015}}].

\bibitem{Franco:2016nwv}
S.~Franco, S.~Lee and R.-K. Seong, \emph{{Brane brick models and 2d (0, 2)
  triality}}, \href{http://dx.doi.org/10.1007/JHEP05(2016)020}{\emph{JHEP} {\bf
  05} (2016) 020}, [\href{https://arxiv.org/abs/1602.01834}{{\tt 1602.01834}}].

\bibitem{Apruzzi:2016iac}
F.~Apruzzi, F.~Hassler, J.~J. Heckman and I.~V. Melnikov, \emph{{UV Completions
  for Non-Critical Strings}},  \href{https://arxiv.org/abs/1602.04221}{{\tt
  1602.04221}}.

\bibitem{Franco:2016qxh}
S.~Franco, S.~Lee, R.-K. Seong and C.~Vafa, \emph{{Brane Brick Models in the
  Mirror}},  \href{https://arxiv.org/abs/1609.01723}{{\tt 1609.01723}}.

\bibitem{Apruzzi:2016nfr}
F.~Apruzzi, F.~Hassler, J.~J. Heckman and I.~V. Melnikov, \emph{{From 6D SCFTs
  to Dynamic GLSMs}},  \href{https://arxiv.org/abs/1610.00718}{{\tt
  1610.00718}}.

\bibitem{Heckman:2013pva}
J.~J. Heckman, D.~R. Morrison and C.~Vafa, \emph{{On the Classification of 6D
  SCFTs and Generalized ADE Orbifolds}},
  \href{http://dx.doi.org/10.1007/JHEP06(2015)017,
  10.1007/JHEP05(2014)028}{\emph{JHEP} {\bf 05} (2014) 028},
  [\href{https://arxiv.org/abs/1312.5746}{{\tt 1312.5746}}].

\bibitem{Ganor:1996mu}
O.~J. Ganor and A.~Hanany, \emph{{Small E(8) instantons and tensionless
  noncritical strings}},
  \href{http://dx.doi.org/10.1016/0550-3213(96)00243-X}{\emph{Nucl. Phys.} {\bf
  B474} (1996) 122--140}, [\href{https://arxiv.org/abs/hep-th/9602120}{{\tt
  hep-th/9602120}}].

\bibitem{Seiberg:1996vs}
N.~Seiberg and E.~Witten, \emph{{Comments on string dynamics in
  six-dimensions}},
  \href{http://dx.doi.org/10.1016/0550-3213(96)00189-7}{\emph{Nucl. Phys.} {\bf
  B471} (1996) 121--134}, [\href{https://arxiv.org/abs/hep-th/9603003}{{\tt
  hep-th/9603003}}].

\bibitem{Witten:1996qb}
E.~Witten, \emph{{Phase transitions in M theory and F theory}},
  \href{http://dx.doi.org/10.1016/0550-3213(96)00212-X}{\emph{Nucl. Phys.} {\bf
  B471} (1996) 195--216}, [\href{https://arxiv.org/abs/hep-th/9603150}{{\tt
  hep-th/9603150}}].

\bibitem{Klemm:1996hh}
A.~Klemm, P.~Mayr and C.~Vafa, \emph{{BPS states of exceptional noncritical
  strings}},  \href{https://arxiv.org/abs/hep-th/9607139}{{\tt
  hep-th/9607139}}.

\bibitem{Minahan:1998vr}
J.~A. Minahan, D.~Nemeschansky, C.~Vafa and N.~P. Warner, \emph{{E strings and
  N=4 topological Yang-Mills theories}},
  \href{http://dx.doi.org/10.1016/S0550-3213(98)00426-X}{\emph{Nucl. Phys.}
  {\bf B527} (1998) 581--623},
  [\href{https://arxiv.org/abs/hep-th/9802168}{{\tt hep-th/9802168}}].

\bibitem{Haghighat:2013gba}
B.~Haghighat, A.~Iqbal, C.~Kozaz, G.~Lockhart and C.~Vafa, \emph{{M-Strings}},
  \href{http://dx.doi.org/10.1007/s00220-014-2139-1}{\emph{Commun. Math. Phys.}
  {\bf 334} (2015) 779--842}, [\href{https://arxiv.org/abs/1305.6322}{{\tt
  1305.6322}}].

\bibitem{Kim:2014dza}
J.~Kim, S.~Kim, K.~Lee, J.~Park and C.~Vafa, \emph{{Elliptic Genus of
  E-strings}},  \href{https://arxiv.org/abs/1411.2324}{{\tt 1411.2324}}.

\bibitem{Haghighat:2014vxa}
B.~Haghighat, A.~Klemm, G.~Lockhart and C.~Vafa, \emph{{Strings of Minimal 6d
  SCFTs}}, \href{http://dx.doi.org/10.1002/prop.201500014}{\emph{Fortsch.
  Phys.} {\bf 63} (2015) 294--322},
  [\href{https://arxiv.org/abs/1412.3152}{{\tt 1412.3152}}].

\bibitem{Gadde:2015tra}
A.~Gadde, B.~Haghighat, J.~Kim, S.~Kim, G.~Lockhart and C.~Vafa, \emph{{6d
  String Chains}},  \href{https://arxiv.org/abs/1504.04614}{{\tt 1504.04614}}.

\bibitem{Haghighat:2015ega}
B.~Haghighat, S.~Murthy, C.~Vafa and S.~Vandoren, \emph{{F-Theory, Spinning
  Black Holes and Multi-string Branches}},
  \href{http://dx.doi.org/10.1007/JHEP01(2016)009}{\emph{JHEP} {\bf 01} (2016)
  009}, [\href{https://arxiv.org/abs/1509.00455}{{\tt 1509.00455}}].

\bibitem{DelZotto:2016pvm}
M.~Del~Zotto and G.~Lockhart, \emph{{On Exceptional Instanton Strings}},
  \href{https://arxiv.org/abs/1609.00310}{{\tt 1609.00310}}.

\bibitem{Mayr:1996sh}
P.~Mayr, \emph{{Mirror symmetry, N=1 superpotentials and tensionless strings on
  Calabi-Yau four folds}},
  \href{http://dx.doi.org/10.1016/S0550-3213(97)00196-X}{\emph{Nucl. Phys.}
  {\bf B494} (1997) 489--545},
  [\href{https://arxiv.org/abs/hep-th/9610162}{{\tt hep-th/9610162}}].

\bibitem{Okazaki:2014sga}
T.~Okazaki, \emph{{Membrane Quantum Mechanics}},
  \href{http://dx.doi.org/10.1016/j.nuclphysb.2014.11.024}{\emph{Nucl. Phys.}
  {\bf B890} (2014) 400--441}, [\href{https://arxiv.org/abs/1410.8180}{{\tt
  1410.8180}}].

\bibitem{Okazaki:2015pfa}
T.~Okazaki, \emph{{Superconformal Quantum Mechanics from M2-branes}}.
\newblock PhD thesis, Caltech, 2015.
\newblock \href{https://arxiv.org/abs/1503.03906}{{\tt 1503.03906}}.

\bibitem{Gates:2002bc}
S.~J. Gates, Jr., W.~D. Linch, III and J.~Phillips, \emph{{When superspace is
  not enough}},  \href{https://arxiv.org/abs/hep-th/0211034}{{\tt
  hep-th/0211034}}.

\bibitem{Bellucci:2005xn}
S.~Bellucci, S.~Krivonos, A.~Marrani and E.~Orazi, \emph{{'Root' action for N=4
  supersymmetric mechanics theories}},
  \href{http://dx.doi.org/10.1103/PhysRevD.73.025011}{\emph{Phys. Rev.} {\bf
  D73} (2006) 025011}, [\href{https://arxiv.org/abs/hep-th/0511249}{{\tt
  hep-th/0511249}}].

\bibitem{Gustavsson:2007vu}
A.~Gustavsson, \emph{{Algebraic structures on parallel M2-branes}},
  \href{http://dx.doi.org/10.1016/j.nuclphysb.2008.11.014}{\emph{Nucl. Phys.}
  {\bf B811} (2009) 66--76}, [\href{https://arxiv.org/abs/0709.1260}{{\tt
  0709.1260}}].

\bibitem{Bagger:2007jr}
J.~Bagger and N.~Lambert, \emph{{Gauge symmetry and supersymmetry of multiple
  M2-branes}}, \href{http://dx.doi.org/10.1103/PhysRevD.77.065008}{\emph{Phys.
  Rev.} {\bf D77} (2008) 065008}, [\href{https://arxiv.org/abs/0711.0955}{{\tt
  0711.0955}}].

\bibitem{Gaberdiel:1997ud}
M.~R. Gaberdiel and B.~Zwiebach, \emph{{Exceptional groups from open strings}},
  \href{http://dx.doi.org/10.1016/S0550-3213(97)00841-9}{\emph{Nucl. Phys.}
  {\bf B518} (1998) 151--172},
  [\href{https://arxiv.org/abs/hep-th/9709013}{{\tt hep-th/9709013}}].

\bibitem{DeWolfe:1998zf}
O.~DeWolfe and B.~Zwiebach, \emph{{String junctions for arbitrary Lie algebra
  representations}},
  \href{http://dx.doi.org/10.1016/S0550-3213(98)00743-3}{\emph{Nucl. Phys.}
  {\bf B541} (1999) 509--565},
  [\href{https://arxiv.org/abs/hep-th/9804210}{{\tt hep-th/9804210}}].

\bibitem{Mikhailov:1998bx}
A.~Mikhailov, N.~Nekrasov and S.~Sethi, \emph{{Geometric realizations of BPS
  states in N=2 theories}},
  \href{http://dx.doi.org/10.1016/S0550-3213(98)80001-1}{\emph{Nucl. Phys.}
  {\bf B531} (1998) 345--362},
  [\href{https://arxiv.org/abs/hep-th/9803142}{{\tt hep-th/9803142}}].

\bibitem{Grassi:2014ffa}
A.~Grassi, J.~Halverson and J.~L. Shaneson, \emph{{Geometry and Topology of
  String Junctions}},  \href{https://arxiv.org/abs/1410.6817}{{\tt 1410.6817}}.

\bibitem{Grassi:2016bhs}
A.~Grassi, J.~Halverson, F.~Ruehle and J.~L. Shaneson, \emph{{Dualities of
  Deformed $\mathcal{N}=2$ SCFTs from Link Monodromy on D3-brane States}},
  \href{https://arxiv.org/abs/1611.01154}{{\tt 1611.01154}}.

\bibitem{MINI}
C.~Lawrie, S.~Schafer-Nameki and T.~Weigand, \emph{{The gravitational sector of
  2d (0, 2) F-theory vacua}},
  \href{http://dx.doi.org/10.1007/JHEP05(2017)103}{\emph{JHEP} {\bf 05} (2017)
  103}, [\href{https://arxiv.org/abs/1612.06393}{{\tt 1612.06393}}].

\bibitem{Benini:2013xpa}
F.~Benini, R.~Eager, K.~Hori and Y.~Tachikawa, \emph{{Elliptic Genera of 2d
  ${\mathcal{N}}$ = 2 Gauge Theories}},
  \href{http://dx.doi.org/10.1007/s00220-014-2210-y}{\emph{Commun. Math. Phys.}
  {\bf 333} (2015) 1241--1286}, [\href{https://arxiv.org/abs/1308.4896}{{\tt
  1308.4896}}].

\bibitem{Bianchi:2011qh}
M.~Bianchi, A.~Collinucci and L.~Martucci, \emph{{Magnetized E3-brane
  instantons in F-theory}},
  \href{http://dx.doi.org/10.1007/JHEP12(2011)045}{\emph{JHEP} {\bf 12} (2011)
  045}, [\href{https://arxiv.org/abs/1107.3732}{{\tt 1107.3732}}].

\bibitem{Gauntlett:2001qs}
J.~P. Gauntlett, N.~Kim, S.~Pakis and D.~Waldram, \emph{{Membranes wrapped on
  holomorphic curves}},
  \href{http://dx.doi.org/10.1103/PhysRevD.65.026003}{\emph{Phys. Rev.} {\bf
  D65} (2002) 026003}, [\href{https://arxiv.org/abs/hep-th/0105250}{{\tt
  hep-th/0105250}}].

\bibitem{Bershadsky:1995vm}
M.~Bershadsky, A.~Johansen, V.~Sadov and C.~Vafa, \emph{{Topological reduction
  of 4-d SYM to 2-d sigma models}},
  \href{http://dx.doi.org/10.1016/0550-3213(95)00242-K}{\emph{Nucl. Phys.} {\bf
  B448} (1995) 166--186}, [\href{https://arxiv.org/abs/hep-th/9501096}{{\tt
  hep-th/9501096}}].

\bibitem{Witten:1995zh}
E.~Witten, \emph{{Some comments on string dynamics}},  in \emph{{Future
  perspectives in string theory. Proceedings, Conference, Strings'95, Los
  Angeles, USA, March 13-18, 1995}}, 1995.
\newblock \href{https://arxiv.org/abs/hep-th/9507121}{{\tt hep-th/9507121}}.

\bibitem{Witten:2007ct}
E.~Witten, \emph{{Conformal Field Theory In Four And Six Dimensions}},  in
  \emph{{Topology, geometry and quantum field theory. Proceedings, Symposium in
  the honour of the 60th birthday of Graeme Segal, Oxford, UK, June 24-29,
  2002}}, 2007.
\newblock \href{https://arxiv.org/abs/0712.0157}{{\tt 0712.0157}}.

\bibitem{Haupt:2008nu}
A.~S. Haupt, A.~Lukas and K.~S. Stelle, \emph{{M-theory on Calabi-Yau
  Five-Folds}},
  \href{http://dx.doi.org/10.1088/1126-6708/2009/05/069}{\emph{JHEP} {\bf 05}
  (2009) 069}, [\href{https://arxiv.org/abs/0810.2685}{{\tt 0810.2685}}].

\bibitem{Witten:1993yc}
E.~Witten, \emph{{Phases of N=2 theories in two-dimensions}},
  \href{http://dx.doi.org/10.1016/0550-3213(93)90033-L}{\emph{Nucl. Phys.} {\bf
  B403} (1993) 159--222}, [\href{https://arxiv.org/abs/hep-th/9301042}{{\tt
  hep-th/9301042}}].

\bibitem{Ohmori:2014kda}
K.~Ohmori, H.~Shimizu, Y.~Tachikawa and K.~Yonekura, \emph{{Anomaly polynomial
  of general 6d SCFTs}},
  \href{http://dx.doi.org/10.1093/ptep/ptu140}{\emph{PTEP} {\bf 2014} (2014)
  103B07}, [\href{https://arxiv.org/abs/1408.5572}{{\tt 1408.5572}}].

\bibitem{Garcia-Etxebarria:2015wns}
I.~García-Etxebarria and D.~Regalado, \emph{{$ \mathcal{N}=3 $ four dimensional
  field theories}},
  \href{http://dx.doi.org/10.1007/JHEP03(2016)083}{\emph{JHEP} {\bf 03} (2016)
  083}, [\href{https://arxiv.org/abs/1512.06434}{{\tt 1512.06434}}].

\bibitem{Sen:1997gv}
A.~Sen, \emph{{Orientifold limit of F theory vacua}},
  \href{http://dx.doi.org/10.1103/PhysRevD.55.R7345}{\emph{Phys. Rev.} {\bf
  D55} (1997) R7345--R7349}, [\href{https://arxiv.org/abs/hep-th/9702165}{{\tt
  hep-th/9702165}}].

\bibitem{Cvetic:2011gp}
M.~Cvetic, I.~Garcia~Etxebarria and J.~Halverson, \emph{{Three Looks at
  Instantons in F-theory -- New Insights from Anomaly Inflow, String Junctions
  and Heterotic Duality}},
  \href{http://dx.doi.org/10.1007/JHEP11(2011)101}{\emph{JHEP} {\bf 11} (2011)
  101}, [\href{https://arxiv.org/abs/1107.2388}{{\tt 1107.2388}}].

\bibitem{Halverson:2016vwx}
J.~Halverson, \emph{{Strong Coupling in F-theory and Geometrically
  Non-Higgsable Seven-branes}},  \href{https://arxiv.org/abs/1603.01639}{{\tt
  1603.01639}}.

\bibitem{Grimm:2010ez}
T.~W. Grimm and T.~Weigand, \emph{{On Abelian Gauge Symmetries and Proton Decay
  in Global F-theory GUTs}},
  \href{http://dx.doi.org/10.1103/PhysRevD.82.086009}{\emph{Phys. Rev.} {\bf
  D82} (2010) 086009}, [\href{https://arxiv.org/abs/1006.0226}{{\tt
  1006.0226}}].

\bibitem{Harvey:2007ab}
J.~A. Harvey and A.~B. Royston, \emph{{Localized modes at a D-brane-O-plane
  intersection and heterotic Alice atrings}},
  \href{http://dx.doi.org/10.1088/1126-6708/2008/04/018}{\emph{JHEP} {\bf 04}
  (2008) 018}, [\href{https://arxiv.org/abs/0709.1482}{{\tt 0709.1482}}].

\bibitem{Blumenhagen:2008zz}
R.~Blumenhagen, V.~Braun, T.~W. Grimm and T.~Weigand, \emph{{GUTs in Type IIB
  Orientifold Compactifications}},
  \href{http://dx.doi.org/10.1016/j.nuclphysb.2009.02.011}{\emph{Nucl. Phys.}
  {\bf B815} (2009) 1--94}, [\href{https://arxiv.org/abs/0811.2936}{{\tt
  0811.2936}}].

\bibitem{Maldacena:1997de}
J.~M. Maldacena, A.~Strominger and E.~Witten, \emph{{Black hole entropy in M
  theory}}, \href{http://dx.doi.org/10.1088/1126-6708/1997/12/002}{\emph{JHEP}
  {\bf 12} (1997) 002}, [\href{https://arxiv.org/abs/hep-th/9711053}{{\tt
  hep-th/9711053}}].

\bibitem{Vafa:1997gr}
C.~Vafa, \emph{{Black holes and Calabi-Yau threefolds}}, {\emph{Adv. Theor.
  Math. Phys.} {\bf 2} (1998) 207--218},
  [\href{https://arxiv.org/abs/hep-th/9711067}{{\tt hep-th/9711067}}].

\bibitem{Minasian:1999qn}
R.~Minasian, G.~W. Moore and D.~Tsimpis, \emph{{Calabi-Yau black holes and
  (0,4) sigma models}}, {\emph{Commun. Math. Phys.} {\bf 209} (2000) 325--352},
  [\href{https://arxiv.org/abs/hep-th/9904217}{{\tt hep-th/9904217}}].

\bibitem{Esole:2011sm}
M.~Esole and S.-T. Yau, \emph{{Small resolutions of SU(5)-models in F-theory}},
  \href{http://dx.doi.org/10.4310/ATMP.2013.v17.n6.a1}{\emph{Adv. Theor. Math.
  Phys.} {\bf 17} (2013) 1195--1253},
  [\href{https://arxiv.org/abs/1107.0733}{{\tt 1107.0733}}].

\bibitem{MS}
J.~Marsano and S.~Schafer-Nameki, \emph{{Yukawas, G-flux, and Spectral Covers
  from Resolved Calabi-Yau's}},
  \href{http://dx.doi.org/10.1007/JHEP11(2011)098}{\emph{JHEP} {\bf 1111}
  (2011) 098}, [\href{https://arxiv.org/abs/1108.1794}{{\tt 1108.1794}}].

\bibitem{Krause:2011xj}
S.~Krause, C.~Mayrhofer and T.~Weigand, \emph{{$G_4$ flux, chiral matter and
  singularity resolution in F-theory compactifications}},
  \href{http://dx.doi.org/10.1016/j.nuclphysb.2011.12.013}{\emph{Nucl. Phys.}
  {\bf B858} (2012) 1--47}, [\href{https://arxiv.org/abs/1109.3454}{{\tt
  1109.3454}}].

\bibitem{Lawrie:2012gg}
C.~Lawrie and S.~Schafer-Nameki, \emph{{The Tate Form on Steroids: Resolution
  and Higher Codimension Fibers}},
  \href{http://dx.doi.org/10.1007/JHEP04(2013)061}{\emph{JHEP} {\bf 04} (2013)
  061}, [\href{https://arxiv.org/abs/1212.2949}{{\tt 1212.2949}}].

\bibitem{Hayashi:2013lra}
H.~Hayashi, C.~Lawrie and S.~Schafer-Nameki, \emph{{Phases, Flops and F-theory:
  SU(5) Gauge Theories}},
  \href{http://dx.doi.org/10.1007/JHEP10(2013)046}{\emph{JHEP} {\bf 10} (2013)
  046}, [\href{https://arxiv.org/abs/1304.1678}{{\tt 1304.1678}}].

\bibitem{Hayashi:2014kca}
H.~Hayashi, C.~Lawrie, D.~R. Morrison and S.~Schafer-Nameki, \emph{{Box Graphs
  and Singular Fibers}},
  \href{http://dx.doi.org/10.1007/JHEP05(2014)048}{\emph{JHEP} {\bf 05} (2014)
  048}, [\href{https://arxiv.org/abs/1402.2653}{{\tt 1402.2653}}].

\bibitem{Morales:1998ux}
J.~F. Morales, C.~A. Scrucca and M.~Serone, \emph{{Anomalous couplings for
  D-branes and O-planes}},
  \href{http://dx.doi.org/10.1016/S0550-3213(99)00217-5}{\emph{Nucl. Phys.}
  {\bf B552} (1999) 291--315},
  [\href{https://arxiv.org/abs/hep-th/9812071}{{\tt hep-th/9812071}}].

\bibitem{Erler:1993zy}
J.~Erler, \emph{{Anomaly cancellation in six-dimensions}},
  \href{http://dx.doi.org/10.1063/1.530885}{\emph{J. Math. Phys.} {\bf 35}
  (1994) 1819--1833}, [\href{https://arxiv.org/abs/hep-th/9304104}{{\tt
  hep-th/9304104}}].

\bibitem{Berman:2004ew}
D.~S. Berman and J.~A. Harvey, \emph{{The Self-dual string and anomalies in the
  M5-brane}},
  \href{http://dx.doi.org/10.1088/1126-6708/2004/11/015}{\emph{JHEP} {\bf 11}
  (2004) 015}, [\href{https://arxiv.org/abs/hep-th/0408198}{{\tt
  hep-th/0408198}}].

\bibitem{Shimizu:2016lbw}
H.~Shimizu and Y.~Tachikawa, \emph{{Anomaly of strings of 6d
  $\mathcal{N}{=}(1,0)$ theories}},
  \href{https://arxiv.org/abs/1608.05894}{{\tt 1608.05894}}.

\bibitem{Kim:2016foj}
H.-C. Kim, S.~Kim and J.~Park, \emph{{6d strings from new chiral gauge
  theories}},  \href{https://arxiv.org/abs/1608.03919}{{\tt 1608.03919}}.

\bibitem{Sadov:1996zm}
V.~Sadov, \emph{{Generalized Green-Schwarz mechanism in F theory}},
  \href{http://dx.doi.org/10.1016/0370-2693(96)01134-3}{\emph{Phys. Lett.} {\bf
  B388} (1996) 45--50}, [\href{https://arxiv.org/abs/hep-th/9606008}{{\tt
  hep-th/9606008}}].

\bibitem{Grassi:2000we}
A.~Grassi and D.~R. Morrison, \emph{{Group representations and the Euler
  characteristic of elliptically fibered Calabi-Yau threefolds}},
  \href{https://arxiv.org/abs/math/0005196}{{\tt math/0005196}}.

\bibitem{Witten:1994tz}
E.~Witten, \emph{{Sigma models and the ADHM construction of instantons}},
  \href{http://dx.doi.org/10.1016/0393-0440(94)00047-8}{\emph{J. Geom. Phys.}
  {\bf 15} (1995) 215--226}, [\href{https://arxiv.org/abs/hep-th/9410052}{{\tt
  hep-th/9410052}}].

\bibitem{Tong:2014yna}
D.~Tong, \emph{{The holographic dual of $AdS_{3} \times S^{3} \times S^{3}
  \times S^{1}$}}, \href{http://dx.doi.org/10.1007/JHEP04(2014)193}{\emph{JHEP}
  {\bf 04} (2014) 193}, [\href{https://arxiv.org/abs/1402.5135}{{\tt
  1402.5135}}].

\bibitem{Putrov:2015jpa}
P.~Putrov, J.~Song and W.~Yan, \emph{{(0,4) dualities}},
  \href{http://dx.doi.org/10.1007/JHEP03(2016)185}{\emph{JHEP} {\bf 03} (2016)
  185}, [\href{https://arxiv.org/abs/1505.07110}{{\tt 1505.07110}}].

\bibitem{MR1078016}
R.~Miranda, \emph{The basic theory of elliptic surfaces}.
\newblock Dottorato di Ricerca in Matematica. [Doctorate in Mathematical
  Research]. ETS Editrice, Pisa, 1989.

\bibitem{MR2732092}
M.~Sch{\"u}tt and T.~Shioda, \emph{Elliptic surfaces},  in \emph{Algebraic
  geometry in {E}ast {A}sia---{S}eoul 2008}, vol.~60 of \emph{Adv. Stud. Pure
  Math.}, pp.~51--160.
\newblock Math. Soc. Japan, Tokyo, 2010.

\bibitem{MR2093043}
D.~Huybrechts, \emph{Complex geometry}.
\newblock Universitext. Springer-Verlag, Berlin, 2005.

\bibitem{Witten:1995ex}
E.~Witten, \emph{{String theory dynamics in various dimensions}},
  \href{http://dx.doi.org/10.1016/0550-3213(95)00158-O}{\emph{Nucl. Phys.} {\bf
  B443} (1995) 85--126}, [\href{https://arxiv.org/abs/hep-th/9503124}{{\tt
  hep-th/9503124}}].

\end{thebibliography}





\def\cprime{$'$} \def\cprime{$'$}
\providecommand{\href}[2]{#2}\begingroup\raggedright\endgroup

\end{document}